\documentclass[aps, prx, reprint, floatfix, superscriptaddress, nobibnotes]{revtex4-2}

\usepackage{float}
\usepackage[table]{xcolor}
\usepackage{graphicx}
\usepackage{braket}
\usepackage{mathtools}
\usepackage{siunitx}
\usepackage{hyperref}
\usepackage{cleveref}
\usepackage{bibunits}

\usepackage{graphicx}
\usepackage{epstopdf}
\usepackage{bm, amsmath, amssymb}
\usepackage{placeins}
\usepackage{xcolor}
\usepackage{siunitx}
\usepackage[T1]{fontenc}
\usepackage{array,booktabs}
\usepackage{comment}
\usepackage{multirow}
\usepackage{enumitem} 

\usepackage{hyperref}
\hypersetup{
    colorlinks=true,
    linkcolor=magenta,
    urlcolor=magenta,
    filecolor=magenta,   
    citecolor=blue
}

\defaultbibliography{lib_clean}
\defaultbibliographystyle{plain}
\crefname{table}{Table}{Tables}
\Crefname{table}{Table}{Tables}
\crefname{figure}{Fig.}{Figures}
\Crefname{figure}{Fig.}{Figures}
\crefname{section}{section}{sections}
\Crefname{section}{Section}{Sections}
\crefname{subsection}{section}{Sections}
\Crefname{subsection}{Section}{Sections}
\usepackage[caption=false]{subfig}

\usepackage{amssymb}
\usepackage{bm}

\makeatletter
\def\maketitle{
	\@author@finish
	\title@column\titleblock@produce
	\suppressfloats[t]}
\makeatother 

\usepackage{xpatch}

\makeatletter
\patchcmd{\@ssect@ltx}
{\addcontentsline{toc}{#1}{\protect\numberline{}#8}}
{}
{}
{}
\makeatother

\newcommand{\figwidth}{3.375in} 
\newcommand{\figwidthDouble}{7in}

\newcommand{\sX}{{\scriptscriptstyle X}}
\newcommand{\sZ}{{\scriptscriptstyle Z}}

\begin{document}

\title{Real-time quantum error correction beyond break-even}

\author{V. V. Sivak}
\email{vladimir.sivak@yale.edu}
\affiliation{Departments of Physics and Applied Physics, Yale University, New Haven, CT 06520, USA}
\affiliation{Yale Quantum Institute, Yale University, New Haven, CT 06511, USA}

\author{A.~Eickbusch}
\affiliation{Departments of Physics and Applied Physics, Yale University, New Haven, CT 06520, USA}
\affiliation{Yale Quantum Institute, Yale University, New Haven, CT 06511, USA}

\author{B.~Royer}
\affiliation{Departments of Physics and Applied Physics, Yale University, New Haven, CT 06520, USA}
\affiliation{Yale Quantum Institute, Yale University, New Haven, CT 06511, USA}
\affiliation{Institut Quantique and Département de Physique, Université de Sherbrooke, Québec, J1K 2R1, Canada}

\author{S.~Singh}
\affiliation{Departments of Physics and Applied Physics, Yale University, New Haven, CT 06520, USA}
\affiliation{Yale Quantum Institute, Yale University, New Haven, CT 06511, USA}

\author{I.~Tsioutsios}
\affiliation{Departments of Physics and Applied Physics, Yale University, New Haven, CT 06520, USA}
\affiliation{Yale Quantum Institute, Yale University, New Haven, CT 06511, USA}

\author{S.~Ganjam}
\affiliation{Departments of Physics and Applied Physics, Yale University, New Haven, CT 06520, USA}
\affiliation{Yale Quantum Institute, Yale University, New Haven, CT 06511, USA}

\author{A.~Miano}
\affiliation{Departments of Physics and Applied Physics, Yale University, New Haven, CT 06520, USA}
\affiliation{Yale Quantum Institute, Yale University, New Haven, CT 06511, USA}

\author{B.~L.~Brock}
\affiliation{Departments of Physics and Applied Physics, Yale University, New Haven, CT 06520, USA}
\affiliation{Yale Quantum Institute, Yale University, New Haven, CT 06511, USA}

\author{A.~Z.~Ding}
\affiliation{Departments of Physics and Applied Physics, Yale University, New Haven, CT 06520, USA}
\affiliation{Yale Quantum Institute, Yale University, New Haven, CT 06511, USA}

\author{L.~Frunzio}
\affiliation{Departments of Physics and Applied Physics, Yale University, New Haven, CT 06520, USA}
\affiliation{Yale Quantum Institute, Yale University, New Haven, CT 06511, USA}

\author{S.~M.~Girvin}
\affiliation{Departments of Physics and Applied Physics, Yale University, New Haven, CT 06520, USA}
\affiliation{Yale Quantum Institute, Yale University, New Haven, CT 06511, USA}

\author{R.~J.~Schoelkopf}
\affiliation{Departments of Physics and Applied Physics, Yale University, New Haven, CT 06520, USA}
\affiliation{Yale Quantum Institute, Yale University, New Haven, CT 06511, USA}

\author{M. H. Devoret}
\email{michel.devoret@yale.edu}
\affiliation{Departments of Physics and Applied Physics, Yale University, New Haven, CT 06520, USA}
\affiliation{Yale Quantum Institute, Yale University, New Haven, CT 06511, USA}

\begin{abstract}
The ambition of harnessing the quantum for computation is at odds with the fundamental phenomenon of decoherence.  The purpose of quantum error correction (QEC) is to counteract the natural tendency of a complex system to decohere. 
This cooperative process, which requires participation of multiple quantum and classical components, creates a special type of dissipation that removes the entropy caused by the errors faster than the rate at which these errors corrupt the stored quantum information.
Previous experimental attempts to engineer such a process faced an excessive generation of errors that overwhelmed the error-correcting capability of the process itself. Whether it is practically possible to utilize QEC for extending quantum coherence thus remains an open question. We answer it by demonstrating a fully stabilized and error-corrected logical qubit whose quantum coherence is significantly longer than that of all the imperfect quantum components involved in the QEC process, beating the best of them with a coherence gain of $G=2.27\pm0.07$. We achieve this  performance by  combining innovations in several domains including the fabrication of superconducting quantum circuits and model-free reinforcement learning.
\end{abstract}

\maketitle
\begin{bibunit}[apsrev_longbib]

Implementing a single correctable logical qubit requires a physical system with a large state space.
It should accommodate the code subspace and its redundant replicas where the logical information will be displaced without distortion when physical errors occur \cite{Knill1997}. 
This redundancy is inextricably associated with an additional operational cost of QEC,  known as the control overhead.  
In the search for an efficient way to alleviate its detrimental effects, bosonic codes \cite{Gottesman2001, Mirrahimi2014,Michael2016, Grimsmo2020} based on the state space of a harmonic oscillator have been proposed as a promising alternative to the standard approach based on registers of physical qubits \cite{Shor1995,Steane1996,Fowler2012}.
In hybrid architectures,  these two approaches are complementary, with qubit-register codes built upon logical qubits dynamically protected with efficient base-layer bosonic QEC \cite{Noh2020,Darmawan2021,Terhal2020}.

Although some aspects of QEC have been demonstrated with superconducting circuits \cite{Ofek2016a,Hu2019a, Campagne-Ibarcq2020, Gertler2020,Krinner2021, Zhao2021, Sundaresan2022,Acharya2022}, trapped ions \cite{DeNeeve2020,Ryan2021,Egan2021}, and spins in solid-state systems \cite{Waldherr2014,Abobeih2022,Xue2022}, the control overhead prevents current-day experiments from getting to the heart of what QEC promises to achieve -- extending the lifetime of quantum information stored in the system.
This extension is quantified by the gain $G$, defined as the ratio between the coherence time of an actively error-corrected logical qubit and the best passive qubit encoding in the same system. The break-even point is reached at $G=1$.
A bosonic cat-code experiment \cite{Ofek2016a} managed to achieved $G=1.1$, but with a code that continuously shrinks to the vacuum state. 
Other experiments with various bosonic codes \cite{Hu2019a, Campagne-Ibarcq2020, Gertler2020} and qubit-register codes  \cite{Krinner2021, Zhao2021, Sundaresan2022,Acharya2022} have achieved $G=0.1-0.9$.

\begin{figure}
    \centering
    \includegraphics[width=\figwidth]{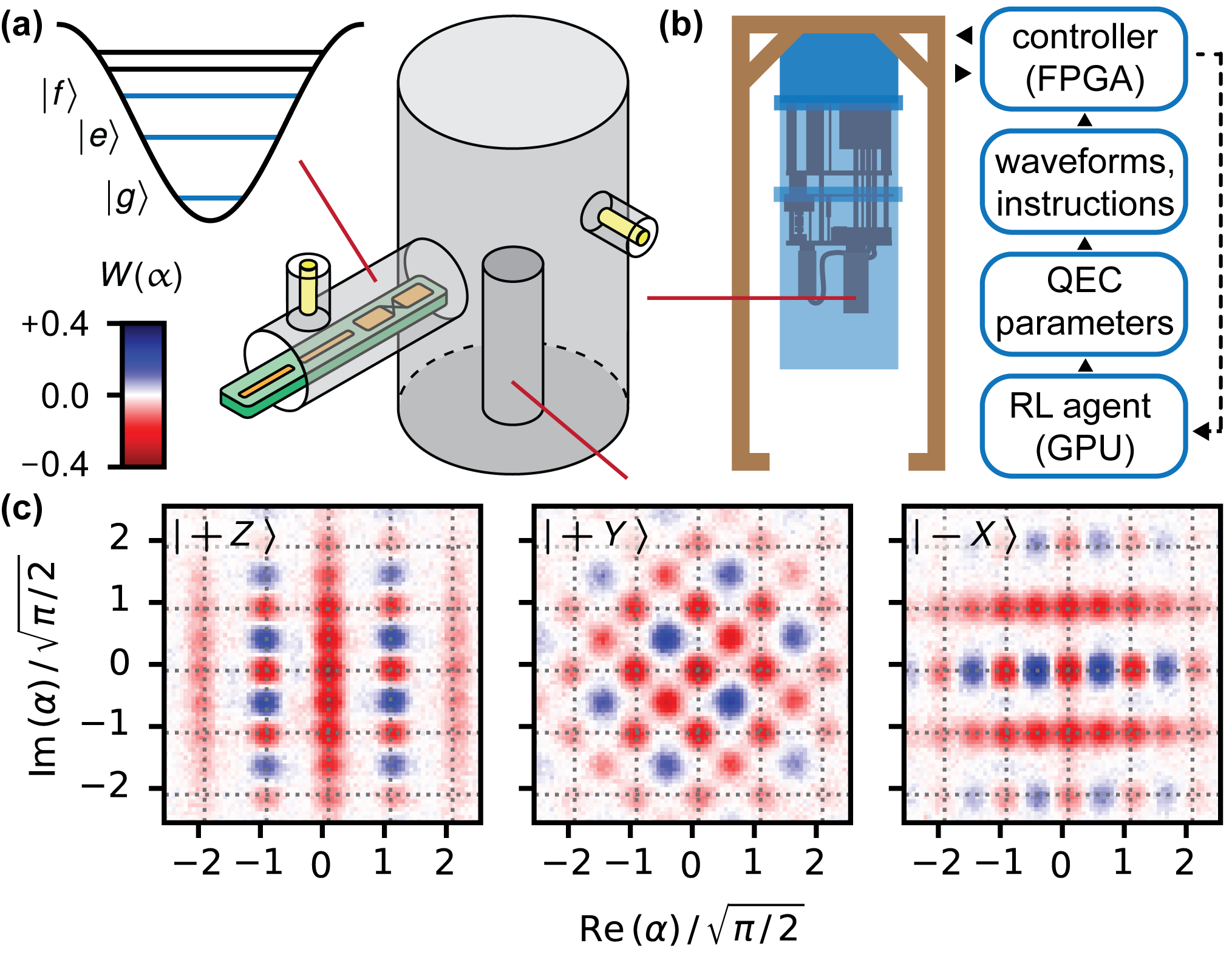}
    \caption[Experimental system]{
    \textbf{Experimental system.}
    {\bf (a)} The sample consists of a superconducting aluminum cavity and a sapphire chip with a transmon circuit, readout resonator and Purcell filter. The electromagnetic mode of the cavity implements a harmonic oscillator, and $\{|g\rangle,|e\rangle\}$ levels of the transmon are used as ancilla qubit to assist in oscillator QEC. 
    {\bf (b)} The sample is cooled in a dilution refrigerator and controlled with microwave and digital electronics. The QEC process is orchestrated by a field-programmable gate array (FPGA), and its  parameters are optimized {\it in-situ} by a reinforcement learning agent implemented on a graphics processing unit (GPU).
    {\bf (c)} Experimental Wigner functions of the Pauli eigenstates of a grid code with $\Delta=0.34$  measured after six QEC cycles.
\label{fig1}}
\end{figure}

We demonstrate full code stabilization and error correction with gain $G=2.27\pm0.07$ using the Gottesman-Kitaev-Preskill (GKP) encoding \cite{Gottesman2001} of a logical qubit into grid states of an oscillator. 
The QEC of this code was previously realized in superconducting circuits \cite{Campagne-Ibarcq2020} and trapped ions \cite{DeNeeve2020}.
In our work, similarly to \cite{Campagne-Ibarcq2020}, the oscillator is an electromagnetic mode of a superconducting cavity whose quantum state is manipulated using a  transmon ancilla, see Fig.~\ref{fig1}(a). 
Our system has an average relaxation and dephasing time of $\overline{T}_1^{\,t}=280\,\mu s$ and $\overline{T}_{2E}^{\,t}=240\,\rm \mu s$ for the tantalum-based transmon \cite{Place2021}, and $\overline{T}_1^{\,c}=610\,\rm \mu s$ and $\overline{T}_2^{\,c}=980\,\mu s$ for the high-purity aluminum cavity \cite{Reagor2016}. 
We implement in this system a ``trickle-down'' QEC scheme based on the proposals in Refs.~\cite{Royer2020,DeNeeve2020}, which includes real-time classical processing and measurement-based feedback.
We train the QEC circuit parameters in-situ with reinforcement learning (RL) \cite{Sutton2017,Schulman2017,TFAgents}, ensuring their adaptation to the real  error channels and control imperfections of our system. At peak performance, the achieved lifetimes of logical Pauli eigenstates are $T_X=T_Z=2.20\pm0.03\,\rm ms$ and $T_Y=1.36\pm0.03\rm \,ms$, and the logical Pauli error probabilities per QEC cycle are $p_Y=(4.3\pm0.4)\times 10^{-4}$ and $p_X=p_Z=(1.81\pm0.04)\times 10^{-3}$.
With such low logical error probabilities, we explore the QEC process on a previously inaccessible time scale of thousands of cycles, subjecting to scrutiny the standard  assumptions of the theory of QEC, such as the stationarity of error rates and absence of leakage-induced correlations.
Finally, we perform error-injection experiments to identify the major factors limiting logical performance and chart the path towards the next-generation logical qubit.

\section*{Engineering error correction}

\begin{figure}
    \centering
    \includegraphics[width=\figwidth]{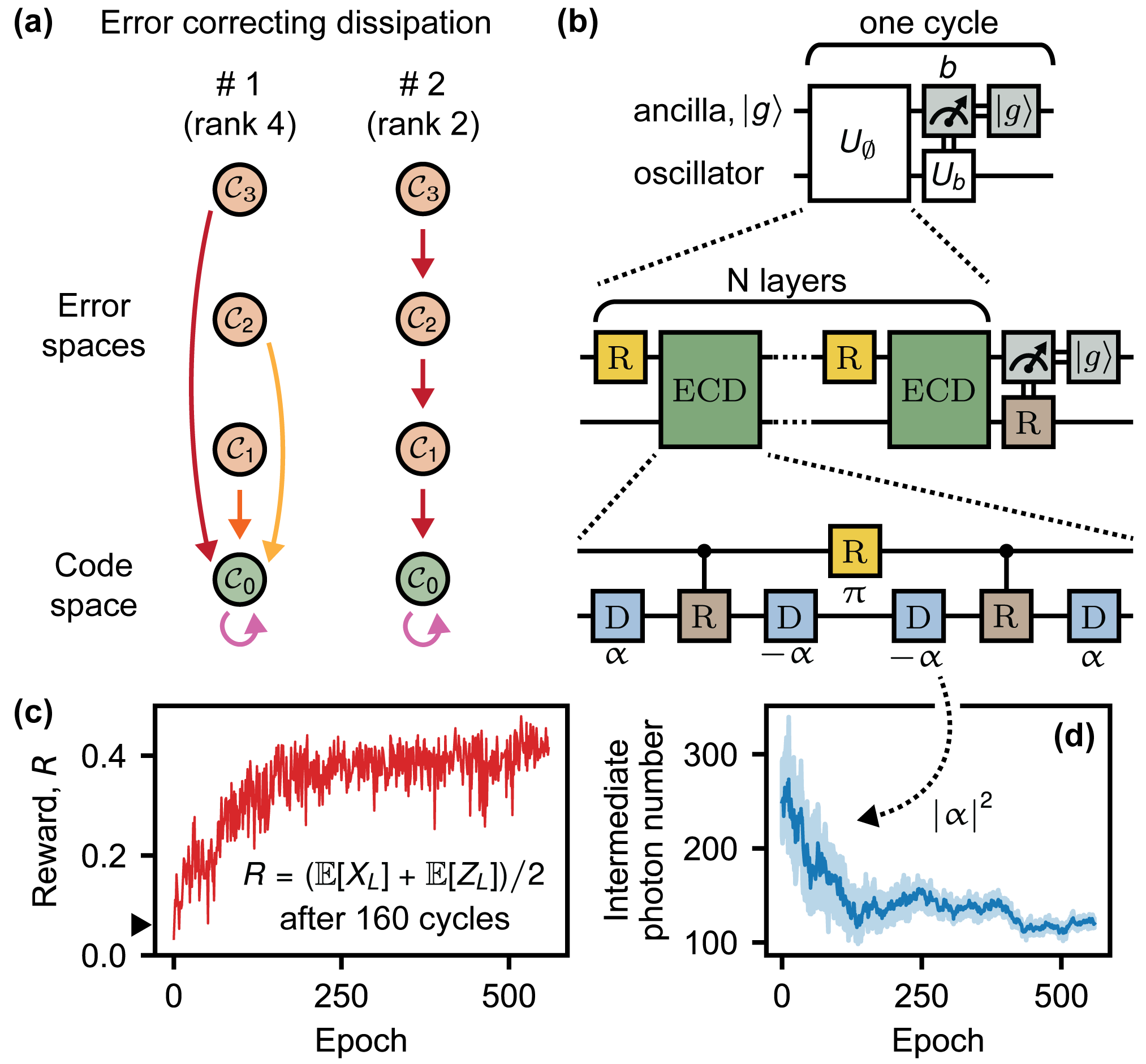}
    \caption[QEC implementation and optimization]{
    \textbf{QEC implementation and optimization.}
    {\bf (a)} Cartoon comparison of error-correcting dissipation channels. The standard dissipation $\#1$  corrects any error in a single step, while the ``trickle-down'' dissipation $\#2$ can be viewed as directional hopping between error spaces that eventually brings the quantum state to the code space ${\cal C}_0$. The colors of the arrows correspond to unique Kraus operators, whose number is equal to the channel rank. Higher-rank dissipation removes entropy more efficiently, but incurs larger control overhead. 
    {\bf (b)} Implementation of a general rank-2 channel on the oscillator using a single ancilla qubit. The unitary $U_\emptyset$ is approximated as a parametrized circuit consisting of $N$ layers of qubit rotations and oscillator conditional displacements. Each conditional displacement gate utilizes a large intermediate displacement of magnitude $|\alpha|$ to enhance the gate speed.
    {\bf (c)} Evolution of reward of the RL agent during the training. The black triangle indicates the start performance based on independent calibrations. Expectations of Pauli operators are taken in their respective eigenstates and include SPAM errors.
    {\bf (d)} One realization of the learning trajectory of the intermediate photon number used to execute the big conditional displacement gate (``B'' in the SBS circuit). Light blue shade shows the variance of the sampled parameter values during the training and dark blue line shows the mean.
\label{fig2}}
\end{figure}

We now explain the principles of our experiment. Its core idea is to realize an artificial error-correcting  dissipation that removes the entropy from the system in an efficient manner by prioritizing the correction of frequent small errors, while not neglecting rare large errors. 
This idea is illustrated in Fig.~\ref{fig2}(a) for a cartoon system in which redundancy is achieved with only 4 orthogonal subspaces in total, where ${\cal C}_0$ is the code subspace and ${\cal C}_1-{\cal C}_3$ are the error subspaces. In this example, the dissipation scheme $\#1$ is maximally efficient from the perspective of entropy removal, since it corrects any error in a single step. 
Such an approach is taken by all qubit-register stabilizer codes, where measurement of the  stabilizers, syndrome decoding, and recovery, when composed, realize a dissipation channel of high Kraus rank. Although this approach can also be applied to the oscillator grid code (see Methods), its implementation entails large control overhead, which in practice might bring more errors than it is designed to correct. 
By contrast, the trickle-down dissipation scheme $\#2$ has the capacity to correct all the same errors, but it is not able to do so in a single step. Importantly, the most probable small errors, corresponding to the error space ${\cal C}_1$, are still corrected in a single step. Owing to this simplification, such an approach reduces control overhead in the grid code, and therefore it was adopted in our work. The continuous-time version of approach $\#2$ was also demonstrated for other bosonic codes in \cite{Lescanne2020,Gertler2020}.

The stabilizer generators of an {\sl ideal} square grid code are $S_0^{\sX}=D(l_S)$ and $S_0^{\sZ}=D(il_S)$, where $l_S=\sqrt{2\pi}$ is the length of a grid unit cell, and $D(\alpha)=\exp(\alpha a^\dagger -\alpha^* a)$ is the displacement operator for an oscillator with creation and annihilation operators $a^\dagger$ and $a$. Logical Pauli operators of the ideal code are defined as $X_L=\sqrt{S_{0}^\sX}$ and $Z_L=\sqrt{S_{0}^\sZ}$. The ideal codewords obey perfect translation symmetry in phase space and thus contain an infinite amount of energy. The finite-energy code is obtained by applying a normalizing envelope operator $N_\Delta=\exp(-\Delta^2 a^\dagger a)$ to the ideal codewords, where $\Delta$ parametrizes the code family that approaches the ideal code in the $\Delta\to0$ limit. In phase space, this parameter controls the extent of the codewords and the squeezing of their probability peaks. Our experimental Wigner functions of the codewords with $\Delta=0.34$ are shown in Fig.~\ref{fig1}(c). The operators of the finite-energy code are obtained through the similarity transformation induced by the envelope operator \cite{Royer2020}, for example, $S^{\sX/\sZ}_{\Delta}=N_{\Delta} S^{\sX/\sZ}_0 N_{\Delta}^{-1}$.

To realize an error-correcting dissipation channel ${\cal R}_{\Delta}$ for the finite-energy code,  there is at our disposal a single ancilla qubit and a classical controller. In principle, with such resources, it is possible to implement arbitrary quantum channels of Kraus rank $2^M$ by recycling the ancilla $M$ times and using feedback operations conditioned on the state of the classical $M$-bit memory of the controller  \cite{Lloyd2001,Shen2017}. Here, we construct a rank-4 error correction channel as a composition of two rank-2 dissipators ${\cal R}_{\Delta}={\cal R}^{\sX}_{\Delta}\circ {\cal R}^{\sZ}_{\Delta}$ that drive the system towards the $+1$ eigenspace of the finite-energy code stabilizers $S^{\sX/\sZ}_{\Delta}$. 
A general rank-2 dissipation can be implemented as a unitary $U_\emptyset$ that entangles the system with the ancilla qubit, followed by ancilla projective measurement with outcome $b$, and a classically-conditioned unitary $U_b$, see Fig.~\ref{fig2}(b).

In our experiment, any unitary is compiled down to a set of primitive operations: qubit rotations around any equatorial axis $R_\varphi(\theta)=\exp[-i(\theta/2)(\cos\varphi \sigma_x+\sin\varphi\sigma_y)]$ implemented as $32\,\rm ns$ Gaussian pulses with spectral corrections \cite{Chen2016};  oscillator displacements $D(\alpha)$ implemented as $40\,\rm ns$ Gaussian pulses; relatively slow conditional rotations ${\rm CR}(\theta)=\exp(i\theta \sigma_z a^\dagger a)$ implemented by waiting a certain amount of time under the dispersive coupling Hamiltonian $H_d/\hbar=\chi\sigma_z a^\dagger a/2$ with $\chi=2\pi\times 46.5\,\rm kHz$; and virtual oscillator rotations $R_V(\vartheta)=\exp(i\vartheta a^\dagger a)$ implemented dynamically on the field-programmable gate array (FPGA) in $448\,\rm ns$.
These primitives are used to construct a fast echoed conditional displacement gate ${\rm ECD}(\beta)=\sigma_x D(\sigma_z\beta/2)$ as shown in Fig.~\ref{fig2}(b), whose speed $\partial_t |\beta| =|\alpha|\chi$ is enhanced compared to the native interaction strength $\chi$ by a large factor $|\alpha|$ -- magnitude of the intermediate displacement in phase space \cite{Campagne-Ibarcq2020, Eickbusch2021}. 

Both rank-2 dissipators are then implemented as follows: the unitary $U_\emptyset$ is decomposed as a parametrized circuit consisting of layers of qubit rotations $R_\varphi(\theta)$ and entangling ${\rm ECD}(\beta)$ gates, while the unitary $U_b$ is realized as only a virtual rotation, see Fig.~\ref{fig2}(b). The role of $U_b$ is twofold: to implement switching between ${\cal R}^{\sX}_{\Delta}$ and ${\cal R}^{\sZ}_{\Delta}$ by changing the quadrature of the oscillator by $\pi/2$, and to compensate for a spurious rotation due to the always-on dispersive coupling $H_d$. The role of $U_\emptyset$ is to approximate the mapping of the finite-energy stabilizer onto the state of the ancilla together with autonomous back-action that pushes the state from the error spaces towards the code space.
Several ans\"{a}tze for decomposition of $U_\emptyset$ were proposed in Ref.~\cite{Royer2020}. We adopt a modified version of the so-called small-big-small (SBS) protocol, named to reflect the relative amplitudes of the three conditional displacement gates that it contains: $\vec{\beta}=l_S\times(i\Delta^2/2,\,1,\,i\Delta^2/2)$, see Supp.~Info.~\ref{sec:qec_details} for further details. 

A single application of the resulting composite dissipator ${\cal R}_{\Delta}$ realizes a {\it QEC cycle}; we refer to applications of constituent dissipators ${\cal R}_{\Delta}^{\sX/\sZ}$ as {\it even$/$odd cycles}. In our implementation, the duration of a QEC cycle is $t_c=2\times 4.924\,\rm \mu s$, which includes execution of unitary gates, ancilla measurements, and real-time processing and decision making by the controller.

\section*{Learning QEC circuit parameters}

While the SBS ansatz and gate calibrations lead to a functioning QEC process, the highest level of performance cannot be achieved with a crude model of the system based on a few independently calibrated parameters -- any such model will inevitably contain unrealistic assumptions. 
Some model inaccuracies and unknown control imperfections can be compensated by closed-loop optimization with direct feedback from the experimental setup.
Previously, pulse-level optimization was successfully utilized to improve gate fidelities \cite{Kelly2014,Rol2017,Werninghaus2021,Baum2021}, but it was never applied to enhance the performance of QEC. Here, for the first time, we apply a real-time reinforcement learning agent to this task, as illustrated in Fig.~\ref{fig1}(b). We use the proximal policy optimization (PPO) algorithm \cite{Schulman2017, TFAgents}, which was shown in simulations to outperform other approaches on high-dimensional problems with stochastic objective that arise in quantum control \cite{Sivak2021}. We parametrize the QEC circuit with $P=45$ parameters that include the amplitudes of various primitive pulses in the circuit decomposition, parameters of the ancilla reset, etc.

The training episodes begin with dissipative pre-cooling of the oscillator followed by feedback cooling to prepare the system ground state $|g\rangle|0\rangle$, see Methods. Then, a logical Pauli eigenstate $|+X\rangle$ or $|+Z\rangle$ is initialized with a method from \cite{Eickbusch2021}, and a candidate QEC protocol is run for $T=160$ cycles. We chose this duration to enhance the signal-to-noise ratio of the reward, similar to the technique used to sample randomized benchmarking cost functions \cite{Kelly2014,Rol2017,Werninghaus2021,Baum2021}.
At the end of the episode, the reward for the RL agent is obtained by measuring the logical Pauli operator $X_L$ or $Z_L$ (depending on the initial state), which provides a proxy for the logical lifetime. This logical measurement is done with one-bit phase estimation of the ideal-code Pauli operators \cite{Terhal2016, Campagne-Ibarcq2020}, and its fidelity is intrinsically limited to $(1+e^{-\pi\Delta^2/4})/2$ \cite{Terhal2020}. Although there exist methods of logical readout adapted to the finite code envelope \cite{Royer2020,DeNeeve2020,Hastrup2021}, we use the phase estimation method to avoid biasing the RL agent towards a particular finite envelope size and to let it pick the optimal size given the error channels of our system.

By construction, the reward incentivizes the RL agent to find a QEC protocol that leads to the longest logical qubit lifetime. The typical evolution of the average reward during the training is shown in Fig.~\ref{fig2}(c). The performance level indicated with a black triangle is achieved with independent calibrations of the system and control parameters, see Supp.~Info.~\ref{sec:cal}. The RL agent significantly improves upon this baseline performance in two stages: in the first hundred training epochs, the agent corrects large errors in the initial parameter values, and in the subsequent few hundreds of epochs, it fine-tunes the circuit parameters to achieve the highest performance. 

Several trends in the learning trajectories highlight the benefits of the model-free RL approach; we elucidate them in more detail in the Supp.~Info.~\ref{sec:parameters}. Here, we only highlight a single illustrative example.
In our implementation of the $\rm ECD$ gate, there exists a nontrivial tradeoff between coherent and incoherent errors: the gate can be implemented faster by displacing the oscillator further in phase space, i.e., populating it with more intermediate photons, but this makes the gate more susceptible to high-order nonlinear effects \cite{Eickbusch2021}. Moreover, some choices of this intermediate photon number can result in a Stark shift of the ancilla into resonance with a spurious degree of freedom, e.g., a two-level defect \cite{Klimov2018,Lisenfeld2019}. How these tradeoffs translate into logical qubit performance is difficult to model, but the RL agent can learn the optimal value of the large intermediate displacement without a model. As shown in Fig.~\ref{fig2}(d), it chose to reduce the intermediate photon number, improving the performance of QEC at the cost of a much slower gate.

\section*{Observing QEC beyond break-even}

\begin{figure}
    \centering
    \includegraphics[width=\figwidth]{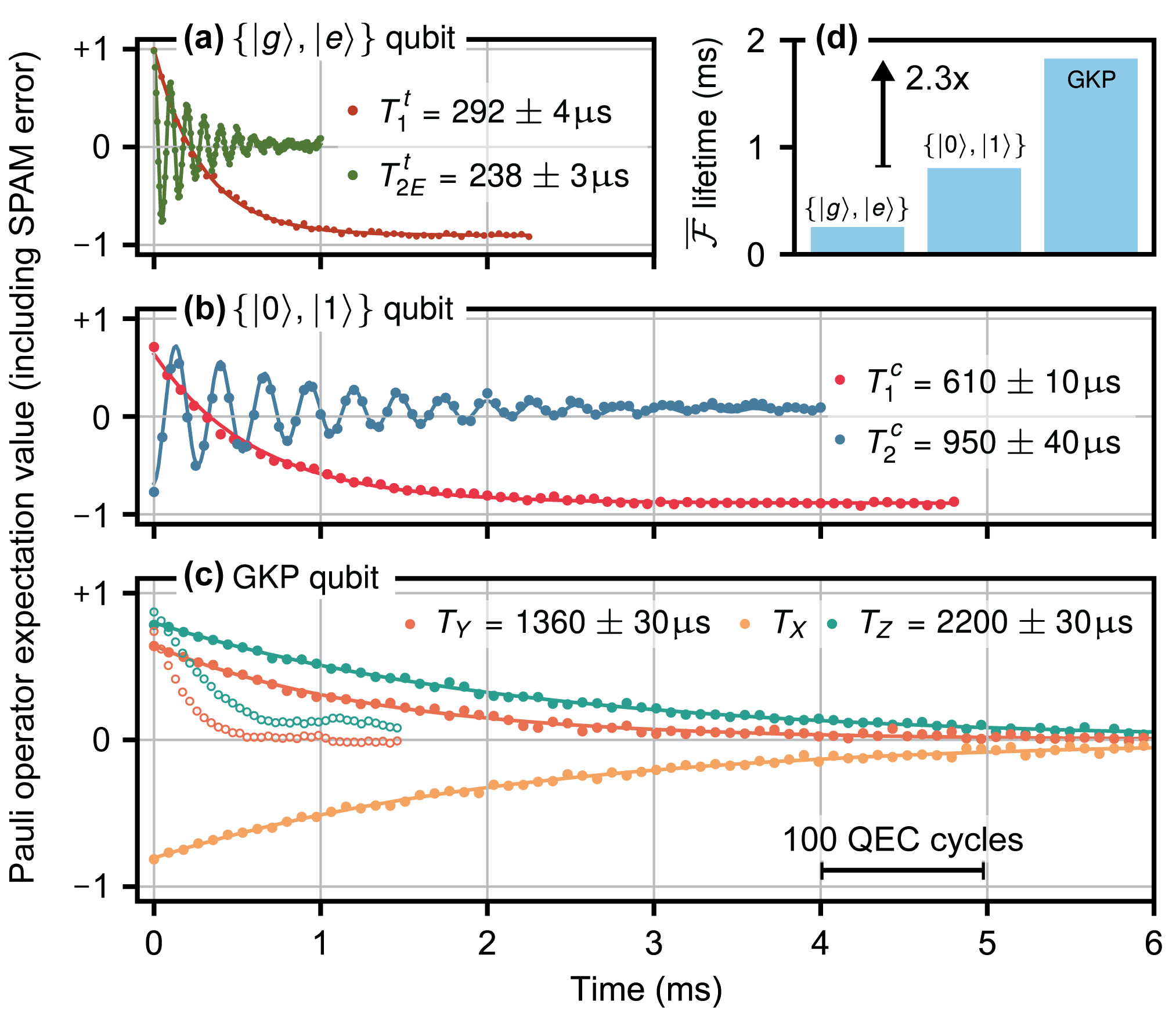}
    \caption[System coherence]{
    \textbf{System coherence.}
    {\bf (a$-$c)} For each qubit, we initialize Pauli eigenstates, let them evolve freely or under QEC for a variable amount of time, and measure the respective Pauli operators. The data for $\{|g\rangle, |e\rangle\}$ and $\{|0\rangle, |1\rangle\}$ qubits is fitted to amplitude damping and white-noise dephasing channel, and data for error-corrected GKP qubit is fitted to a Pauli channel. In (c), the $|+X\rangle$ data is symmetrically reflected with respect to $0$ for better visibility. Empty circles represent evolution in absence of QEC, when grid states decay towards vacuum.
    {\bf (d)} Lifetime of average channel fidelity for these three qubits. 
\label{fig3}}
\end{figure}

After the training is finished, we pick the best performing QEC circuit for further characterization.
Here, we focus on the ability of QEC to create a good quantum memory, i.e. to convert the effect of passage of time into an identity channel ${\cal I}:\rho\to\rho$ that preserves {\it all} qubit states. 

A metric quantifying the deviation of any quantum channel ${\cal E}:\rho\to{\cal E}(\rho)$ from the identity is the {\sl average channel fidelity},
$
\overline{{\cal F}}[{{\cal E}}] = \int d\psi \langle \psi | {\cal E}(|\psi\rangle\langle \psi|)|\psi\rangle,
$
where the integral is over the uniform measure on the qubit state space, normalized so that $\int d\psi=1$. 
In general, this fidelity decays over time in a nontrivial way, but to leading order it evolves as $\overline{{\cal F}}(t)\approx 1-\frac12\Gamma\, t$, where the decay rate $\Gamma$ is equivalent to an average decoherence rate of all pure states on the qubit Bloch sphere. Conveniently, it suffices to  average across the six Pauli eigenstates alone \cite{Nielsen2002}, leading to an experimental procedure for extracting $\Gamma$ that can be applied to any kind of qubit irrespective of its error channel. In Fig.~\ref{fig3}, we show the results of such an experiment, conducted for three different qubit encodings in our system: the $\{|g\rangle, |e\rangle\}$ subspace of the transmon, the $\{|0\rangle, |1\rangle\}$ subspace of the oscillator, and grid code of the oscillator (with and without QEC).

Both the $\{|0\rangle, |1\rangle\}$ and $\{|g\rangle, |e\rangle\}$ qubits are subject to amplitude damping and white-noise dephasing channels, captured by their respective $T_1$ and $T_2$ times, with  fidelity decay constant given by $\Gamma=(1/T_1+2/T_2)/3$. From the perspective of a quantum memory, the best uncorrectable physical qubit in our system is $\{|0\rangle, |1\rangle\}$, shown in Fig.~\ref{fig3}(b), which achieves $\Gamma_{\rm \{01\}}=(800\, {\rm \mu s})^{-1}$. The $\{|g\rangle, |e\rangle\}$ qubit, shown for completeness in Fig.~\ref{fig3}(a), only achieves $\Gamma_{\rm \{ge\}}=(250\, {\rm \mu s})^{-1}$.

\begin{figure*}
    \centering
    \includegraphics[width=\figwidthDouble]{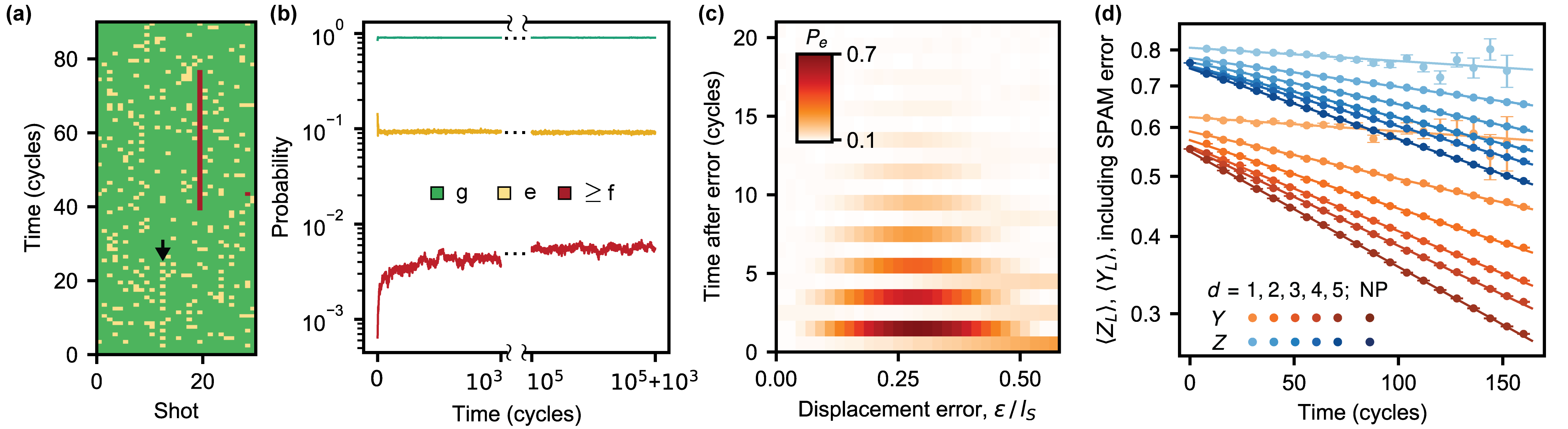}
    \caption[Analysis of error syndromes]{
    \textbf{Analysis of error syndromes.}
    {\bf (a)} A sample of ancilla measurement outcomes during QEC. The $e$ outcome (yellow) indicates correction of physical errors. The black arrow points to a syndrome string of the type $eg/eg/...$ likely left by a large error in one oscillator quadrature. Red indicates transmon leakage out of the $\{|g\rangle,|e\rangle\}$ qubit subspace.
    {\bf (b)} Average probability of each measurement outcome as a function of time. After correcting state initialization errors, QEC settles into a steady state that persists for at least a hundred thousand cycles.
    {\bf (c)} Probability of $e$ outcome as a function of time after injecting position displacement errors of varying amplitude. Since a logical gate in the ideal code comprises a displacement of amplitude $l_S/2$, a displacement of amplitude $l_S/4$ makes a large-distance error, which takes several cycles to correct with our low-rank QEC channel.
    {\bf (d)} Decay of Pauli eigenstates after eliminating by post-selection the experimental shots with strings of $\geq d$ consecutive $e$ outcomes in the same-quadrature cycles. Data for $X_L$ eigenstates is not shown; it is expected to behave similarly to $Z_L$. The improvement in lifetime indicates that $e$ outcomes are indeed correlated with occurrence of errors. NP in the legend stands for ``no post-selection''.
\label{fig4}}
\end{figure*}

Higher excited states of the oscillator have shorter lifetime due to bosonic enhancement of spontaneous emission. Therefore, as with any QEC code, encoding a qubit using grid states incurs an immediate penalty in the  fidelity decay rate. Moreover, this natural decay, shown in Fig.~\ref{fig3}(c) with empty circles, takes the grid states outside the logical manifold and eventually towards the vacuum state $|0\rangle$. 

Our error correcting dissipation stabilizes the grid code manifold and, together with naturally occurring dissipation, leads to a logical Pauli channel within this manifold, with 
the lifetimes of logical Pauli eigenstates of $T_X=T_Z=2.20\pm0.03\,\rm ms$ and $T_Y=1.36\pm0.03\rm \,ms$. Under the Pauli channel, the fidelity decay constant is given by $\Gamma=(1/T_X+1/T_Y+1/T_Z)/3$, which in our experiment amounts to $\Gamma_{\rm GKP}=(1.82\, {\rm ms})^{-1}$.

The principal metric characterizing the quality of QEC from the perspective of quantum memory is the coherence gain of an actively error-corrected logical qubit over the best passive qubit encoding. In our experiment, the highest achieved gain is $G=\Gamma_{\rm \{01\}}/\Gamma_{\rm GKP}=2.27\pm0.07$, confidently beyond break-even.

\section*{QEC process characterization}

Having characterized the logical qubit as a quantum memory, we next examine the properties of the QEC process. Ancilla measurement outcomes, referred to as syndromes, inform us which stochastic path the QEC process has taken in each cycle.  In Fig.~\ref{fig4}(a) we show a (statistically unrepresentative) sample of these outcomes that comprise trajectories of different experimental shots. Such a dataset contains an immense amount of information about the QEC process, not available in previous experiments with the grid code QEC \cite{Campagne-Ibarcq2020,DeNeeve2020}.

To interpret this dataset, we adopt here a simplified model of trickle-down dissipation such as depicted in Fig.~\ref{fig2}(a), which captures the essence of our QEC process. The caveats of this model and the exact Kraus decomposition of our QEC circuit are provided in Supp.~Info.~\ref{sec:sbs}. In this simplified model, the $g$ outcome indicates that the state was projected onto the code space, while an $e$ outcome indicates that the state was transferred one level down the error hierarchy, partially or completely correcting an error.

From the dataset in Fig.~\ref{fig4}(a), we observe that the vast majority of outcomes are $g$ (green), which means that errors are rare. The stochastic pattern of $e$ outcomes (yellow) reflects  randomly occurring errors. Most errors are small and, when corrected, leave single isolated $e$ outcomes. An example syndrome string likely generated by a large error in {\it one} quadrature is indicated with an arrow: it has a characteristic $eg/eg/...$ pattern.
We also observe isolated ancilla leakage events (red). Leakage to $|f\rangle$ is reset in the same cycle with high probability. Sometimes, leakage persists for multiple cycles (streak of red), due to the transmon escaping to a state higher than $|f\rangle$, which is not addressed in our reset scheme.

The average probability of each outcome as a function of time is shown Fig.~\ref{fig4}(b), where the process starts from a $|+X\rangle$ state. After about $10$ cycles of initial state correction, the process settles into a dynamical equilibrium which
persists for at least a hundred thousand cycles (the longest measured here) without any notable increase of the error rates over time. Detailed analysis reveals that the QEC process is 
nearly stationary, with residual deviations from stationarity caused  by the transmon leakage to states higher than $|f\rangle$ at a rate $1.3\times 10^{-4}$ per cycle, see Supp.~Info.~\ref{sec:leakage}. 

In this dynamical equilibrium, physical errors excite the quantum state out of the code space with probability $p_{\rm err}=0.13\pm0.02$ per QEC cycle, as deduced from the statistics of syndrome outcomes. 
The competition between physical errors and error-correcting dissipation results in a ``thermal'' distribution across the subspaces with probability $\langle\Pi_0\rangle=0.82\pm0.02$ of occupying the code space, see Methods. 
Having $p_{\rm err}\ll1$ justifies the use of low-rank error-correcting dissipation in our system, which is sufficient to prevent physical errors from accumulating and causing logical errors. 
At the highest achieved QEC gain, the logical Pauli error probabilities per QEC cycle are $p_Y=(4.3\pm0.4)\times 10^{-4}$ and $p_X=p_Z=(1.81\pm0.04)\times 10^{-3}$.
By comparing the total logical error probability, $p_X+p_Y+p_Z$, to the physical error probability $p_{\rm err}$, we conclude that $97\%$ of the errors are successfully corrected by our process.

Since rare large errors require several cycles to be corrected, the QEC process is weakly time-correlated with a correlation length of $3.9\pm0.1$ cycles, see Supp.~Info.~\ref{sec:leakage}. To understand these correlations, in Fig.~\ref{fig4}(c) we inject displacement errors along the position quadrature and monitor the syndromes that they produce as a function of time. 
Such errors leave traces of $e$ outcomes in proportion to their distance to the closest logical operation. 
For example, a displacement of length $0$, equivalent to a logical identity, leaves no syndrome trace; a displacement of length $l_S/2$ is close to a logical bit-flip of the finite-energy code, and hence it leaves only a small syndrome trace; on the other hand, a midway displacement of length $l_S/4$ makes a large-distance error which takes the longest time to correct with a low-rank dissipator, generating a lasting trace of $e$ outcomes.

This displacement error injection experiment confirms that errors indeed generate the $e$ syndromes, but do these syndromes herald the occurrence of errors? 
To verify this, we perform post-selection of trajectories with different syndrome patterns. 
In particular, we discard trajectories that have $\geq d$ consecutive $e$ outcomes in the same-quadrature cycles, with resulting post-selected decay of Pauli eigenstates shown in Fig.~\ref{fig4}(d). In the case $d=5$, post-selection eliminates rare large-distance errors and improves the fidelity lifetime only by a factor $1.2$, but at the cost of rejection probability of $7\times10^{-4}$ per cycle. On the other hand, in the case $d=1$, post-selection eliminates relatively frequent small errors that are close to identity, as well as rare large uncorrectable errors that are close to a  logical operation. It is because of the latter that the fidelity lifetime in this setting improves by a factor $6.3$, but with a more severe rejection probability of $6\times10^{-2}$ per cycle. These favorable post-selection results indicate that such a method can be used for probabilistic preparation of high-fidelity logical states, including the magic states required for universal quantum computing \cite{Bravyi2005}, which is left for future investigation.

\section*{Conclusion and Outlook}

In this work, we used real-time error correction to realize a fully stabilized logical qubit whose lifetime is more than doubled compared to the best passive qubit encoding in the system, marking the transition of QEC from proof-of-principle studies to a practical tool for enhancing quantum memories.  Our work improves upon previous QEC experiments, which do not protect the logical identity operator $I_L$ \cite{Ofek2016a}, protect only one of the logical Pauli operators $X_L$ or $Z_L$\cite{Grimm2019,Lescanne2020,Chen2021}, implement correction in post-processing \cite{Krinner2021,Zhao2021,Acharya2022}, require post-selection \cite{Andersen2019}, and do not reach break-even \cite{Hu2019a,Gertler2020,Campagne-Ibarcq2020,Krinner2021,Zhao2021,Acharya2022}.
Instrumental for this achievement, among other factors, was the adoption of a model-free learning framework, improved fabrication technique for the ancilla transmon, and a novel grid-code QEC protocol. 

Performing additional experiments, we identified the core challenges that need to be addressed to ensure future progress of grid-code QEC. 
In particular, by studying long-time system stability, we found that occasional collapses of the logical performance are strongly correlated with appearance of spurious degrees of freedom in the system. Their resonant interaction with the Stark-shifted transmon ancilla degrades the fidelity of our operations, see Supp.~Info.~\ref{sec:stability}.
In the short term, this effect could be mitigated by adopting a tunable ancilla and periodically re-training the QEC circuit to find better spectral locations. In the long term, the behavior of these defects needs to be understood, as they pose even greater danger for scaled-up quantum devices \cite{Krinner2021, Zhao2021, Acharya2022}.

In addition, we expect that considerable enhancement can be gained by tailoring the QEC process not only to error channels of the oscillator, but also those of the ancilla. Our QEC circuit is fault-tolerant with respect to ancilla phase-flip errors by design \cite{Royer2020}. With the transmon ancilla used here, the sensitivity of the logical lifetime to ancilla phase flips is $65$ times smaller than to ancilla bit flips, as found with noise injection experiments, see Supp.~Info.~\ref{sec:noise injection}. Future development should incorporate robustness against ancilla bit flips, either through path-independent control \cite{Ma2020,Rosenblum2018} or by adopting a biased-noise ancilla \cite{Puri2019}.

\section*{Acknowledgments}

We acknowledge discussions with R.~Corti\~nas, J.~Claes, and A.~Mi. This research was supported by the U.S. Army Research Office (ARO) under grants  W911NF-18-1-0212 and W911NF-16-1-0349, and by the U.S. Department of Energy, Office of Science, National Quantum Information Science Research Centers, Co-design Center for Quantum Advantage (C2QA) under contract number DE-SC0012704. The views and conclusions contained in this document are those of the authors and should not be interpreted as representing official policies, either expressed or implied, of the U.S. Government. The U.S. Government is authorized to reproduce and distribute reprints for Government purpose notwithstanding any copyright notation herein. Fabrication facilities use was supported by the Yale Institute for Nanoscience and Quantum Engineering (YINQE) and the Yale SEAS Cleanroom.

\section*{Author contributions}

V.S., A.M. and A.E. built the experimental setup.
R.J.S. contributed to experimental apparatus.
I.T., S.G. and L.F. fabricated the ancilla transmon chip.
B.R., S.S. and S.M.G. developed the theory.
B.R., V.S., A.E. and B.B. developed dissipative oscillator cooling.
A.E., V.S. and A.Z.D. developed state initialization technique.
V.S. implemented reinforcement learning, performed the experiments, and analyzed data.
V.S., A.E., B.R. and M.H.D. regularly discussed the project and provided insight.
M.H.D. supervised the project.
V.S. and M.H.D. wrote the manuscript with feedback from all authors.

\section*{Competing interests}

R.J.S., L.F. and M.H.D. are founders, and R.J.S. and L.F. are shareholders of Quantum Circuits, Inc.

\section*{Methods \label{Methods}}

\subsection*{QEC of the ideal grid code}

To understand error correcting properties of the ideal code, consider an error channel $\cal E$ decomposed in the displacement basis. An ideal grid code with code projector $\Pi_0$ satisfies Knill-Laflamme conditions \cite{Knill1997} $\Pi_0 D^\dagger (\varepsilon_\alpha)D(\varepsilon_\beta) \Pi_0 \propto \delta(\varepsilon_\alpha-\varepsilon_\beta) \Pi_0$ for all errors in a correctable set $E_+=\{D(\varepsilon)\, :\, |{\rm Re}(\varepsilon)|, |{\rm Im}(\varepsilon)|<l_S/4 \}$. A displacement error of amplitude $\varepsilon$ creates an error state $|\psi_\varepsilon\rangle = D(\varepsilon)|\psi\rangle$, where $|\psi\rangle$ is any state from the code space. Since a displaced grid state is still translationally invariant, it remains an eigenstate of the ideal code stabilizers, and the phase of its eigenvalue encodes a continuous error syndrome: $S_0^{\sZ}|\psi_\varepsilon\rangle=\exp(2il_S{\rm Re}[\varepsilon])|\psi_\varepsilon\rangle$ and $S_0^{\sX}|\psi_\varepsilon\rangle=\exp(-2il_S{\rm Im}[\varepsilon])|\psi_\varepsilon\rangle$. 
Error correction of an ideal grid code can be done in a similar manner to any stabilizer code: first, measure the stabilizers to obtain the error syndrome, which here corresponds to phase estimation of  $S^{\sX/\sZ}_0$ that yields the error amplitude $\varepsilon$. This step projects the state onto one of the orthogonal error spaces. Then, apply the recovery operation, here a simple displacement $D(-\varepsilon)$, to correct the error. Such procedure realizes an artificial dissipation ${\cal R}$ of an infinite rank which corrects any error from $E_+$ in a single cycle, $({\cal R}\circ{\cal E})(\rho)\propto\rho$, analogously to the cartoon high-rank dissipation in Fig.~\ref{fig2}(a). In contrast to this approach, our experiment realizes low-rank dissipation that asymptotically satisfies $([{\cal R}]^{n\to\infty}\circ{\cal E})(\rho)\propto\rho$.

\subsection*{Dissipative cooling to vacuum}

We utilize the dissipation engineering framework \cite{Gross2018} to design fast cooling of the oscillator to vacuum state $|0\rangle$ in the weak-coupling regime where previous known cooling methods \cite{Pfaff2017} fail; we also expect this novel method to be applicable to cooling of trapped ions \cite{DeNeeve2020}. Similarly to error-correcting dissipation, we realize this cooling as a composition of two rank-2 channels that shrink the oscillator state in the orthogonal quadratures. The unitary $U_\emptyset$ in this case is realized as a three-layer circuit obtained from first-order Trotter decomposition of $U=\exp[-i\varepsilon(a\sigma_++a^\dagger\sigma_-)]$, where $\varepsilon\ll1$ controls the cooling rate. This unitary swaps the excitations of the oscillator into the ancilla, which is reset in every cycle. The duration of one full cooling cycle (including both quadratures) is $t_c=2\times 3.38\,\rm us$. With $\varepsilon=0.4$, we achieve cooling at a rate $20$ times faster than natural energy damping rate of the oscillator. In our experiment, $25$ full cycles of such a dissipative cooling are then followed with a feedback cooling protocol adapted from \cite{Ofek2016a} to remove any residual thermal population. See Supp.~Info.~\ref{sec:oscillator cooling} for more details.

\subsection*{Reinforcement learning implementation}
The QEC circuit is parametrized with a vector $\vec{p}$. Instead of optimizing $\vec{p}$ directly, the RL agent learns parameters of the probability distribution from which $\vec{p}$ is stochastically sampled during the training to ensure adequate exploration of parameter space. To this end, we use a factorized multivariate Gaussian distribution ${\cal N}({\vec{\mu}}, {\vec{\sigma}})$ with mean $\vec{\mu}$ and covariance matrix ${\rm diag}[{\vec{\sigma}}]^2$. To capture the pattern of  relations between different components of $\vec{p}$, the mean and covariance are represented as parametrized functions $\vec{{\mu}}(\theta)$ and $\vec{{\sigma}}(\theta)$ of common hidden variables $\theta$. In this work, $\vec{\mu}$ and $\vec{\sigma}$ are produced at the output of a neural network with two fully connected layers of 50 and 20 rectifier linear unit (ReLU) neurons. Starting with initial vector of parameters $\vec{{\mu}}_i$ found with independent calibrations, during the course of learning the agent gradually deforms the distribution and localizes it on the new vector $\vec{{\mu}}_f$, the final result of the optimization. Typically, as it proceeds, the agent also reduces the covariance of the distribution to have a finer control over the mean. These features of learning are observed in the example evolution of one component of $\vec{p}$ in Fig.~\ref{fig2}(d). During one training epoch, we evaluate 10 QEC circuit candidates with 300 episodes (i.e. experimental shots) per candidate. The collected information is used to update the neural network parameters $\theta$ according to the PPO algorithm, which completes the epoch. One epoch takes approximately $16$ seconds, with the majority of time spent on re-compilation of FPGA instruction sequences and its re-initialization. See Supp.~Info.~\ref{sec:RL} for more details.

\subsection*{Steady state of the QEC process}
We perform Wigner tomography of the logical states after varying duration of the QEC process, reconstruct the density matrix, and from its spectral decomposition extract the expectation value of the code projector $\langle\Pi_0\rangle=0.825\pm0.003$, where error bar represents the standard deviation with respect to different process durations of $100,200,400$, and $800$ cycles. In addition to the code space, only one error space is occupied in the steady state with an appreciable probability of $0.170\pm0.005$. The logical decoherence within this error space happens at the same rate as within the code space. For more details, see Supp.~Info.~\ref{sec:wigner_tomo}.

The expectation value of the code projector in the steady state can be estimated independently,  using the statistics of syndrome outcomes. Under the approximations discussed in Supp.~Info.~\ref{sec:msmt stat}, the probability that a syndrome string of length $2n$ consists only of $g$ outcomes asymptotically approaches $\langle\Pi_0\rangle (1-p_{\rm err})^{n-1}$ for large $n$. Using this method, we extract $\langle\Pi_{0}\rangle=0.81\pm0.02$ and $p_{\rm err}=0.13\pm0.02$. The error bar in this case represents the accuracy of the model for the string probability, which is valid to first order in $p_{\rm err}$. The value of $\langle\Pi_{0}\rangle$ quoted in the main text is the average of the two methods. Constructing a detailed error budget of the aggregate error probability $p_{\rm err}$ based on the system-level simulation of the known error processes is an avenue  left for future work.

\let\oldaddcontentsline\addcontentsline
\renewcommand{\addcontentsline}[3]{}
\putbib[lib_clean]
\let\addcontentsline\oldaddcontentsline
\end{bibunit}

\setcounter{equation}{0}
\setcounter{figure}{0}
\setcounter{table}{0}
\setcounter{section}{0}
\makeatletter

\renewcommand{\theequation}{S\arabic{equation}}
\renewcommand{\thefigure}{S\arabic{figure}}
\renewcommand{\thetable}{S\arabic{table}}
\renewcommand{\thesection}{S\arabic{section}}

\crefname{table}{Table}{Tables}
\Crefname{table}{Table}{Tables}
\crefname{figure}{Fig.}{figs.}
\Crefname{figure}{Fig.}{Figures}
\crefname{section}{section}{sections}
\Crefname{section}{Section}{Sections}
\crefname{subsection}{section}{sections}
\Crefname{subsection}{Section}{Sections}

\title{Supplementary Information \\[1ex] \large
``Real-time quantum error correction beyond break-even''}

\clearpage
\maketitle

\onecolumngrid
\tableofcontents
\clearpage

\begin{bibunit}[apsrev_longbib]

\section{Experimental setup and sample parameters\label{sec:setup}}

\begin{figure}[h]
    \centering
    \includegraphics[width=\figwidthDouble]{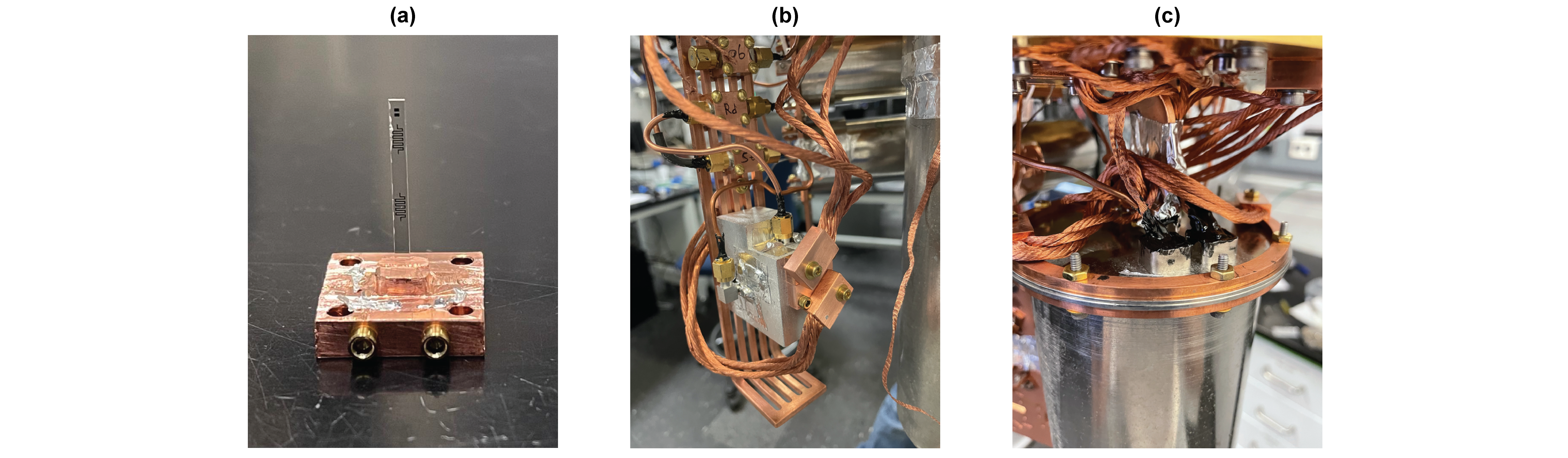}
    \caption[Sample assembly]{
    \textbf{Sample assembly.} 
    {\bf (a)} Clamped ancilla chip. 
    {\bf (b)} Thermalization.
    {\bf (c)} Shielding.
\label{fig:sample}}
\end{figure}

{\bf Assembly.} Our system design follows the hybrid planar-3D circuit QED architecture developed in \cite{Axline2016}. The storage oscillator is realized as an electromagnetic mode hosted by a seamless superconducting coaxial stub cavity made of high-purity aluminum and treated with a chemical etch to improve surface quality. This is the same physical cavity as used in \cite{Rosenblum2018}, although with lower coherence time due to aging during the storage time of $\sim$ 2 years. The cavity is anchored to a copper bracket inside a cryoperm shield. The ancilla chip is inserted in a tunnel waveguide that connects to the storage cavity, and is secured at one end with a copper clamp.   Thermalizing copper braids run from the clamp to the base plate of the dilution refrigerator, see Fig.~\ref{fig:sample}.

{\bf Ancilla chip fabrication.}
The ancilla chip contains a transmon qubit, a stripline readout resonator, and a stripline bandpass Purcell filter. The resonators and transmon are tantalum-based devices, with Josephson junction of the transmon made of a small aluminum section; they are fabricated with a process similar to \cite{Place2021}. Adopting a tantalum-based platform results in improved qubit coherence relative to an all-aluminum platform; however, the reasons for this are still under active investigation. Possible explanations include, but are not limited to: 1) The corrosion resistance allows for the use of rigorous acid-based cleaning techniques to be employed during the fabrication process that improves surface dielectric quality and minimizes the presence of organic residues; 2) The high melting point of tantalum allows for deposition to occur at higher temperatures, where atomic mobility is high enough to enable epitaxial film growth with a high degree of crystalline order; 3) Tantalum has a higher superconducting transition temperature than aluminum, which may lead to increased resistance to quasiparticle loss. 

We use a C-plane sapphire wafer produced using the heat exchanger method (HEM), as it was shown to have  smaller dielectric loss \cite{Read2022}. The wafer was initially cleaned with a piranha solution ($2:1$ $\rm H_2SO_4:H_2O_2$) for 20 minutes and rinsed with DI water. The wafer was then annealed at $1200^{\,\circ}\rm C$ in an oxygen-rich environment for 1 hour. After cooling down to room temperature, the wafer was immediately transferred to a sputtering system for tantalum deposition. $150\,\rm nm$ of tantalum was deposited by DC magnetron sputtering with the substrate temperature being held at $800^{\,\circ}\rm C$. The Purcell filter, readout resonator, and transmon pads were subtractively patterned using a positive photoresist mask and reactive ion etching. After tantalum patterning, the Josephson junction was patterned using electron-beam lithography and defined using the Dolan bridge method. The junction was deposited using electron-beam evaporation of aluminum at 2 angles with an interleaved static oxidation step to construct the tunnel barrier. Liftoff was performed in NMP heated to $90^{\,\circ}\rm C$, followed by sonication in acetone, isopropanol, and DI water. Finally, the wafer was protected with a layer of photoresist before dicing into individual chips, followed by additional cleaning with NMP, acetone, and isopropanol to remove the protective photoresist.

{\bf System parameters.}
The measured parameters of this system are summarized in Table~\ref{tab:Measured-system-parameters}.

\begin{table}[h]
\begin{tabular}{|c|l|l|}
\hline 
\multirow{6}{*}{\textbf{$\,\,\,$Cavity mode$\,\,\,$}}
 & Frequency & $\omega_{c}=2\pi\times4.479\,\rm GHz$ \tabularnewline 
 & 1st order dispersive shift & $\chi=2\pi\times46.5\,\rm kHz$ \tabularnewline
 & 2nd order dispersive shift & $\chi'=2\pi\times5.8\,\rm Hz$ \tabularnewline
 & Kerr nonlinearity & $K=-2\pi\times 4.8\,\rm Hz$ \tabularnewline
 & Relaxation & $\overline{T}_{1}^{\,c}=606\pm 10\,\rm us$\tabularnewline
 & Dephasing & $\overline{T}_{2}^{\,c}=980\pm 30\,\rm us$ \tabularnewline
\hline 
\multirow{6}{*}{\textbf{$\,\,\,$Ancilla transmon$\,\,\,$}}
 & Frequency & $\omega_{t}=2\pi\times5.921\,\rm GHz$ \tabularnewline 
 & Anharmonicity & $\alpha=-2\pi\times222\,\rm MHz$ \tabularnewline
 & Relaxation & $\overline{T}_{1}^{\,t}=280\pm 30\,\rm us$ \tabularnewline
 & Equilibrium population & $\overline{n}_{{\rm th}}^{\,t}=0.043\pm 0.008$\tabularnewline
 & Dephasing (Ramsey) & $\overline{T}_{2R}^{\,t}=62\pm 5\,\rm us$ \tabularnewline
 & Dephasing (Echo) & $\overline{T}_{2E}^{\,t}=238\pm 8\,\rm us$ \tabularnewline
\hline 
\multirow{4}{*}{\textbf{$\,\,\,$Readout resonator$\,\,\,$}} 
 & Frequency & $\omega_{r}=2\pi\times9.107\,\rm GHz$ \tabularnewline 
 & Dispersive shift & $\chi_{qr}=2\pi\times0.60\,\rm MHz$ \tabularnewline
 & Coupling strength & $\kappa_{r(c)}=2\pi\times0.47\,\rm MHz$\tabularnewline
 & Internal loss & $\kappa_{r(i)}=2\pi\times0.03\,\rm MHz$\tabularnewline
\hline 
\end{tabular}

\caption{\textbf{Measured system parameters.} \label{tab:Measured-system-parameters} For transmon  parameters ($\overline{T}_{1}^{\,t}$, $\overline{T}_{2R}^{\,t}$, $\overline{T}_{2E}^{\,t}$, $\overline{n}_{{\rm th}}^{t}$) and cavity parameters ($\overline{T}_{1}^{\,c}$, $\overline{T}_{2}^{\,c}$) that appreciably fluctuate in time, we provide the mean and standard deviation over a week-long period. The definition of the Hamiltonian parameters can be found in Section~\ref{sec:hamiltonian}.}

\end{table}


\begin{figure}
    \centering
    \includegraphics[width=\figwidthDouble]{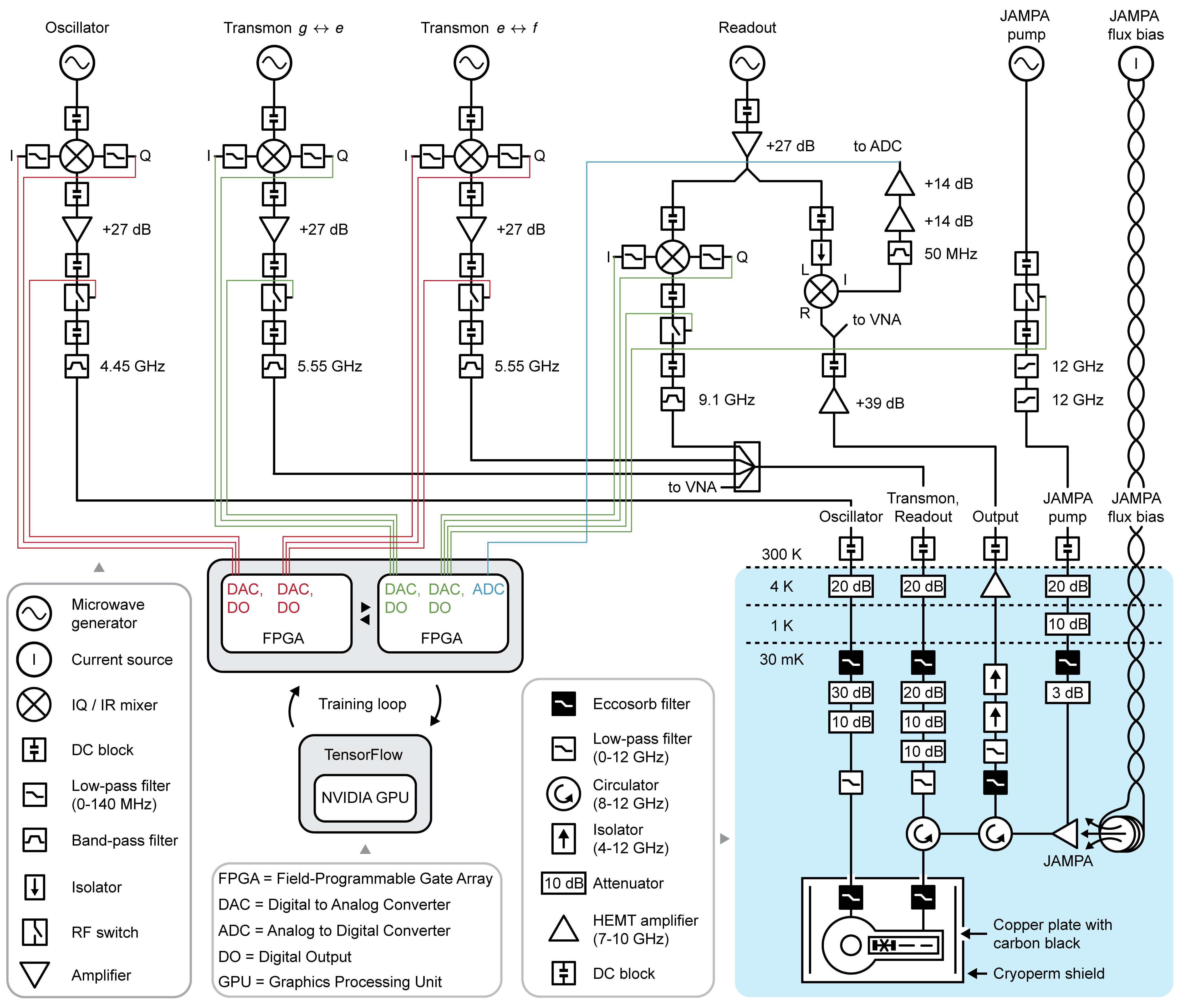}
    \caption[Experimental setup]{
    \textbf{Experimental setup.} For simplicity, the diagram omits DC ground connections of all active components, and attenuators placed at different locations at $300\,\rm K$ stage to ensure power levels within specs for amplifiers and switches. 
\label{Wiring diagram}}
\end{figure}

{\bf Control wiring.}
As shown in Fig.~\ref{Wiring diagram}, the quantum system is controlled by a classical computer (VPXI-ePC) that hosts two control cards (X6-1000M) from Innovative Integration. Each card integrates digital-to-analog converters (DAC), analog-to-digital converters (ADC), digital inputs and outputs (DIO), and a Xilinx VIRTEX-6 field-programmable gate array (FPGA). This controller was developed in \cite{Ofek2016a} and used in prior bosonic QEC experiments \cite{Ofek2016a,Hu2019a}. The baseband control signals are sampled from the DACs at $500\, \rm MS/s$ rate with 16-bit resolution and upconverted to the oscillator, qubit, and readout frequencies through single-sideband modulation of the local oscillators (Agilent N5183A). After amplification, the signals are gated with fast RF switches ($9\,\rm ns$ rise time, $5\,\rm ns$ fall time) and filtered before entering the dilution refrigerator. The signals are further attenuated and filtered in the cryogenic environment. A crucial component of the filtering is the eccosorb CR-110 infrared absorber filter \cite{Serniak2019} located inside the cryoperm shield, and the copper plate, painted with stycast epoxy mixed with black carbon powder, that wraps around the sample. On the output side, the reflected readout signal is amplified at $30\,\rm mK$ stage with a near-quantum-limited Josephson array-mode parametric amplifier (JAMPA) \cite{Sivak2021}, followed at $4\,\rm K$ stage by a low-noise HEMT amplifier. Upon further amplification at $300\,\rm K$ stage and down-conversion to $50\,\rm MHz$, the readout signal is digitized, demodulated, and integrated with a filter function to obtain $I$ and $Q$ quadratures. Their values are compared to the decision boundaries $I_{th}$ and $Q_{th}$ to obtain two bits of information $s_0=\Theta(I-I_{th})$ and $s_1=\Theta(Q-Q_{th})$, where $\Theta$ is the Heaviside step function. This allows to classify the measurement outcome as $``g"$ if $s_0=0$, $``e"$ if $(s_0, s_1)=(1,0)$, and $``f"$ if $(s_0, s_1)=(1,1)$. The bits $s_0$ and $s_1$ are redistributed to all control cards which run independent but synchronized control flows that include conditional branching on these bits. Further details of the readout subroutine are described in Section~\ref{sec:readout}.


\section{Calibration and characterization experiments\label{sec:cal}}

\subsection{Primitive pulses \label{sec:primitive pulses}}

\begin{figure}
    \centering
    \includegraphics[width=\figwidthDouble]{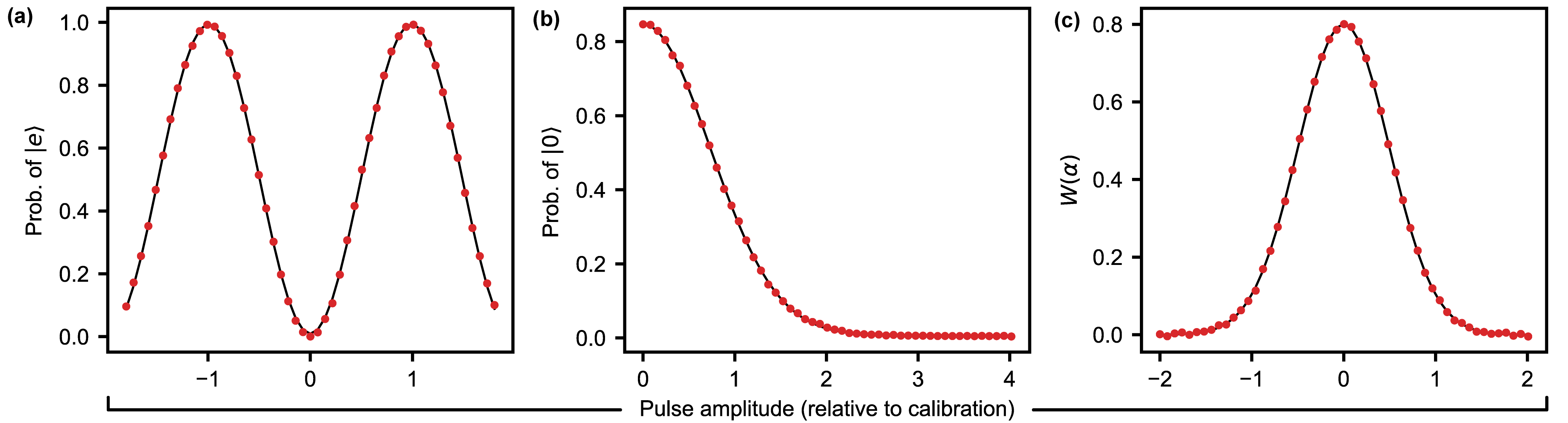}
    \caption[Calibration of primitive pulses]{
    \textbf{Calibration of primitive pulses.} 
    {\bf (a)} Amplitude Rabi experiment to calibrate qubit rotations. 
    {\bf (b)} First-step calibration of displacement: probability of $|0\rangle$ in a coherent state,  $P(0)=e^{-|\alpha|^2}$. 
    {\bf (c)} Second-step calibration of displacement: Wigner function of vacuum, $W(\alpha)=(2/\pi)e^{-2|\alpha|^2}$. 
\label{fig:calibration of primitive pulses}}
\end{figure}

{\bf Qubit rotation.} The waveform for transmon $g\leftrightarrow e$ and $e\leftrightarrow f$ rotations is a Gaussian with $\sigma=8\,\rm ns$ and symmetric chop at $2\sigma$. The pulse amplitude is calibrated with a standard amplitude Rabi experiment, shown in Fig.~\ref{fig:calibration of primitive pulses}(a). We find that finite negative detuning of a few MHz is needed to maximize the Rabi contrast in both cases. In a similar manner, we calibrate a selective square pulse of duration $2\pi/\chi\approx 22\,\rm \mu s$ that performs $g\leftrightarrow e$ rotations conditioned on the oscillator in $|0\rangle$.

{\bf Oscillator displacement.} The waveform for oscillator displacements is a Gaussian with $\sigma=10\,\rm ns$ and symmetric chop at $2\sigma$. It is calibrated in several steps, refining the accuracy at each step. First, before the precise value of $\chi$ is known, we use a rough calibration by creating a coherent state of unknown amplitude $\alpha$ and measuring the probability of $|0\rangle$ via a selective qubit pulse, with the results shown in Fig.~\ref{fig:calibration of primitive pulses}(b). 
Fitting the data to $P(0)=e^{-|\alpha|^2}$ allows us to calibrate the DAC amplitude for displacement of $|\alpha|=1$. 
This first-stage calibration enables us to use active oscillator cooling, see Section~\ref{sec:oscillator cooling}, which is important for the next calibration step that relies on a vacuum state.
Next, after determining the value of $\chi$ (using number-resolved qubit spectroscopy, see Section~\ref{sec:hamiltonian}), we measure the Wigner function of vacuum $W(\alpha)=(2/\pi)e^{-2|\alpha|^2}$ and adjust the DAC amplitude calibration to obtain the variance of $1/4$, with the results shown in Fig.~\ref{fig:calibration of primitive pulses}(c). We find that these two calibrations  typically agree within $2\%$.


\subsection{Hamiltonian parameters \label{sec:hamiltonian}}

Our system is well described with the following Hamiltonian
\begin{align}
H/\hbar=\Delta\, (a^\dagger a) + \frac{1}{2}\chi\, (a^\dagger a)\, \sigma_z + \frac{1}{2}K\,(a^{\dagger}a)^2 + \frac{1}{4}\chi'\,(a^{\dagger}a)^2\,\sigma_z, \label{Hamiltonian}
\end{align}
where $\Delta$ is the oscillator frequency detuning in the chosen rotating frame, $\chi$ is the dispersive shift, $\chi'$ is the second-order dispersive shift, and $K$ is the Kerr nonlinearity.

\begin{figure}
    \centering
    \includegraphics[width=\figwidthDouble]{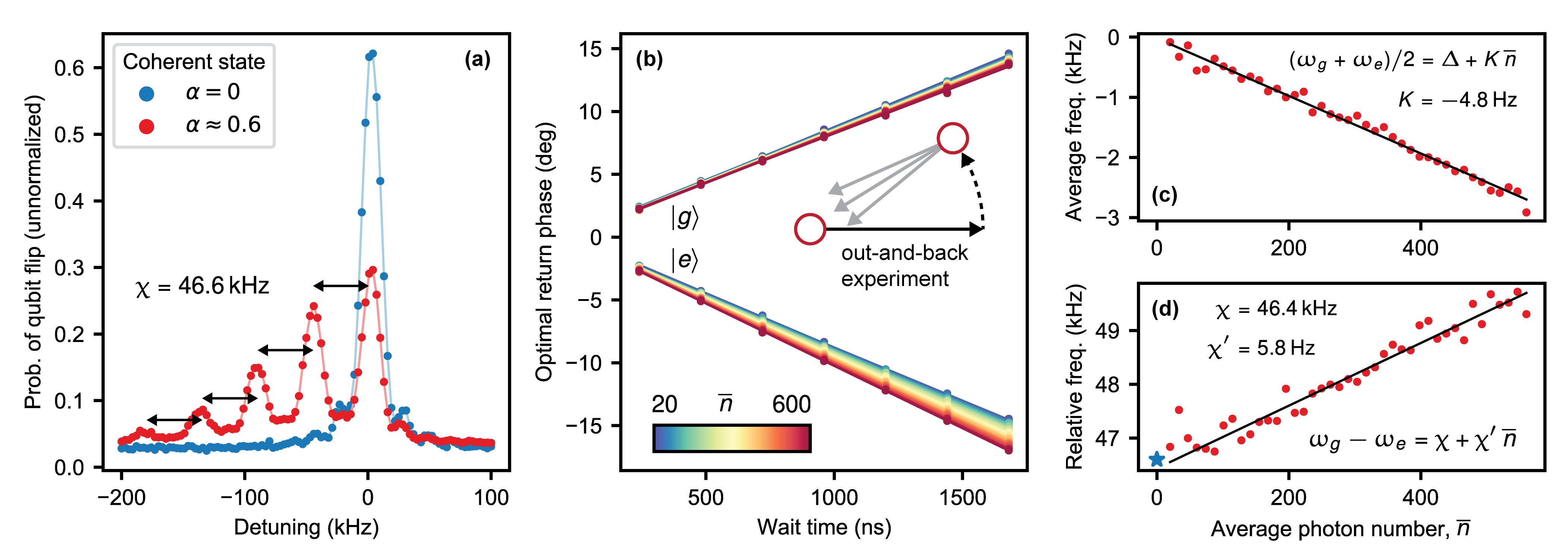}
    \caption[Calibration of Hamiltonian parameters]{
    \textbf{Calibration of Hamiltonian parameters.} 
    {\bf (a)} Number-resolved qubit spectroscopy with a selective square pulse of duration $\sim 50\,\rm \mu s$ when the oscillator is in the vacuum state (blue) and a coherent state (red). 
    {\bf (b)} Optimal return phase in the out-and-back experiment (inset) with the qubit in $|g\rangle$ and $|e\rangle$. As seen from the phase dispersion with $\overline{n}$, the effective oscillator nonlinearity is larger when the qubit is in $|e\rangle$. 
    {\bf (c)} Average oscillator rotation frequency $(\omega_g+\omega_e)/2$, and a linear fit to extract $\Delta$ and $K$. 
    {\bf (d)} Relative oscillator rotation frequency $\omega_g-\omega_e$, and a linear fit to extract $\chi$ and $\chi'$. The star indicates $\chi$ obtained in (a).
\label{fig:hamiltonian_calibration}}
\end{figure}

We calibrate $\chi$ with number-resolved qubit spectroscopy \cite{Schuster2007a} in the presence of a coherent state of small  amplitude $\alpha\approx0.6$ in the oscillator. The spectroscopy data, shown in Fig.~\ref{fig:hamiltonian_calibration}(a) in red, is fitted to a 5-component equal-spacing mixture of the spectroscopy lineshapes with the oscillator in vacuum, shown in blue, which results in $\chi=46.6\,\rm kHz$. After additionally performing the cavity mode spectroscopy (data not shown), we set the LO frequency to work in the rotating frame with $\Delta=0$.

After calibrating the displacement amplitude, as described in Section~\ref{sec:primitive pulses}, we perform an out-and-back experiment \cite{Eickbusch2021} to determine the higher order nonlinearities $K$ and $\chi'$.  In this experiment, shown in the inset of Fig.~\ref{fig:hamiltonian_calibration}(b), we create a coherent state $|\alpha\rangle$ with an average number of $\overline{n}=|\alpha|^2$ photons, wait for some time while it rotates in phase space, and attempt to return it back to the origin with a displacement of variable phase. The optimal return phase for qubit in $|g\rangle$ and $|e\rangle$ is shown in Fig.~\ref{fig:hamiltonian_calibration}(b). Performing this experiment for different wait times allows to extract the effective oscillator rotation frequencies $\omega_g(\overline{n})$ and $\omega_e(\overline{n})$. The linear fit of the average rotation frequency $(\omega_g+\omega_e)/2=\Delta + K\, \overline{n}$ yields the values of the detuning $\Delta$ and Kerr nonlinearity $K$, as shown in Fig.~\ref{fig:hamiltonian_calibration}(c). The linear fit of the relative  rotation frequency $\omega_g-\omega_e=\chi + \chi'\,\overline{n}$ yields the values of the dispersive shift $\chi$ and the second-order dispersive shift $\chi'$, as shown in Fig.~\ref{fig:hamiltonian_calibration}(d). We find that the value of $\chi$ predicted with this method typically agrees with the value obtained via number-resolved spectroscopy to within $1\%$.


\subsection{Readout and reset \label{sec:readout}}

\begin{figure}
    \centering
    \includegraphics[width=\figwidthDouble]{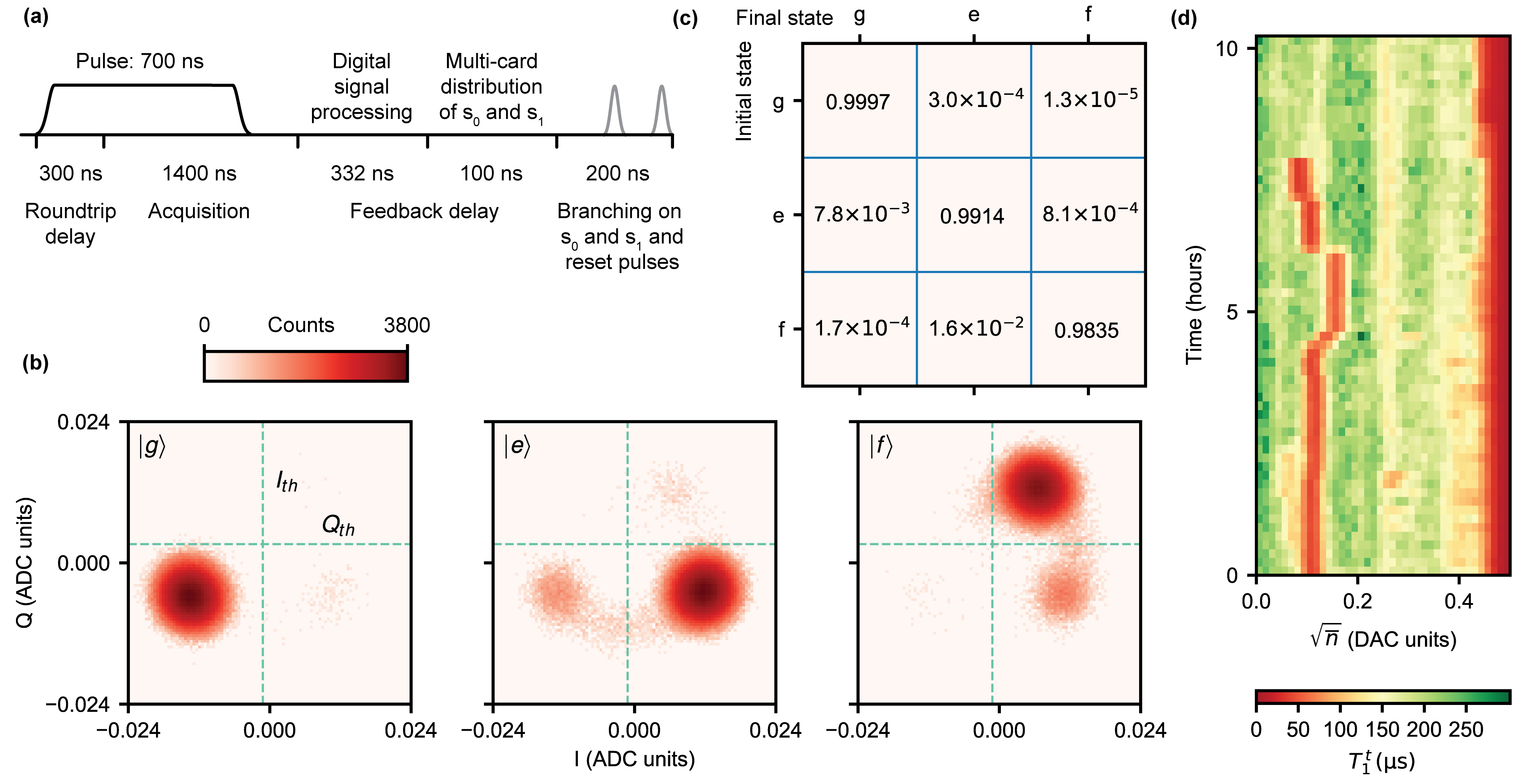}
    \caption[Ancilla measurement]{
    \textbf{Ancilla measurement.}
    {\bf (a)} Timing of different components of readout and reset.
    {\bf (b)} Logarithmic histogram of the integrated readout signal for different transmon initial states. 
    {\bf (c)} Markov transition matrix derived from the histogram in (b). It shows the probability of transition from any initial state to any final state during the measurement.
    {\bf (d)} Time trace of $|e\rangle$-state lifetime (color coded). The resonator field amplitude in the steady state is proportional to the DAC amplitude (horizontal axis). The amplitude used for the actual readout corresponds to $\sqrt{\overline{n}}=0.22$. When spurious resonance around $\sqrt{\overline{n}}=0.1$ reappears, the readout fidelity significantly reduces.
\label{fig:readout}}
\end{figure}

The transmon measurement consists of a readout pulse of duration $700\,\rm ns$ with $40\,\rm ns$ ramp-up and ramp-down. The reflected microwave signal is acquired (after $300\,\rm ns$ delay to account for signal travel time) for the duration of $1400\,\rm ns$. 
After acquiring the readout signal, FPGA performs digital signal processing, which consists of demodulation, integration of the signal with a filter function, and thresholding, all of which takes $332\,\rm ns$. Next, the bits $s_0$ and $s_1$ that encode the measurement outcome are distributed to all control cards, which takes $100\, \rm ns$. For a schematic of this measurement process, see Fig.~\ref{fig:readout}(a).
When the readout is used to realize ancilla reset, additional time is spent on branching on the $s_0$ and $s_1$ signals to apply appropriate feedback pulses. The branching is done as shown in Fig.~\ref{fig:cavity_cooling}(a), taking additional $200\,\rm ns$ to complete the reset. Due to the slow ringdown of the readout photons on a time scale of $1/(\kappa_{r(c)}+\kappa_{r(i)})\approx 320\,\rm ns$, the readout resonator is not fully empty when the feedback pulses are applied, partly limiting their fidelity through measurement-induced dephasing mechanism \cite{Gambetta2006}. This limitation could be addressed in the future by using a strongly coupled readout resonator with photon lifetime on the order of ten nanoseconds \cite{Walter2017}, or, alternatively, by using an active resonator depletion protocol \cite{McClure2016} as was done in the grid-code experiment \cite{Campagne-Ibarcq2020}.

To characterize the readout, we perform a two-measurement experiment in which the ancilla state is prepared with post-selection on the outcome of the first measurement \cite{Touzard2018}. The second measurement follows with a $500\,\rm ns$ delay after the first one. Its outcome is histogrammed, as shown in Fig.~\ref{fig:readout}(b) for $|g\rangle$, $|e\rangle$ and $|f\rangle$ initial states. The SNR of our readout is large enough to not be a dominant cause of the readout infidelity. Instead, the fidelity is limited by state transitions during the finite readout time. Some transitions are expected due to the finite lifetime of the ancilla excited states, and the excess is induced by the readout pulse itself \cite{Sank2016}. 

The Markov matrix in Fig.~\ref{fig:readout}(c) describes the transition probabilities in this characterization experiment. It is obtained by integrating the parts of a histogram on various sides of two thresholds. The diagonal elements of this matrix can be interpreted as readout fidelities of different transmon states, with precision of about $\sim10^{-3}$ for $|e\rangle$ and $|f\rangle$ states due to possible decay during the $500\,\rm ns$ delay between the two measurements.
The readout fidelity of the ground state ${\cal F}_r^{(g)}=0.9997$ is significantly better than that of the excited state ${\cal F}_r^{(e)}=0.9914$ -- a crucial feature exploited in our QEC protocol, where the dominant ``no error'' syndrome is mapped to the $g$ outcome 

The readout fidelity of $|g\rangle$ is also more stable over time, see Section~\ref{sec:stability}.  
The readout fidelity of $|e\rangle$ fluctuates over time due to fluctuations of $T_1^{\,t}(\overline{n})$ (qubit lifetime in the presence of $\overline{n}$ readout photons). A sample drift of $T_1^{\,t}(\overline{n})$ is shown in Fig.~\ref{fig:readout}(d). The exact reason for this effect is still not well understood. 
It is possible that the dependence $T_1^{\,t}(\overline{n})$ comes from drive-induced hybridization of the transmon energy levels  \cite{Shillito2022}. Higher levels are sensitive to offset charge, and thus fluctuations of environmental charges can affect the hybridization strength and lead to fluctuations of $T_1^{\,t}(\overline{n})$. 
Another possible explanation is that fluctuating $T_1^{\,t}(\overline{n})$ dependence comes from a spectral overlap of the Stark-shifted qubit frequency with a spurious degree of freedom (not necessarily charged), e.g. a two-level defect, which itself fluctuates \cite{Klimov2018}.
The correlation between the logical qubit performance, the readout infidelity of $|e\rangle$ state, and the fluctuating $T_1^{\,t}(\overline{n})$ is further discussed in Section~\ref{sec:stability}.


\subsection{Conditional displacement \label{sec:conditional displacement}}

\begin{figure}
    \centering
    \includegraphics[width=\figwidthDouble]{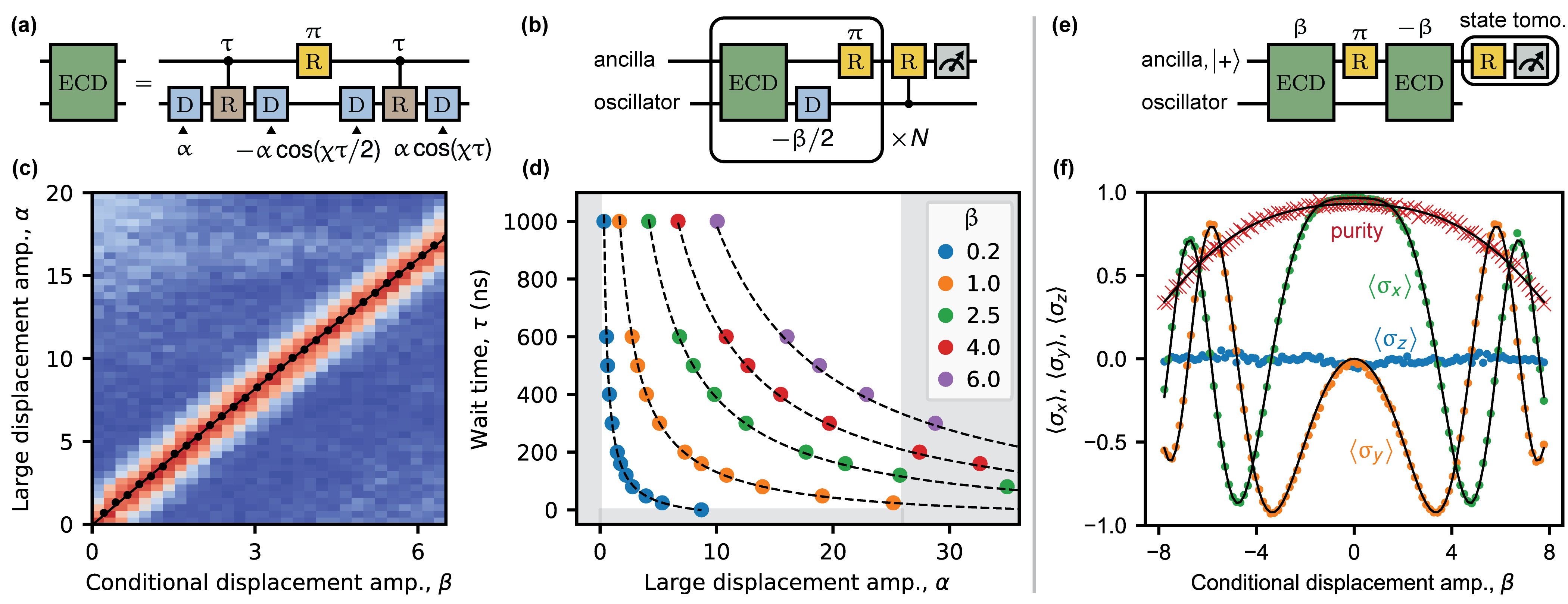}
    \caption[Calibration of the ECD gate]{
    \textbf{Calibration of the ECD gate.}
    {\bf (a)} Realization of ECD gate using the approach from \cite{Eickbusch2021}.
    {\bf (b)} Variation of the out-and-back experiment for calibration of the amplitude $\alpha$ of the large displacement required to achieve a conditional displacement amplitude $\beta$. For a fixed delay $\tau$ between out and back displacements, and a given value of $\beta$, we sweep $\alpha$ to find the optimum.  The out-and-back sequence is repeated $N$ times to increase the resolution. At the end of the experiment, the qubit is probed with a  selective pulse conditioned on oscillator in $|0\rangle$.
    {\bf (c)} Data from the amplitude calibration experiment shown in (b), using $\tau=600\rm\, ns$ and $N=4$.
    {\bf (d)} Simultaneous fit of the collection of fixed-$\tau$ datasets, such as the one shown in (c), to the empirical function in Eq.~\eqref{tau_empirical}. The shaded region indicates the prohibited parameter values. 
    {\bf (e)} Cat-and-back experiment. Starting with a pure qubit state $|+\rangle$ and arbitrary oscillator state, this experiment results in a phase accumulation in the equatorial plane on a qubit Bloch sphere, which is detected with qubit state tomography. 
    {\bf (f)} Results of the qubit state tomography in cat-and-back experiment with the $\rm ECD$ gate with wait time $\tau=600\rm\, ns$. In this experiment, the oscillator was initially prepared in the $|+Z\rangle$ grid state. The data is fitted to the model in Eq.~\eqref{cat-and-back model}, shown with black solid lines.
\label{fig:ECD_calibration}}
\end{figure}

We create an echoed conditional displacement gate ${\rm ECD}(\beta)=\sigma_x \,D(\sigma_z\,\beta/2)$ using the approach described in Ref.~\cite{Eickbusch2021}. As illustrated in Fig.~\ref{fig:ECD_calibration}(a), this gate consists of the following steps: (i) the oscillator is displaced out in phase space by large amplitude $\alpha$;  (ii) the conditional rotation is accumulated during the time interval $\tau$ along the arc of a large radius $|\alpha|$ -- this is equivalent to accumulation of the conditional displacement in the direction orthogonal to $\alpha$ at an enhanced rate $\chi|\alpha|$; (iii) the oscillator is returned back towards the origin of phase space with displacement of amplitude $-\alpha\cos(\chi\tau/2)$; (iv) the qubit state is flipped with an echo $\pi$-pulse; (v) an analogous large displacement sequence is repeated in the symmetrically opposite direction in phase space. Under the dispersive coupling model and in the limit of instantaneous rotation and displacement pulses, this protocol results in a net conditional displacement of amplitude $\beta=-2i\alpha\sin(\chi\tau)$. Due to deviations from this idealized scenario, such as finite pulse durations and higher order Hamiltonian terms, we need to experimentally calibrate the amplitude $\alpha(\beta)$ of the large displacement required to achieve a desired conditional displacement.

{\bf Calibration of amplitude.}
Starting with qubit in $|g\rangle$ (or $|e\rangle$, with similar results) and oscillator in $|0\rangle$, we apply the $\rm ECD$ gate with fixed delay $\tau$ in the displaced state and varying amplitude $\alpha$, and then attempt to undo the effect of the gate and return the oscillator to vacuum with a simple displacement $D(-\beta/2)$. This out-and-back sequence is repeated $N$ times to increase the resolution. At the end of the experiment, the qubit is probed with a  selective pulse conditioned on oscillator in $|0\rangle$. The complete sequence is illustrated in Fig.~\ref{fig:ECD_calibration}(b), and the experimental data for the gate with delay $\tau=600\,\rm ns$ is shown in Fig.~\ref{fig:ECD_calibration}(c). From this calibration measurement, we fit the dependence $\alpha(\beta|\tau)$, and we perform this calibration for a set of different wait times $\tau$.

During the optimization of the QEC performance, our RL agent is asked to pick the optimal values of the large displacement amplitude $\alpha$ and of the conditional displacement amplitude $\beta$. Therefore, we need to have a calibrated inversion function $\tau(\alpha, \beta)$ that predicts the wait time $\tau$ to realize a gate with these parameters. We find that the empirical relation
\begin{align}
\tau_e(\alpha,\beta) = \beta\left(p_0+\frac{p_1}{2\alpha}\right)-p_2, \label{tau_empirical}
\end{align}
with fit parameters $\vec{p}=\{p_0,p_1,p_2\}$, is able to simultaneously fit all ECD calibration datasets, such as the one shown in Fig.~\ref{fig:ECD_calibration}(c), sufficiently well to be used with the training of the RL agent. Note that in the idealized model, we would have $\vec{p}=\{0,1/\chi,0\}$. The empirical fit results are shown in Fig.~\ref{fig:ECD_calibration}(d), where the shaded region indicates the prohibited parameter values, including the limited dynamic range of the DAC that allows $\alpha\in [0, 26]$ given our choice of fixed-duration displacement pulses.

{\bf Calibration of qubit phase.}
As explained in Ref.~\cite{Eickbusch2021}, this experimental implementation of the ${\rm ECD}(\beta)$ gate results in additional qubit phase accumulation $\Theta[\beta]=\xi|\beta|^2$, i.e. we implement $\overline{{\rm ECD}}(\beta) = \exp(-i\sigma_z\,\Theta[\beta]/2) \, {\rm ECD}(\beta)$. The amplitude calibration experiment described above is not sensitive to this phase, because the qubit always remains in the eigenstate of $\sigma_z$. However, this phase is important when conditional displacements are concatenated, e.g. in the ECD control unitaries, as described in Section~\ref{sec:ECD control optimization}.

To calibrate this phase $\Theta[\beta]$, we perform the following ``cat-and-back'' experiment
\begin{align}
\overline{{\rm ECD}}(-\beta) \, R_x(\pi) \, \overline{{\rm ECD}}(\beta),
\end{align}
also shown in Fig.~\ref{fig:ECD_calibration}(e), which is ideally equivalent to $\sigma_x\exp(i\Theta[\beta]\sigma_z)$. Starting with a qubit in $|+\rangle$, the final state will satisfy $\langle\sigma_y\rangle=\sin(2\xi|\beta|^2)$ and $\langle\sigma_x\rangle=\cos(2\xi|\beta|^2)$ irrespective of the initial oscillator state. 

However, due to decoherence and control imperfections in the $\rm ECD$ implementation, we find that the ancilla qubit also experiences loss of purity. Under the assumption that the losses of purity during the two conditional displacement gates $\overline{{\rm ECD}}(\beta)$ and $\overline{{\rm ECD}}(-\beta)$ are uncorrelated and independent of the direction in phase space, we model it as a uniform contraction of the Bloch vector by $\sqrt{1-p[\beta]}$ per $\rm ECD$ gate, where $p[\beta]=\eta_0+\eta_2\,|\beta|^2+\eta_4\,|\beta|^4$. Hence, we fit the cat-and-back experiment to the following model:
\begin{align}
\left(\begin{array}{c}
\langle\sigma_{x}\rangle\\
\langle\sigma_{y}\rangle
\end{array}\right)=(1-p[\beta])\left(\begin{array}{c}
\cos(2\Theta[\beta])\\
\sin(2\Theta[\beta])
\end{array}\right), \label{cat-and-back model}
\end{align}
with fit parameters $\{\eta_0,\eta_2,\eta_4,\xi\}$, of which only $\xi$ is used in the $\rm ECD$  control compilation method, see Section~\ref{sec:ECD control optimization}.

The results of qubit state tomography together with the fit to the model in Eq.~\eqref{cat-and-back model} are shown in Fig.~\ref{fig:ECD_calibration}(f) for the same $\rm ECD$ gate as in Fig.~\ref{fig:ECD_calibration}(c). As explained in Ref.~\cite{Eickbusch2021}, the value of $\xi$ depends on the shape of the phase space trajectory during the $\rm ECD$ gate, and thus we calibrate it independently for every choice of delay time $\tau$. 


\subsection{Oscillator error channels \label{sec:Fock encoding characterization}}

\begin{figure}
    \centering
    \includegraphics[width=\figwidthDouble]{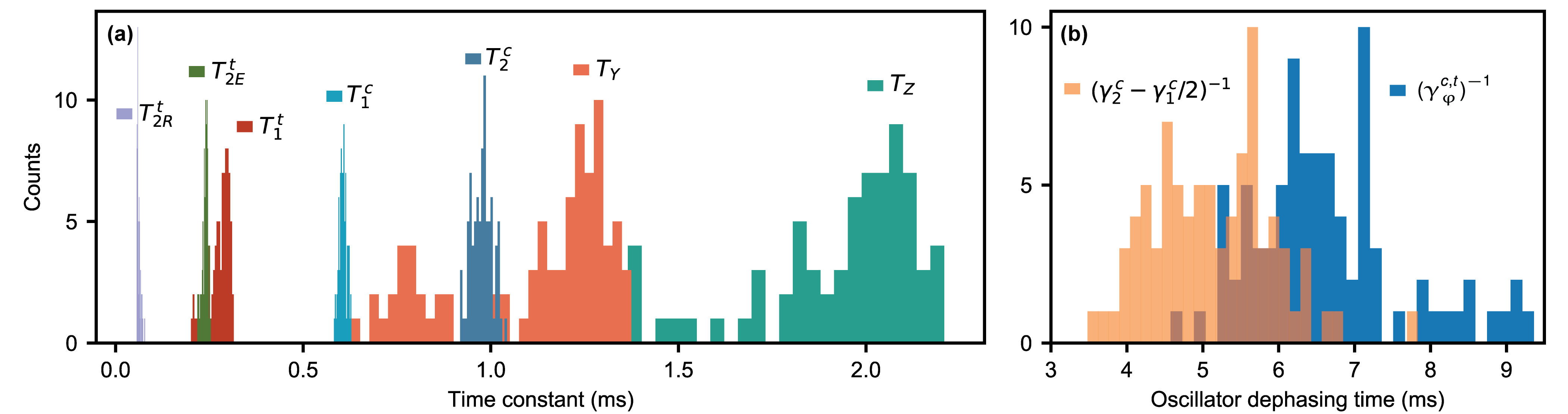}
    \caption[Fluctuating error channels]{
    \textbf{Fluctuating error channels.}
    {\bf (a)} Histogram of $T_1$ and $T_2$ times of the transmon and the oscillator, and logical lifetimes of error-corrected grid states. The histogram is derived from a week-long scan described in Chapter~\ref{sec:stability}. 
    {\bf (b)} Oscillator pure dephasing time extracted from the measured oscillator parameters and predicted from the dispersive coupling model.
\label{fig:histogram_lifetimes}}
\end{figure}

{\bf Relaxation and excitation.} To measure the oscillator relaxation rate $\gamma_1^{\,c}=\gamma_\downarrow^{\,c}+\gamma_\uparrow^{\,c}$, we first prepare Fock state $|1\rangle$ using a unitary control circuit with 5 layers, see Section~\ref{sec:ECD control optimization}. After a time delay of varying length, we measure the remaining occupation of $|1\rangle$ and fit it to an exponential decay with time constant $T_1^{\,c}=1/\gamma_1^{\,c}$. To measure this occupation, we apply a spectrally selective ancilla qubit pulse which flips the qubit conditioned on one photon in the oscillator. Monitoring the oscillator over a week-long period, we find the mean and standard deviation of $\overline{T}_1^{\, c}=606\pm10\,\rm \mu s$. As seen from the histogram in Fig.~\ref{fig:histogram_lifetimes}(a), the relative fluctuations of $T_1^{\, c}$ are small compared to the relative fluctuations of other error channels in the same time frame. We attribute this stability to the fact that most of the electromagnetic field of this mode resides in the vacuum of the cavity.

To bound the rate of thermal excitation $\gamma_\uparrow^{\,c}$, we apply the feedback cooling technique described in Section~\ref{sec:oscillator cooling}, to the oscillator in its steady state. Since we find no detectable difference in the qubit number-resolved spectroscopy contrast of the zeroth peak after feedback cooling, the  resolution of this measurement of $\sim 1\,\%$ provides a bound on the oscillator excitation rate of $\gamma_\uparrow^{\,c}<1/(60{\,\rm ms})$. This rate is negligible compared to all other rates in the system and is ignored in the rest of the analysis.


{\bf Dephasing.} To measure the rate of dephasing $\gamma_2^{\,c}$ within the $\{|0\rangle,|1\rangle\}$ manifold, we prepare a superposition $|0\rangle + |1\rangle$ using the $Y90$ gate realized with a unitary control circuit with 8 layers, see Table~\ref{tab:ECDC-Parameters}. After a time delay of varying length we apply the $Y90$ gate again and  measure the occupation of $|0\rangle$. In the  reference frame of the LO, the oscillator state rotates with angular frequency $\chi/2$ during the time delay, which results in Ramsey oscillations modulating the exponential decay with decay time constant $T_2^{\,c}=1/\gamma_2^{\,c}$. We adjust the sampling rate to make the oscillations appear slow. We find the one-week mean and standard deviation of $\overline{T}_2^{\, c}=980\pm30\,\rm \mu s$.

One possible source of oscillator dephasing is stochastic rotations acquired due to dispersive coupling with the transmon combined with transmon stochastic excitation and relaxation events \cite{Reagor2016}. The  dephasing rate due to this effect was predicted to be $\gamma_\varphi^{\,c,t}\approx n_{\rm th}^{\,t}\gamma_\downarrow^{\,t}$  in the limit $\chi\gg\gamma_1^{\,t}$ and $\gamma_\downarrow^{\,t}\gg\gamma_\uparrow^{\,t}$, where $n_{\rm th}^{\,t}$ is the steady-state population of $|e\rangle$. In our system, the correlation between $\gamma_\varphi^{\,c}=\gamma_2^{\,c}-\gamma_1^{\,c}/2$ and $\gamma_\varphi^{\,c,t}$ is difficult to measure because these rates are small and  their estimators are subject to strong relative fluctuations. 
By comparing the medians of their marginal distributions, $\overline{\gamma}_\varphi^{\,c,t}=1/(6.5\,\rm ms)$ and $\overline{\gamma}_\varphi^{\,c}=1/(5.1\,\rm ms)$, shown in Fig.~\ref{fig:histogram_lifetimes}(b), we find the remaining unexplained contribution to dephasing at a rate $\gamma^{\, c}_{?}=1/(24\,\rm ms)$ whose source is not yet identified. It is plausibly related to second-order excitations from $|e\rangle$ to $|f\rangle$ \cite{Rosenblum2018}.



\subsection{Active oscillator cooling \label{sec:oscillator cooling}}

Given the long relaxation time $\overline{T}_1^{\,c}=606\,\rm \mu s$ of our oscillator, passive cooling that relies on the natural interaction with the cold environment is impractically long. For example, starting with a Fock state $|1\rangle$, it would take approximately $4.6\,\overline{T}_1^{\,c}=2.8\, \rm ms$ to reduce the average population to $0.01$ photons. In practice, since we work with the grid states, the required cooling time is even longer. Therefore, the goal of our active cooling routine is to reduce the experimental duty cycle time and also to remove any residual thermal population. We achieve these goals via a two-step procedure which consists of an engineered dissipative pre-cooling and subsequent feedback cooling. 


\begin{figure}
    \centering
    \includegraphics[width=\figwidthDouble]{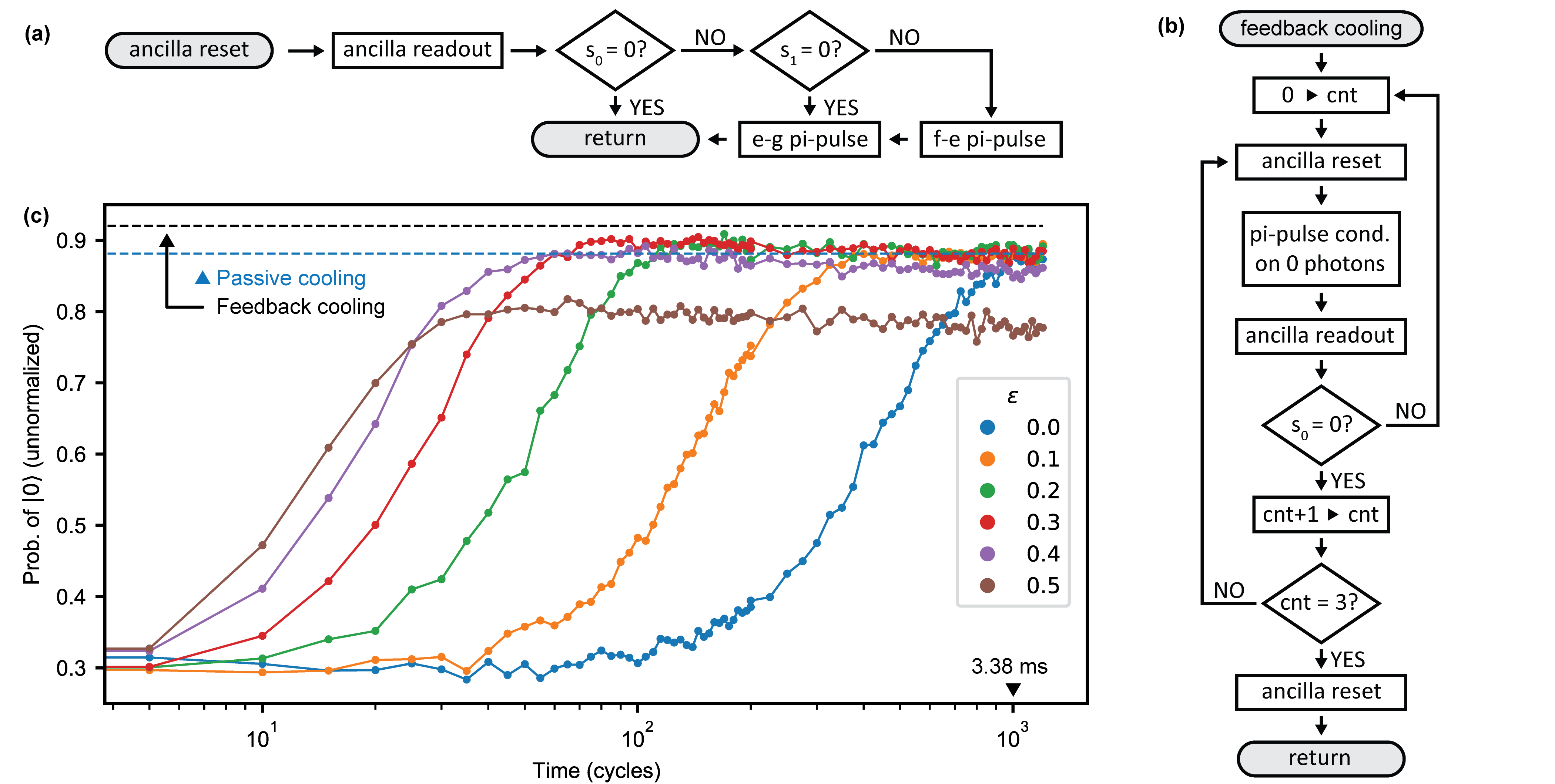}
    \caption[System cooling]{
    \textbf{System cooling.}
    {\bf (a)} Ancilla reset subroutine with measurement-based three-state feedback.
    {\bf (b)} Oscillator feedback cooling subroutine adapted from Ref.~\cite{Ofek2016a}.
    {\bf (c)} Demonstration of dissipative cooling of the oscillator starting from GKP $|+Z\rangle$ state with $\Delta=0.3$. A single cooling cycle consists of a pulse sequence in Eq.~\eqref{cooling_ecdc}, ancilla qubit reset as in (a), and virtual rotation gate to the orthogonal quadrature for the next cycle. The duration of a single such cycle is $3.38\,\rm us$. The case $\varepsilon=0$ is equivalent to passive cooling. Dashed lines represent the contrast of the zeroth photon number peak in qubit spectroscopy after passive cooling of $5\,\rm ms$ and after feedback cooling with $Y=3$. 
\label{fig:cavity_cooling}}
\end{figure}

{\bf Dissipative pre-cooling.}
We introduce a novel oscillator cooling method based on the conditional displacements, ancilla rotations, and ancilla resets. This protocol can also be realized in trapped ions, as was hinted in Ref.~\cite{DeNeeve2020}.

To derive this protocol, we apply the same dissipation engineering framework \cite{Gross2018} as used in Ref.~\cite{Royer2020} to derive the SBS stabilization of the GKP manifold. The dissipator $\gamma\,{\cal D}[a]$ can be approximated with a sequence of discrete entangling interactions $U(t)$ between the ancilla and the oscillator, and ancilla resets. For $\gamma\,{\cal D}[a]$, the  interaction should be of the form $U(t)=\exp[-i\sqrt{\gamma t}(a\sigma_++a^\dagger\sigma_-)]$, where the constraint $\langle a^\dagger a\rangle \gamma t\ll 1$ controls the validity of this discrete approximation. To further approximate this unitary as a multi-layer circuit with gates from our gate set, we perform the first order Trotter decomposition:
\begin{align}
U &=\exp\left(-i\sqrt{\frac{\gamma t}{2}}(x\sigma_x+p\sigma_y)\right) \\
&= \exp\left(-i\sqrt{\frac{\gamma t}{2}}x\sigma_x\right) \exp\left(-i\sqrt{\frac{\gamma t}{2}}p\sigma_y\right) + O(\gamma t) \\
&\approx R_y^\dagger(\pi/2)\, {\rm ECD}(-i\varepsilon)\, R_x^\dagger(\pi/2)\, {\rm ECD}(\varepsilon)\, R_y(\pi/2)R_z (\pi/2),
\end{align}
where we defined the ``trimming amplitude'' $\varepsilon=\sqrt{\gamma t}$. Furthermore, since the ancilla qubit is assumed to always start in $|g\rangle$, we can omit the first gate $R_z(\pi/2)$. The resulting unitary part of the dissipative cooling circuit is:
\begin{align}
R_y^\dagger(\pi/2)\,\, {\rm ECD}(-i\varepsilon)\, R_x^\dagger(\pi/2)\,\, {\rm ECD}(\varepsilon)\, R_y(\pi/2),\label{cooling_ecdc}
\end{align}
also summarized in Table~\ref{tab:ECDC-Parameters}. To achieve uniform cooling in all directions in phase space, the orientation of the $\rm ECD$ gates needs to cycle between position and momentum quadratures. A single cycle, including the pulse sequence in \eqref{cooling_ecdc}, ancilla reset, and subsequent virtual rotation gate on the FPGA, has a duration of $3.38\,\rm \mu s$.

To demonstrate the performance of this cooling protocol, we start with a $|+Z\rangle$ grid state with $\Delta=0.3$ and apply varying number of cooling cycles, monitoring the population of $|0\rangle$ with a selective qubit pulse. As seen in Fig.~\ref{fig:cavity_cooling}(c), dissipative cooling allows the state to shrink towards vacuum significantly faster than passive cooling. With $\varepsilon=0.4$, the cooling rate is 20 times faster than energy relaxation time of the oscillator. For small $\varepsilon\leq 0.3$ the steady-state thermal occupation after dissipative cooling is similar to passive cooling of this state of duration $5\rm \, ms$. Larger $\varepsilon$ allows for faster cooling, but at the expense of  significant residual thermal occupation.

{\bf Feedback cooling.}
To remove the residual thermal photons, we further apply the feedback cooling protocol introduced in Ref.~\cite{Ofek2016a} and shown in Fig.~\ref{fig:cavity_cooling}(b). With the help of a selective qubit pulse conditioned on $|0\rangle$ and qubit measurement, the protocol repetitively asks the question ``Is the oscillator in vacuum?'' and terminates only when it receives $Y$ consecutive ``yes'' answers. It would be inefficient to run this feedback protocol starting with an arbitrary initial oscillator state, since the probability $p_y$ of obtaining ``yes'' can be very small. The dissipative pre-cooling quickly boosts this probability to a level essentially limited by the fidelity of the selective qubit pulse, and thereby decreases the run time of the subsequent feedback cooling step. The run time of feedback cooling is non-deterministic, but the expected number of rounds in a model with constant $p_y$ is 
\begin{align}
\overline{N}_{fc}(p_y, Y) = \frac{p_y^{-Y}-1}{1-p_y}
\end{align}

Our final routine, called ``active cooling'' throughout this work, consists of $25$ cycles of dissipative pre-cooling ($50$ cycles, if counting each quadrature individually) with $\varepsilon=0.4$ followed by the feedback cooling with $Y=3$. We estimate that with $p_y=0.87$, achieved after the pre-cooling, the expected run time of the whole routine is approximately $50\times 3.38{\rm\, \mu s}+\overline{N}_{fc}(0.87, 3)\times 25{\rm\, \mu s}=270 {\,\rm \mu s}$, which in our system corresponds to $0.45\, T_1^{\,c}$ (and could potentially be reduced further). 

From the contrast of the zeroth photon number peak in the qubit spectroscopy [dashed lines in Fig.~\ref{fig:cavity_cooling}(c)], we see that passive cooling of duration $5\rm\, ms$ starting from the $|+Z\rangle$ grid state still leaves a residual thermal population larger than what our protocol achieves in a much shorter time. However, when active cooling is applied to an oscillator in its steady state (nominally, vacuum) we find no resolvable improvement of the spectroscopy contrast, which leads us to conclude that the residual thermal population after active cooling is at the sub-percent level where it cannot be resolved with our spectroscopy. This observation is used in Section~\ref{sec:Fock encoding characterization} to derive an upper bound on the  oscillator thermal excitation rate.

\section{Quantum control optimization \label{sec:quantum control}}

\begin{table}
\begin{centering}
\begin{tabular}{|c|>{\centering}p{1cm}>{\centering}p{1cm}>{\centering}p{1cm}|>{\centering}p{1cm}>{\centering}p{1cm}>{\centering}p{1cm}|>{\centering}p{2.25cm}>{\centering}p{1cm}>{\centering}p{1cm}|>{\centering}p{2.25cm}>{\centering}p{1cm}>{\centering}p{1cm}|}
\hline 
 & \multicolumn{3}{c|}{Dissipative cooling} & \multicolumn{3}{c|}{SBS protocol} & \multicolumn{3}{c|}{$|+Z\rangle$ grid state prep.} & \multicolumn{3}{c|}{$Y90$ gate for $\{|0\rangle,|1\rangle\}$ qubit}\tabularnewline
\hline
\hline
$t$ & \multicolumn{1}{>{\centering}p{1cm}|}{$\boldsymbol{\beta}$} & \multicolumn{1}{>{\centering}p{1cm}|}{$\boldsymbol{\varphi}$} & $\boldsymbol{\theta}$ & \multicolumn{1}{>{\centering}p{1cm}|}{$\boldsymbol{\beta}$} & \multicolumn{1}{>{\centering}p{1cm}|}{$\boldsymbol{\varphi}$} & $\boldsymbol{\theta}$ & \multicolumn{1}{>{\centering}p{2.25cm}|}{$\boldsymbol{\beta}$} & \multicolumn{1}{>{\centering}p{1cm}|}{$\boldsymbol{\varphi}$} & $\boldsymbol{\theta}$ & \multicolumn{1}{>{\centering}p{2.25cm}|}{$\boldsymbol{\beta}$} & \multicolumn{1}{>{\centering}p{1cm}|}{$\boldsymbol{\varphi}$} & $\boldsymbol{\theta}$\tabularnewline
\hline 
1 & $+0.4\,$ & $+\pi/2$ & $+\pi/2$ & $+0.2i$ & $+\pi/2$ & $+\pi/2$ & $+0.52+2.54i$ & $-1.28$ & $+1.57$ & $+0.64+0.11i$ & $-1.06$ & $+1.58$\tabularnewline
2 & $-0.4i$ & $0$ & $-\pi/2$ & $+\sqrt{2\pi}$ & $0$ & $-\pi/2$ & $-0.83-0.36i$ & $+2.85$ & $-2.76$ & $-0.15-1.00i$ & $+2.64$ & $-1.44$\tabularnewline
3 & $0$ & $+\pi/2$ & $-\pi/2$ & $+0.2i$ & $0$ & $+\pi/2$ & $-0.36+0.85i$ & $+0.29$ & $+0.55$ & $+1.02+0.05i$ & $+0.58$ & $-1.97$\tabularnewline
4 &  &  &  & $0$ & $+\pi/2$ & $-\pi/2$ & $-0.86+1.61i$ & $-0.29$ & $+1.43$ & $+1.55+1.02i$ & $-1.84$ & $-1.55$\tabularnewline
5 &  &  &  &  &  &  & $-2.16+0.12i$ & $+0.29$ & $+0.92$ & $+0.34+1.06i$ & $+2.74$ & $+0.26$\tabularnewline
6 &  &  &  &  &  &  & $-0.09+1.73i$ & $+2.85$ & $-1.56$ & $-0.26-0.92i$ & $-0.01$ & $-1.25$\tabularnewline
7 &  &  &  &  &  &  & $+2.05+0.73i$ & $+0.29$ & $+1.08$ & $-0.61-0.05i$ & $-2.91$ & $-1.75$\tabularnewline
8 &  &  &  &  &  &  & $+0.22-0.66i$ & $-0.29$ & $-2.71$ & $+0.02+0.05i$ & $-1.79$ & $-1.67$\tabularnewline
9 &  &  &  &  &  &  & $-0.08-1.56i$ & $+0.29$ & $+2.06$ &  &  & \tabularnewline
10 &  &  &  &  &  &  & $+0.19+0.04i$ & $+2.85$ & $+1.60$ &  &  & \tabularnewline
11 &  &  &  &  &  &  & $0$ & $+1.86$ & $+1.57$ &  &  & \tabularnewline
\hline 
\end{tabular}
\par\end{centering}

\caption{\textbf{Circuit parameters.} Parameters for dissipative cooling and SBS protocol are created based on the models described in Section~\ref{sec:oscillator cooling} and Ref.~\cite{Royer2020}  respectively. Parameters for $|+Z\rangle$ grid state preparation and Fock $Y90$ gate on $\{|0\rangle,|1\rangle\}$ qubit are numerically optimized with Keras. \label{tab:ECDC-Parameters}}
\end{table}

\subsection{Model-based optimization of control circuits \label{sec:ECD control optimization}}

{\bf Circuit decomposition.} Our control gate set consists of two parametrized gates: (i) echoed conditional displacement of the oscillator ${\rm ECD}(\beta)=\sigma_x \, D(\sigma_z\,\beta/2)$, where $D(\alpha)=\exp [\alpha a^\dagger-\alpha^* a]$ is the displacement operator, and (ii) rotation of the qubit $R(\varphi, \theta)=\exp\left[-i(\theta/2)(\sigma_x\,\cos\varphi+\sigma_y\,\sin\varphi)\right]$. Recently, it was shown that this gate set is well suited for the universal control of an oscillator with weak dispersive coupling to a qubit \cite{Eickbusch2021}. Most unitary operations in our experiment are decomposed as parametrized multilayer circuits of the form
\begin{align}
{\rm circuit}(\boldsymbol{\beta}, \boldsymbol{\varphi}, \boldsymbol{\theta}) = 
\underbrace{{\rm ECD}(\beta_{T}) R(\varphi_{T}, \theta_{T})}_{\mbox{layer }T} \,\, \cdots \,\, \underbrace{{\rm ECD}(\beta_1) R(\varphi_1, \theta_1)}_{\mbox{layer 1}},
\label{parametrized_circuit}
\end{align}
where $\boldsymbol{\beta}\in \mathbb{C}^T$ is a vector of conditional displacement amplitudes, and $\boldsymbol{\varphi}, \boldsymbol{\theta}\in\mathbb{R}^T$ are vectors of qubit rotation phases and angles respectively. For example, we utilize this decomposition as part of the following operations:
\begin{itemize}
\item[$-$] Dissipative cooling of the oscillator, see Section~\ref{sec:oscillator cooling}.
\item[$-$] Preparation of the GKP states, see Section~\ref{sec:wigner_tomo}.
\item[$-$] Small-Big-Small protocol, see Section~\ref{sec:qec_details}.
\item[$-$] Preparation of the Fock state $|1\rangle$, see Section~\ref{sec:Fock encoding characterization}.
\item[$-$] $Y90$ gate on Fock $\{|0\rangle, |1\rangle\}$ encoding, see Section~\ref{sec:Fock encoding characterization}.
\end{itemize}

\begin{figure}
    \centering
    \includegraphics[width=\figwidthDouble]{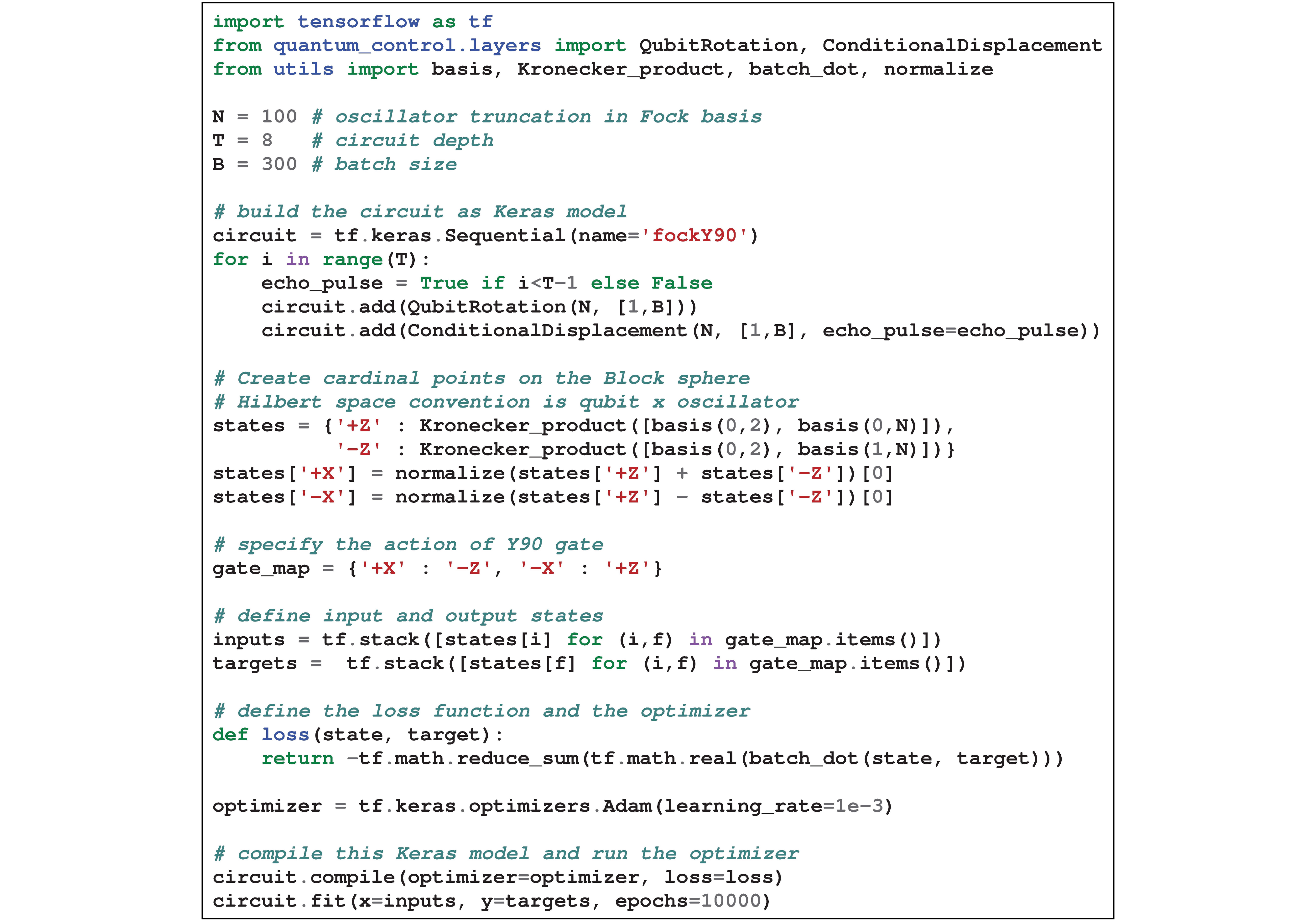}
    \caption[Circuit optimization]{
    \textbf{Circuit optimization.} Example of a Python script for optimization of the circuit parameters for $Y90$ gate on the $\{|0\rangle,|1\rangle\}$ qubit. Gates are represented as custom Keras layers, and the circuit is compiled as sequential model. Optimization utilizes TensorFlow backend for automatic differentiation of the model. 
\label{circuit_optimiztion}}
\end{figure}


{\bf Circuit optimization.} A circuit optimization method for this gate set was developed in Ref.~\cite{Eickbusch2021}. Here, we present a simplified modular framework based on the Keras library \cite{Chollet2015}, which allows to optimize circuit parameters in a manner similar to training of the  neural networks. 
The parametrized control circuits \eqref{parametrized_circuit} are created as instances of the \texttt{tf.keras.Sequential} class which is commonly used to concatenate multiple neural network layers. Here, we instead use custom layers that represent the parametrized gates ${\rm ECD}(\beta)$ and $R(\varphi,\theta)$ as subclasses of \texttt{tf.keras.layers.Layer}. This allows us to exploit flexible and user-friendly application-programming interface of the Keras library to optimize the circuit parameters and automatically monitor various aspects of the optimization progress. 
To illustrate the accessibility of such an approach, in Fig.~\ref{circuit_optimiztion} we provide an example code for optimization of the $Y90$ gate on the $\{|0\rangle,|1\rangle\}$ qubit.  Complete code with dependencies and further examples is available in Ref.~\cite{Sivak2022github}. Such optimization, which is performed for a batch of $B=300$ circuit candidates in parallel on a graphics processing unit (GPU), takes about $10$ minutes to finish. In Table~\ref{tab:ECDC-Parameters}, we list circuit parameters for some of the control operations in our experiment. Curiously, some of the numerically optimized parameter values are clearly interpretable, e.g. in GKP state preparation circuit the rotations at steps $t=1,6,10,11$ seem to be by  an angle $\pi$. Detailed inspection of these circuits can lead to improved analytic constructions, which is left for future research. 


{\bf Pulse compilation.} Having obtained the circuit parameters, we compile the waveforms to be played on the qubit and oscillator control lines. Such compilation requires prior calibration of the rotation $R(\varphi, \theta)$ gate, described in Section~\ref{sec:primitive pulses}, and the ${\rm ECD}(\beta)$ gate, described in Section~\ref{sec:conditional displacement}. 

As explained in Ref.~\cite{Eickbusch2021} and in Section~\ref{sec:conditional displacement}, our experimental implementation of the ${\rm ECD}(\beta)$ gate results in additional qubit phase accumulation $\Theta[\beta]\propto|\beta|^2$, i.e. we implement $\overline{{\rm ECD}}(\beta) = \exp(-i\sigma_z\,\Theta[\beta]/2) \, {\rm ECD}(\beta)$. We use the experimental calibration of this phase to adjust the numerically optimized vector $\boldsymbol{\varphi}$ according to the rule
\begin{align}
\varphi_t \leftarrow \varphi_{t}-\sum_{\tau=1}^{t-1} (-1)^{t-\tau} \Theta[\beta_\tau], \quad t>1.
\end{align}

In addition, in many cases of interest the ancilla qubit at the end of the circuit returns to $|g\rangle$ and disentangles from the oscillator. In such cases, the last conditional displacement ${\rm ECD}(\beta_{T})$ can be realized as a simple displacement $D(\beta_T/2)$. We use this simplification in state preparation circuits and in the SBS protocol.

\begin{figure}
    \centering
    \includegraphics[width=\figwidthDouble]{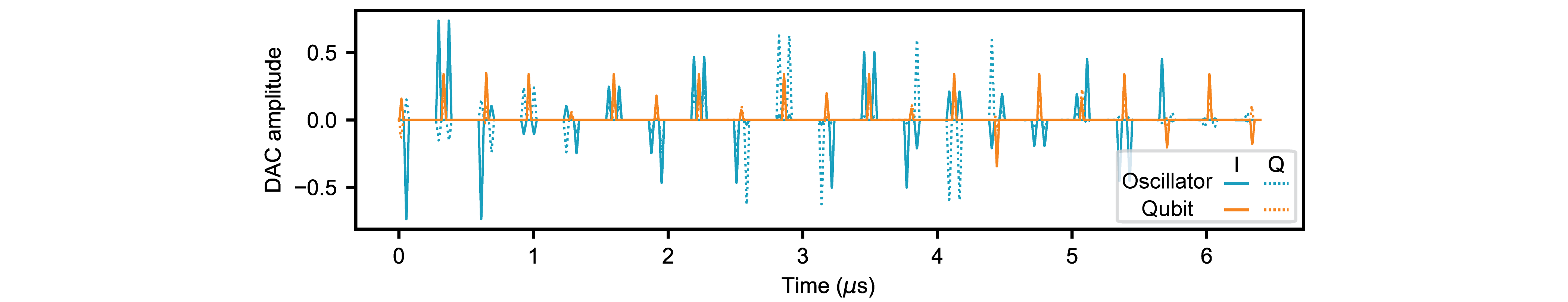}
    \caption[Waveform for grid state preparation]{
    \textbf{Waveform for $|+Z\rangle$ grid state preparation.} 
    The parametrized control circuit is decomposed into primitive gates: qubit rotations, oscillator displacements, and conditional rotations. The waveform is compiled from this sequence of gates using experimental calibrations. Each qubit rotation and oscillator displacement is replaced with a corresponding Gaussian pulse, and the conditional rotation is replaced with a delay of certain length during which the system freely evolves under the dispersive coupling Hamiltonian. 
\label{ECDC_waveform}}
\end{figure}

In Fig.~\ref{ECDC_waveform}, we show an example waveform for unitary preparation of the $|+Z\rangle$ grid state using a parametrized circuit with $T=11$ layers. Each $\rm ECD$ gate is decomposed via large displacements and conditional rotations. For clarity, in this example all conditional rotations are implemented with a constant wait time $\tau=200\,\rm ns$; hence, the whole compiled waveform has a duration of $6.4\,\rm \mu s$. Faster implementations are possible if the wait time is adapted to the magnitude of the conditional displacement, as described in Section~\ref{sec:conditional displacement}. For example, in our system the conditional displacement of amplitude $|\beta|<0.5$ could, in principle, be implemented with zero wait time, see Fig.~\ref{fig:ECD_calibration}(d).


\begin{figure}
    \centering
    \includegraphics[width=\figwidthDouble]{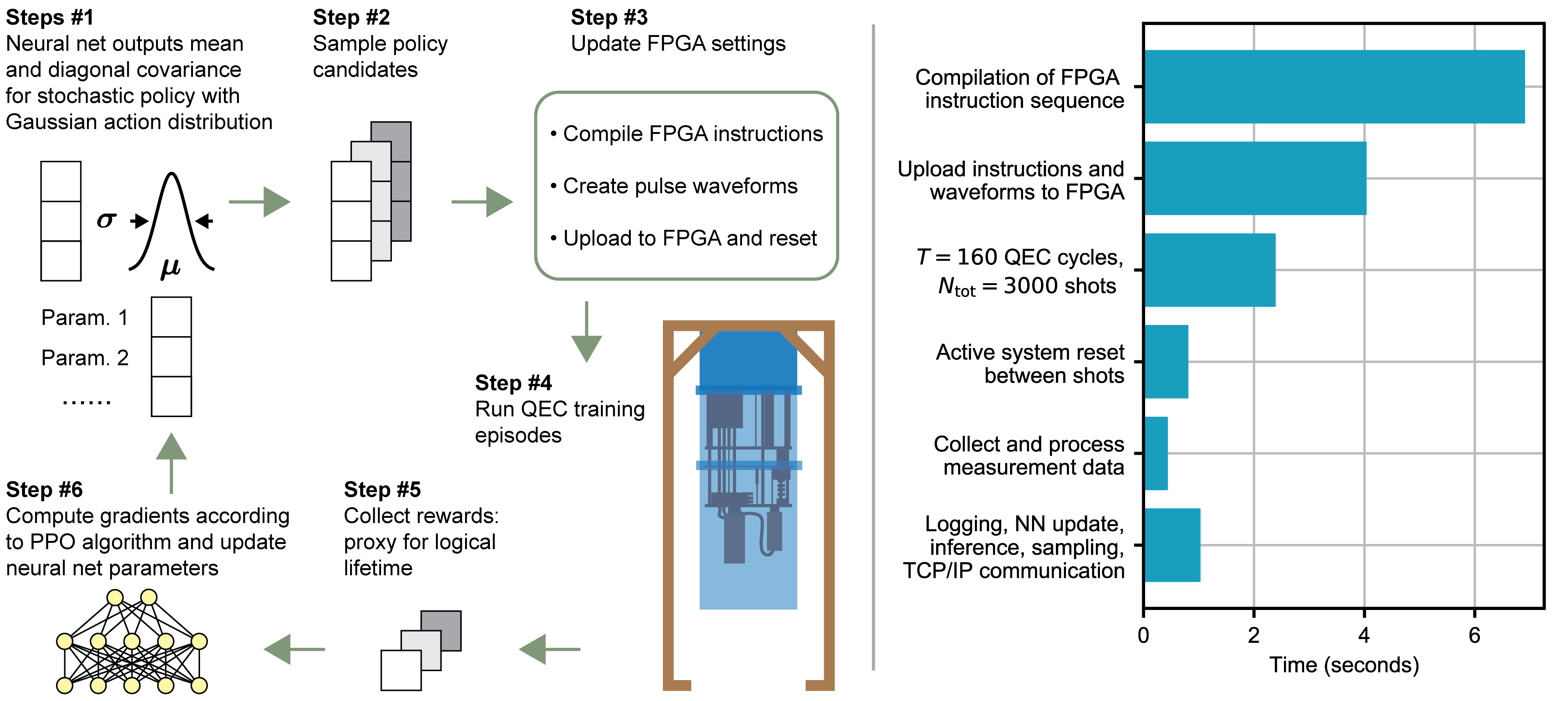}
    \caption[Reinforcement learning]{
    \textbf{Reinforcement learning. }
    {\bf (a)} Experimental training loop. 
    {\bf (b)} Training time budget per epoch.
\label{fig:RL_pipeline}}
\end{figure}

\subsection{Model-free reinforcement learning for QEC \label{sec:RL}}

While most quantum operations in our experiment are optimized with a model-based approach described above, for quantum error correction we deploy a more powerful framework of model-free optimization. We use a reinforcement learning algorithm called proximal policy optimization (PPO) \cite{Schulman2017, TFAgents}. For a detailed description of this algorithm in the context of quantum control we refer to Ref.~\cite{Sivak2021}; here, we only provide a basic high-level picture. The complete training loop of our experiment is illustrated in Fig.~\ref{fig:RL_pipeline}; it is structured as follows:


{\bf Step 1.} On training epoch $t$, neural network produces a probability distribution ${\cal N}({\vec{\mu}_t}, {\vec{\sigma}_t})$, where ${\vec{\mu}_t}=\vec{\mu}(\theta_t)$, ${\vec{\sigma}_t}=\vec{\sigma}(\theta_t)$, and $\theta_t$ summarizes the values of all weights and biases of the neural network in the current epoch.

{\bf Step 2.} We sample a batch of $B=10$ parameter vectors from this distribution.  They correspond to different QEC circuit candidates that should be evaluated in experiment.  The neural network and sampling are implemented on NVIDIA 2080Ti graphics processing unit (GPU) in a separate computer. The sampled vectors are sent to the control computer via a local area network with negligible communication time. 

{\bf Step 3.} Based on these parameter vectors, we compile QEC circuit candidates, translated into FPGA instructions and DAC waveforms. All circuit candidates follow the same program execution flow, but the control waveforms and the content of FPGA registers is different for every candidate. The FPGA is reset and its wave memory is updated.  This time-consuming step is the bottleneck of the training loop.

{\bf Step 4.} Each candidate is evaluated in experiment. To this end, we initialize logical Pauli eigenstates $|+Z\rangle$ and $|+X\rangle$, run the QEC for $T=160$ cycles, and then perform one-bit phase estimation of the corresponding logical Pauli operators. To suppress the sampling noise, we repeat this $N_{\rm avg}=150$ times per Pauli and per circuit candidate. In total, one epoch of training consists of $N_{\rm tot}=2BN_{\rm avg}=3000$ experimental shots. 

{\bf Step 5.} To produce the reward, we treat the measurement of a Pauli operator after $T$ cycles as a proxy for logical lifetime. While averaging the measurement outcomes, we mask the experimental shots that started with incorrect state initialization, as flagged by a verification ancilla measurement after the state initialization. 

{\bf Step 6.} Once the rewards are available, PPO algorithm updates the neural network parameters $\theta_t\to\theta_{t+1}$ for the next epoch. The gradients of these parameters are computed with automatic differentiation via back-propagation. After updating the neural network, the new training epoch begins.

The time budget of this training is shown in Fig.~\ref{fig:RL_pipeline}. All steps outlined above amount to $15.6\rm \, s$ per epoch. In the current implementation, the major bottleneck is Python-to-FPGA transition (step 3). Because of this, the implementation is less optimal in terms of sample efficiency than the proposal in Ref.~\cite{Sivak2021}. The optimal approach would be to spend the total sample budget per epoch to evaluate more circuit candidates  with minimal accuracy, instead of evaluation only a few candidates with high accuracy (achieved through  averaging). In other words, based on the results of Ref.~\cite{Sivak2021}, we expect that a training with $(B,N_{\rm avg})=(1000,1)$ would require fewer experimental shots to reach a given performance level than a training with $(B,N_{\rm avg})=(10,100)$. However, considering the total run time of the training, we had to compromise between bare sample efficiency (number of shots) and the overhead in step 3 of the pipeline. The overhead is independent of $N_{\rm avg}$ but increases with $B$, and due to a limited FPGA instruction sequence length we can only evaluate $B\leq10$ candidates per compilation. After paying the compilation overhead in step 3, a certain amount of averaging comes essentially for free and does not considerably affect the run time, hence the choices made here.

In Section~\ref{sec:parameters}, we describe the QEC circuit parametrization, show the evolution of parameter values during the course of training, and provide interpretation of the observed trends.



\section{Quantum error correction of the grid code \label{sec:grid code stuff}}

\subsection{Brief introduction to grid code \label{sec:intro to grid code}}


\begin{figure}[b]
    \centering
    \includegraphics[width=\figwidthDouble]{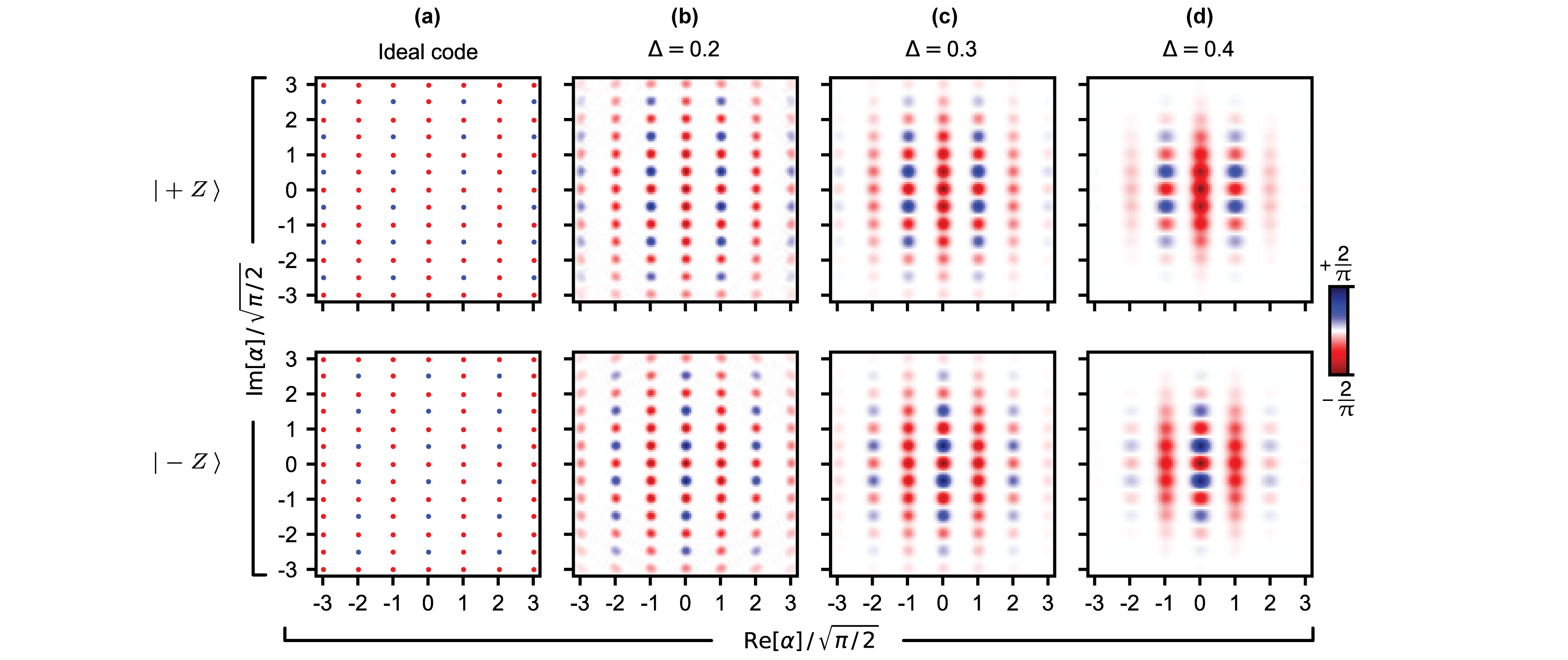}
    \caption[Wigner functions of grid states]{
    \textbf{Wigner functions (numerical) of grid states.}
\label{fig:gkp_intro}}
\end{figure}

Qubit-register stabilizer codes are based on the group of Pauli operators; consider instead a stabilizer code based on the group of oscillator displacement operators. By definition, the $+1$ eigenstates of a displacement operator $D(\alpha)$ are displacement-invariant in phase space along the direction of $\alpha$ with a period $|\alpha|$. Having two code stabilizers $S_{0}^\sX=D(\alpha_{X})$ and $S_{0}^\sZ=D(\alpha_{Z})$ imposes displacement invariance along two non-equivalent directions, which means that all codewords are grids in phase space with a unit cell defined by $\{\alpha_{X},\alpha_{Z}\}$. The requirement of commutativity of $S_{0}^\sX$ and $S_{0}^\sZ$ imposes a constraint
\begin{align}
\alpha_{X}^{*}\alpha_{Z}-\alpha_{X}\alpha_{Z}^{*}=2l_s^2 ni,\quad n\in\mathbb{Z},
\label{eq:constraint stabilizers}
\end{align}
where $l_S=\sqrt{2\pi}$. Here, we consider encoding of a single logical qubit into an oscillator, which corresponds to $n=1$. 
By parametrizing the complex-valued displacement amplitudes as $\alpha_{X}=l_S[M_{22}-iM_{12}]$
and $\alpha_{Z}=l_S[iM_{11}-M_{21}]$, we obtain a grid code with the following stabilizers:
\begin{align}
S_{0}^\sZ & =D(l_S[iM_{11}-M_{21}]),\\
S_{0}^\sX & =D(l_S[M_{22}-iM_{12}]).
\end{align}

From the constraint (\ref{eq:constraint stabilizers}) we derive a single requirement that a real matrix 
$M=\left[\begin{array}{cc}
M_{11} & M_{12}\\
M_{21} & M_{22}
\end{array}\right]$
has a determinant $\det M=1$. This matrix defines the structure of the grid in phase space. Here, we only consider the square grid code, which is obtained with
$M=\left[\begin{array}{cc}
1 & 0\\
0 & 1
\end{array}\right]$. The hexagonal code with 
$M=\sqrt{\frac{2}{\sqrt{3}}}\left[\begin{array}{cc}
1 & 1/2\\
0 & \sqrt{3}/2
\end{array}\right]$
was realized in Ref.~\cite{Campagne-Ibarcq2020}.

The Pauli operators of the logical qubit are defined as
\begin{align}
X_L & =\sqrt{S_{0}^\sX}=D(l_S[M_{22}-iM_{12}]/2),\\
Z_L & =\sqrt{S_{0}^\sZ}=D(l_S[iM_{11}-M_{21}]/2).
\end{align}
They satisfy the standard algebraic properties $X_L^{2}=I$, $Z_L^{2}=I$, and $X_LZ_L=-Z_LX_L$, inside the code space. Using the identity $Y_L=-i Z_L X_L$, we find the third Pauli operator $Y_L=-iD(l_S[iM_{11}-iM_{12}+M_{22}-M_{21}]/2)$. 

The eigenstates of Pauli $Z_L$ of the ideal grid code are shown in Fig.~\ref{fig:gkp_intro}(a).  Finite-energy code families can be obtained by regularizing the ideal code through application of an envelope operator \cite{Tzitrin2019,Royer2020}, with a common choice being $N_\Delta=\exp(-\Delta^{2}n)$. We show several members of this code family in Fig.~\ref{fig:gkp_intro}(b-d). Note that such a regularization leads to non-orthogonal states, with fidelity loss due to the finite state overlap that scales as $\propto\exp[-(3/8)\pi/\Delta^2]/(1-\exp(-2\Delta^2))$ (see Eq. S28 in Ref.~\cite{Royer2020}), which is negligible for our choice of $\Delta=0.34$.

\subsection{Small-Big-Small (SBS) protocol \label{sec:sbs}}


\begin{figure}
    \centering
    \includegraphics[width=\figwidthDouble]{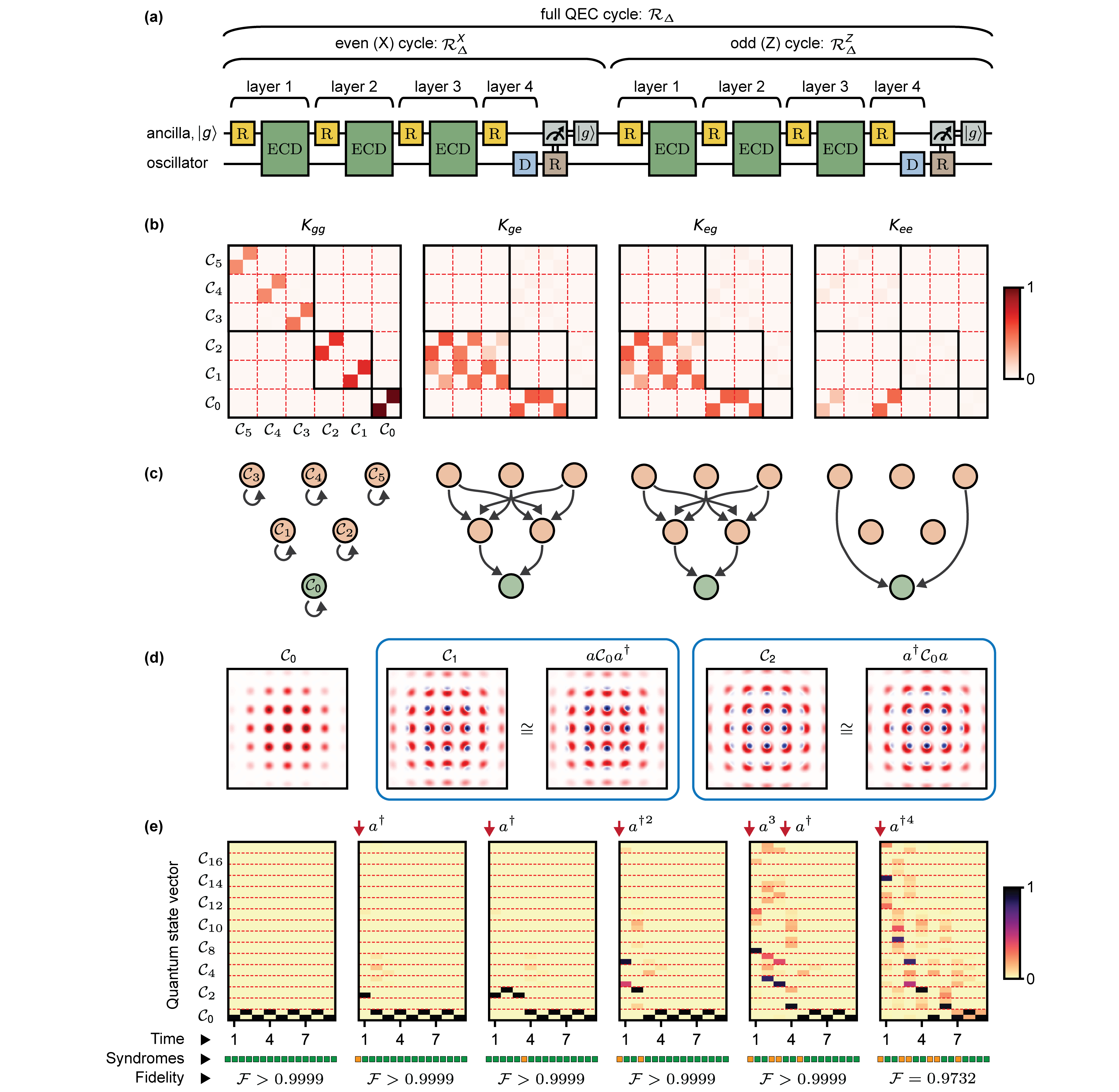}
    \caption[SBS protocol]{
    \textbf{SBS protocol.}
    {\bf (a)} Circuit structure of one QEC cycle.
    {\bf (b)} Kraus operators of the QEC cycle with $\Delta=0.34$, written in the eigenbasis of $K_{gg}^\dagger K_{gg}$. This eigenbasis splits into pairs of states ${\cal C}_i=\{|0^L_i\rangle,|1^L_i\rangle\}$, $i\in \mathbb N$, that define replicas of the logical subspace ${\cal C}_0$. Color encodes the absolute value of the matrix elements.
    {\bf (c)} Flow diagram corresponding to each Kraus operator. Circles represent error spaces, and arrows show the most relevant matrix elements. The dynamics within the subspaces is discarded in this representation. 
    {\bf (d)} Numerical Wigner functions of the projectors onto the subspaces ${\cal C}_0$, ${\cal C}_1$, and ${\cal C}_2$. Comparison to the subspaces generated from the code space by the errors $a$ and $a^\dagger$ reveals that the errors in the first level of hierarchy approximately correspond to $a$ and $a^\dagger$.
    {\bf (e)} Quantum state trajectories with errors and QEC. The state is represented in the same basis as in (b); color encodes the absolute value of the state components in this basis. Red dotted lines are guides to the eye that separate the error subspaces. The occurrence of errors is indicated with red arrows at the top. The time axis is measured in QEC cycles. The Kraus operators are applied between the time steps, and the syndrome string encodes which Kraus operator was applied on every step. The state transfer fidelity, shown at the bottom, measures the squared overlap of the final and initial state vectors. 
\label{fig:SBS_circuit}}
\end{figure}

Here, we describe the SBS protocol, first proposed in Ref.~\cite{Royer2020,DeNeeve2020} from a new angle. The full  QEC circuit in this protocol is shown in Fig.~\ref{fig:SBS_circuit}(a) with nominal parameter values listed in Table~\ref{tab:ECDC-Parameters}; it implements a channel ${\cal R}_\Delta(\rho) = ({\cal R}_\Delta^\sZ \circ {\cal R}_\Delta^\sX)(\rho)$. Let $(K_g^{\sX/\sZ}, K_e^{\sX/\sZ})$ denote the Kraus operators of the constituent rank-2 channels ${\cal R}_\Delta^{\sX/\sZ}$ (we omit the $\Delta$ subscript from the Kraus operators for simplicity). These operators read:
\begin{align}
K_g^{\sX} &= \cos(\sqrt{\pi}p)\cos(\sqrt{\pi}\Delta^2x)+\sin(\pi\Delta^2/2)\cos(\sqrt{\pi}p),  \\
K_e^{\sX} &= -\cos(\pi\Delta^2/2)\sin(\sqrt{\pi}p) +i\cos(\sqrt{\pi}p)\sin(\sqrt{\pi}\Delta^2x),
\end{align} 
where $x=(a+a^\dagger)/\sqrt{2}$ and $p=i(a^\dagger-a)/\sqrt{2}$, and $(K_g^{\sZ}, K_e^{\sZ})$ are obtained with a substitution $(x,p)\to (-p,x)$. Then, the Kraus operators of a composite rank-4 channel are:
\begin{align}
K_{gg} = K_g^{\sZ}K_g^{\sX},\quad
K_{ge} = K_g^{\sZ}K_e^{\sX},\quad
K_{eg} = K_e^{\sZ}K_g^{\sX},\quad
K_{ee} = K_e^{\sZ}K_e^{\sX}.\quad
\end{align}

For $\Delta=0.34$, these Kraus operators are shown as matrices in the truncated eigenbasis of $K_{gg}^\dagger K_{gg}$ in Fig.~\ref{fig:SBS_circuit}(b). This eigenbasis splits into pairs of states ${\cal C}_i=\{|0^L_i\rangle,|1^L_i\rangle\}$, $i\in \mathbb N$, that define orthogonal replicas of the logical subspace ${\cal C}_0$ generated by the errors. We show the Wigner functions of the projectors $\Pi_{0}$, $\Pi_{1}$, and $\Pi_{2}$ onto the first three subspaces in Fig.~\ref{fig:SBS_circuit}(d). 
Note that $\Pi_{1}\approx a\,\Pi_{0}\, a^\dagger$ and $\Pi_{2}\approx a^\dagger\, \Pi_{0}\, a$, hence  the errors in the first level of hierarchy resemble photon loss ($a$) and gain ($a^\dagger$) errors. While $a$ and $a^\dagger$ only approximately satisfy the Knill-Laflamme conditions \cite{Knill1997} for the finite-energy grid code, the actual error operators that define the subspaces ${\cal C}_1$ and ${\cal C}_2$ satisfy these conditions exactly (since the eigenspaces of a Hermitian operator $K_{gg}^\dagger K_{gg}$ are orthogonal). Similarly, by inspecting the Wigner functions of the projectors onto higher subspaces, we find that the second level of error hierarchy resembles $a^2$, $a^\dagger a$ and $a^{\dagger 2}$. The number of error subspaces in each level is given by the number of  unique combinations of $a$ an $a^\dagger$: two subspaces ($a$ and $a^\dagger$) in the first level, and three subspaces ($a^2$, $a^\dagger a$ and $a^{\dagger 2}$) in the second level, leading to the blocks of size $4\times4$ and $6\times6$ in the Kraus matrices in Fig.~\ref{fig:SBS_circuit}(b).
Further understanding the structure of the error hierarchy is the subject of ongoing research.

Unlike in the standard stabilizer formalism of QEC \cite{Gottesman1996}, Kraus operators here do not correspond to a projection of a state onto a single error subspace and its subsequent transfer to the code space. Instead, the transfer here is realized gradually, following an error hierarchy imposed by the QEC circuit. To clarify the action of the Kraus operators, their reduced representation using directional flow of a quantum state between error subspaces is shown in Fig.~\ref{fig:SBS_circuit}(c) [this representation ignores the dynamics within each subspace].  We now briefly discuss the interpretation of the processes corresponding to each of the $gg$, $ge$, $eg$, $ee$ outcomes of a QEC cycle. Outcome $gg$ heralds a process in which the state has remained in the same subspace.  The probability of emitting $gg$ from within the code space is nearly $1$. This property is exploited in Section~\ref{sec:msmt stat} to extract the expectation value of the code projector $\langle \Pi_{0}\rangle$ from the statistics of long strings of the  $gg/gg/...$ type.
Both $ge$ and $eg$ outcomes herald the process in which the quantum state was transferred one level down the error hierarchy. Strings like $eg/eg/eg/...$ therefore correspond to processes in which the state directionally hops level by level towards the code space. Finally, the $ee$ outcome heralds a  transfer two levels down the error hierarchy. 

Besides the transfer between the error spaces, the Kraus operators apply a deterministic logical flip: $X_L$ in the ${\cal R}_\Delta^\sX$ cycles, and $Z_L$ in the ${\cal R}_\Delta^\sZ$ cycles. This flip is visible in the off-diagonal structure of the sub-blocks in the Kraus matrices, see Fig.~\ref{fig:SBS_circuit}(b). For example, the lower right $2\times 2$ block in $K_{gg}$ represents the code subspace, and the off-diagonal structure represents the combined effect of $X_L Z_L=-i Y_L$ on the codewords.  Due to this effect, the lifetime of $+1$ and $-1$ logical Pauli eigenstates in our QEC protocol are exactly equal. We track the Pauli frame in software, and undo its change in the data reported in Fig.~3 of the main text.

To demonstrate how the errors are corrected by this QEC scheme, we show several examples of quantum state trajectories in Fig.~\ref{fig:SBS_circuit}(e). In the first trajectory, the state is initialized as one of the logical basis states, and then evolved for several QEC cycles without any errors. The Pauli frame switching is apparent here from the oscillating pattern within the code space (in this picture, the phase information is not shown, but the QEC process also protects the phase of the logical qubit). In the second trajectory, an error $a^\dagger$ was applied to the state prior to QEC, and then it was almost perfectly corrected, accompanied by the emission of $eg/gg/...$ syndrome string. In the third trajectory, this error was instead corrected during the third QEC cycle, and the quantum state spent extra time in the error space ${\cal C}_2$.  This example explicitly demonstrates that the Pauli frame update is applied correctly irrespective of the subspace, hence Pauli gates done in this manner are transversal. The subsequent trajectories demonstrate that even higher-order errors, such as $a^{\dagger 2}$ or $a^{\dagger 4}$, can be corrected with high fidelity. Moreover, as seen in the fifth trajectory, the state can be recovered even if additional errors happen while the previous errors have not yet been fully corrected. The latter example highlights that the ``slowness'' of the low-rank error-correction dissipation is not a problem, as long as the error rate is sufficiently small compared to the correction rate.

A few remarks with regards to the simplified interpretation of the QEC process in the main text are in order: (i) The correct interpretation of the action of a QEC cycle requires considering pairs of outcomes, like $ge$, instead of isolated outcomes, like $g$ or $e$. We adopted the latter approach in the main text for simplicity of exposition. (ii) The $gg$ outcome does not herald the projection onto the code space, as mentioned in the main text, but rather a process in which ``no error was corrected''.  Conditioned on the state residing in the code subspace, this outcome will be emitted with probability nearly 1. However, if the state is in one of the error spaces this outcome can still occur with smaller probability starting from about $0.47$ at the lowest level in the error hierarchy and reducing for higher levels. (iii) When one of the outcome $eg$, $ge$ or $ee$ is obtained, there is a small chance that the QEC process has added an error, leading to a random walk among the error spaces that is heavily biased towards the code space.


\subsection{QEC cycle: implementation details \label{sec:qec_details}}

In this section, we describe implementation details of a QEC cycle, whose schematic is shown in Fig.~\ref{fig:SBS_circuit}. The various datasets in this work were taken with several different versions of the QEC circuit. All these versions have the same overall structure, but different parameter values obtained from re-training after the system drift has appreciably affected the logical performance (this happens on a time scale of 1-2 weeks, see Section~\ref{sec:stability}). Below, the quoted durations of various components of a QEC cycle refer to the circuit version that we used to collect the system lifetimes dataset and that achieved the highest reported QEC gain.

{\bf SBS unitary.} We refer to the unitary part of the circuit $U_\emptyset$ prior to ancilla measurement as ``SBS unitary'' since it is based on the ansatz from Ref.~\cite{Royer2020}. The SBS unitary is compiled as a four-layer parametrized circuit with nominal parameters shown in Table~\ref{tab:ECDC-Parameters}, and is further translated into the pulse sequence with the method  described in Section~\ref{sec:ECD control optimization}. The last circuit layer does not contain an $\rm ECD$ gate, and instead only contains a qubit rotation and oscillator displacement. Since the qubit is reset after the SBS, the function of the latter rotation is to choose the ``reset axis'', which can be an arbitrary axis on the qubit Bloch sphere. 


As shown in Ref.~\cite{Royer2020}, without any special asymmetries between $|g\rangle$ and $|e\rangle$ it would not matter along which axis the ancilla reset is done -- all choices result in the same completely positive trace-preserving map after averaging over the measurement outcomes. However, in practice the asymmetry comes from the ancilla relaxation channel that degrades the readout fidelity of the $|e\rangle$ state. Hence, it is advantageous to choose the reset axis that preferentially returns the $|g\rangle$ outcome. The parameter sequence for SBS unitary in Table~\ref{tab:ECDC-Parameters} takes this choice into account. The choice of reset axis also results in different unraveling of state trajectories and different Kraus operators. The choice made here enabled the interpretation of $e$ outcomes as syndromes that signal occurrence and correction of errors, which is utilized in the post-selection experiments, described in Section~\ref{sec:post-selection}. This is in contrast with Ref.~\cite{Campagne-Ibarcq2020}, where $g$ and $e$ outcomes are interpreted as left or right displacement of the grid.

The duration of the SBS unitary is not fixed, because its constituent $\rm ECD$ gates can be implemented with different choices of the speed enhancement factor $\alpha$ (amplitude of the intermediate displacement). Since $\alpha$ is included in the action space of the RL agent, all circuit candidates during the training have different durations of the SBS unitary (we will soon comment on how this affects the reward comparison among them). In the final circuit that achieved the highest reported QEC gain, the duration of SBS unitary is $t_{\rm SBS}=1546\,\rm ns$. 

\begin{table}[b]
\begin{tabular}{|c|c|c|}
\hline 
\textbf{Component} & \textbf{Subcomponent} & \textbf{Duration (ns)}\tabularnewline
\hline 
\hline 
 & Enter cycle & 24\tabularnewline
\hline 
\multirow{6}{*}{SBS} & Enter SBS & 24\tabularnewline
 & Circuit layer 1 & 502\tabularnewline
 & Circuit layer 2 & 708\tabularnewline
 & Circuit layer 3 & 262\tabularnewline
 & Circuit layer 4 & 76\tabularnewline
 & Exit SBS & 24\tabularnewline
\hline 
\multirow{7}{*}{Reset} & Enter reset & 24\tabularnewline
 & Roundtrip delay & 300\tabularnewline
 & Acquisition & 1400\tabularnewline
 & Signal processing & 332\tabularnewline
 & Distribution of $s_0$ and $s_1$ & 100\tabularnewline
 & Branching and feedback & 200\tabularnewline
 & Exit reset & 24\tabularnewline
\hline 
\multirow{2}{*}{Virtual rotation} & Mixer matrix calculation & 400\tabularnewline
 & Mixer update & 48\tabularnewline
\hline 
Idle & Delay & 452\tabularnewline
\hline 
 & Exit cycle & 24\tabularnewline
\hline 
\end{tabular}
\caption{{\bf Timing of the cycle components.}
\label{tab:cycle_components}}
\end{table}

{\bf Ancilla reset.} In principle, error correction with SBS protocol could be fully autonomous (without a classical feedback loop) as was envisioned in the  proposal \cite{Royer2020} and realized in a trapped ion system \cite{DeNeeve2020}. The autonomous scheme has an advantage of significantly simplifying the demands on the classical co-processor (in our case, the FPGA). Moreover, there exist various dissipative reset protocols for the transmon \cite{Murch2012,Geerlings2013,Magnard2018}. However, the disadvantage of a fully autonomous implementation in our system is that it is not able to compensate for a spurious rotation of the oscillator due to the always-on dispersive coupling with the ancilla. The back-action of discarding the ancilla state during the reset is the dephasing of the oscillator -- a particularly harmful error channel for the GKP code \cite{Royer2020}. Partly because of this reason, we chose to implement ancilla reset through measurement and classical feedback, as described in Section~\ref{sec:readout}, with the total duration of ancilla reset subroutine of $t_{\rm reset}=2332\,\rm ns$. 



{\bf Virtual rotation.} Due to the always-on dispersive coupling, the oscillator acquires a spurious rotation during the ancilla readout time. In experiment \cite{Campagne-Ibarcq2020}, a simple echo sequence was used to cancel this rotation. With such an approach, ancilla spends half of the time in $|g\rangle$ and half in $|e\rangle$ regardless of the actual syndrome measurement outcome, which is detrimental to the code due to additional error sources associated with the $|e\rangle$ state. Here, we instead chose the reset axis which results in $0.9$ probability of detecting $|g\rangle$. Therefore,  the ability to compensate for the spurious oscillator rotation without echoing the state back to $|e\rangle$ is crucial to maintain this advantage.

We achieve this by dynamically tracking the oscillator phase that stochastically changes due to random ancilla measurement outcomes, and compensating for it with a virtual counter-rotation. 
The spurious oscillator rotation angle accumulates during the reset time $t_{\rm reset}$, during the time $t_{\rm VR}$ that it takes to execute the virtual rotation on the FPGA, and during the idle time $t_{\rm idle}$ when ancilla is nominally in $|g\rangle$ (the latter will be explained shortly). Therefore, in the idealistic dispersive coupling model, the oscillator would rotate by $\vartheta_g=\chi(t_{\rm VR}+t_{\rm idle}+t_{\rm reset})/2$ if the ancilla is found in $|g\rangle$, and $\vartheta_e=\chi(t_{\rm VR}+t_{\rm idle}-t_{\rm reset})/2$ if it is found in $|e\rangle$. Although the $|f\rangle$ state is not computational, our controller is able to reset it with an accompanying virtual rotation by angle $\vartheta_f$. Instead of relying on the simple dispersive coupling model, in experiment we independently calibrate the angles $\vartheta_{g/e/f}$ with a variation of out-and-back experiment \cite{Eickbusch2021} to account for additional minor timing contributions related to FPGA program entering or exiting a subroutine, etc. These calibrated angles are used to initialize the QEC circuit for training.

Another important aspect of the virtual rotation is the switching between momentum and position quadratures of the oscillator to realize ${\cal R}_\Delta^{\scriptscriptstyle X}$ and ${\cal R}_\Delta^{\scriptscriptstyle Z}$ dissipators. Such switching can be achieved with a rotation of the SBS unitary by $\pi/2$ in phase space. This results in additional deterministic contribution $\vartheta_{\rm SBS}$ in every virtual rotation gate. The value of $\vartheta_{\rm SBS}$ is $\pi/2$ for the square grid code, and $\pi/3$ for hexagonal grid code. 

The virtual rotation gate utilizes a floating point register $\vartheta$ on the FPGA. During this gate, the FPGA performs the calculation $\vartheta \leftarrow \vartheta+\vartheta_{g/e/f}+\vartheta_{\rm SBS}$ with subsequent reconfiguration of the dynamic mixer matrix which applies a rotation transformation to the oscillator pulses before they are being streamed at the DAC. The total duration of the virtual rotation gate that includes all these steps is independent of $\vartheta$ and is equal to $t_{\rm VR}=448\rm \,ns$.



{\bf Idle section.} During the agent training, the reward is measured after a fixed number of $T=160$ cycles, and not after some fixed physical duration of time. Therefore, it is necessary to keep the duration of a QEC cycle constant across different protocol candidates to ensure a fair reward comparison. At the same time, the agent is able to affect the physical duration of the SBS unitary by changing the speed enhancement factor $\alpha$ in the $\rm ECD$ gates. To reconcile these two requirements, after the virtual rotation gate we add a section of idle time that is calculated based on  the duration of the SBS unitary in each circuit candidate. We constrain the combined duration of SBS unitary and the idle section to be $2\,\rm \mu s$. In the circuit that achieved the highest reported QEC gain, the duration of the idle section was  $t_{\rm idle}=452\,\rm ns$. 
Note that in the previous QEC experiments that tried to create a long-lived quantum memory \cite{Ofek2016a,Hu2019a}, the idle section was inserted intentionally to avoid frequently entangling the high-quality oscillator with low-quality ancilla. Here, we find that inserting any additional idle time degrades the performance, hence we kept it nearly to a minimum while still leaving some room for change of the SBS duration by the RL agent.


In Table~\ref{tab:cycle_components}, we provide a detailed timing breakdown of all components of the cycle.

\subsection{Learned and scripted parameters \label{sec:parameters}}


\begin{figure}
    \centering
    \includegraphics[width=\figwidthDouble]{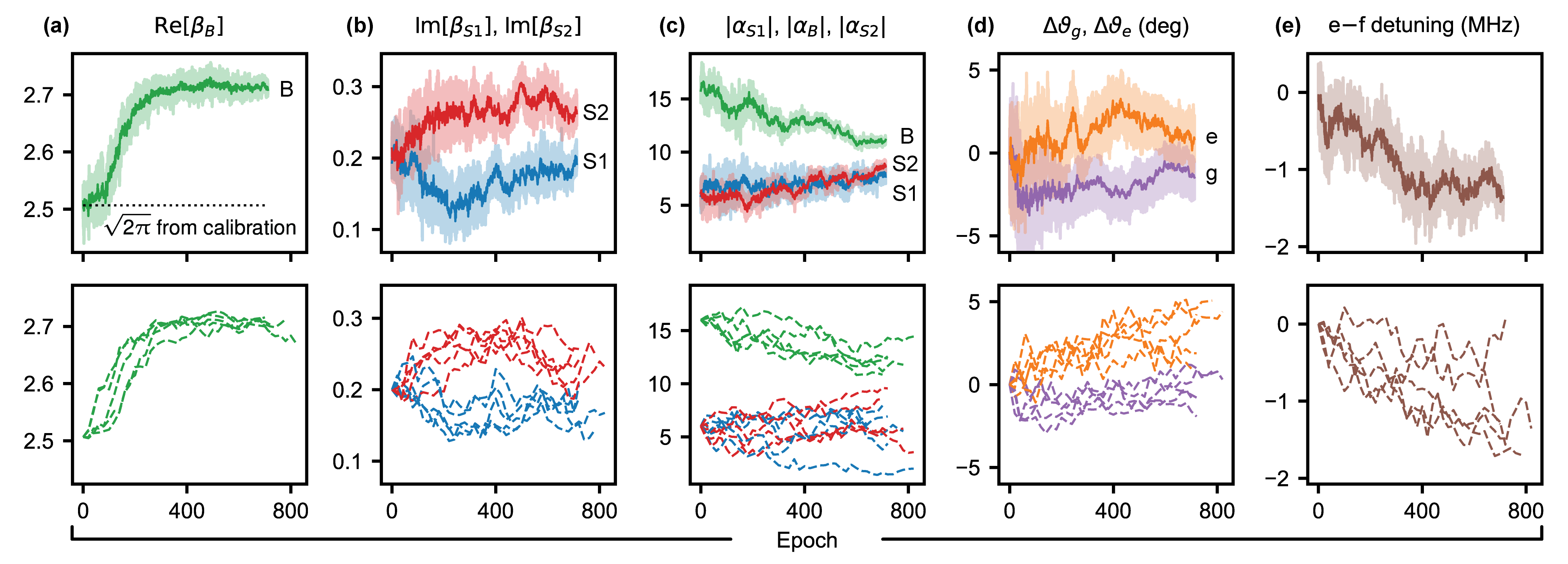}
    \caption[Evolution of QEC circuit parameters during the training]{
    \textbf{Evolution of QEC circuit parameters during the training.} Top row: example from one particular training run. The shaded region bounds the minimal and maximal sampled parameter values. Solid lines indicate the mean. Bottom row: evolution of the mean parameter values in several independent training runs performed during a two-day period, showcasing the reproducibility of the training results. 
\label{fig:action_evolution}}
\end{figure}

Our QEC protocol has multiple parameters which could be optimized to improve its performance. Some of these parameters are difficult to incorporate into our optimization framework in its current form, and therefore their values are chosen as an approximate compromise between various tradeoffs and then held constant. The rest $P=45$ of them are optimized with reinforcement learning, given a reasonable starting point obtained from independent  calibrations. Here, we briefly explain the meaning of these parameters.


{\bf Scripted parameters:}


$\circ$ Duration, shape, and amplitude of the readout pulse. 

$\circ$ State classification thresholds. 


$\circ$ Timing of all components of the reset subroutine. 


$\circ$ Durations of primitive pulses (qubit rotations and oscillator displacements).




$\circ$ Combined duration of the SBS unitary and the idle section.


{\bf Learned parameters:}


$\circ$ Virtual rotation angle $(\vartheta_{g}, \vartheta_{e}, \vartheta_{f})$ for each measurement outcome. It is initialized with a result of independent calibration using a variation of out-and-back experiment.


$\circ$ Detuning of transmon $|g\rangle\leftrightarrow|e\rangle$ and $|e\rangle\leftrightarrow|f\rangle$  pulses (same parameter for all pulses). It is initialized with a result of independent calibration, when oscillator is in the vacuum state (i.e. when there is no Stark shift).

$\circ$ Spectral corrections to $|g\rangle\leftrightarrow|e\rangle$ and $|e\rangle\leftrightarrow|f\rangle$ pulses based on derivative reduction by adiabatic gate (DRAG) scheme \cite{Chen2016} (same parameter for all pulses). It is initialized with 0.


$\circ$ Complex-valued amplitudes of conditional displacement gates in the first three layers of the SBS unitary, and a complex-valued amplitude of the unconditional displacement in the fourth layer. Two small amplitudes are initialized with $\beta_{\rm S1}=\beta_{\rm S2}=0.2i$, and the big amplitude is initialized with $\beta_{\rm B}=\sqrt{2\pi}$. The unconditional displacement is initialized with 0.


$\circ$ Magnitudes of the intermediate large displacements used to execute the $\rm ECD$ gates in the first three layers of the SBS unitary. They are initialized with $\alpha_{\rm S1}=\alpha_{\rm S2}=6$ for small conditional displacements and $\alpha_{\rm B}=16$ for the big conditional displacement. Note that changing these parameters also influences the duration of the $\rm ECD$ gates. 


$\circ$ Angular corrections to intermediate large displacements in the first three layers of the SBS unitary. These heuristic parameters compensate for effect of second-order dispersive shift and for the fact that conditional displacement accumulates along an arc of small curvature instead of a straight  line. These corrections are initialized with 0.


$\circ$ Phases and angles of all ancilla rotations in the SBS circuit layers (including the echo pulses inside the $\rm ECD$ gates), and in the ancilla reset subroutine. These parameters are initialized with nominal values from Table~\ref{tab:ECDC-Parameters}.

$\circ$ Detuning of the local oscillator (LO) frequency for the cavity mode. This LO is calibrated with spectroscopy and set to be half-way between the number-split oscillator frequencies when qubit is in the states $|g\rangle$ and $|e\rangle$, corresponding to $\Delta=0$ in Eq.~\eqref{Hamiltonian}.

{\bf Evolution of parameters during training.}
In Fig.~\ref{fig:action_evolution}, we show the evolution of several QEC circuit parameters during the  training. Most parameters, when initialized well, merely exhibit small fluctuations around the mean. However, some parameters undergo systematic and reproducible changes, as observed in the provided examples. For instance, in Fig.~\ref{fig:action_evolution}(a), the big conditional displacement amplitude ${\rm Re}[\beta_{B}]$, which we expect to be equal to the size of the grid unit cell, changes from a calibrated value of $\sqrt{2\pi}$ by about $8\%$, likely indicating the presence of a miscalibration error (the last calibration was done several weeks prior to this training). Similarly, the trend in Fig.~\ref{fig:action_evolution}(e) towards the negative detuning of the $|e\rangle\leftrightarrow|f\rangle$ pulses could be compensating for an additional Stark shift that was not present at the initial calibration stage (calibration was performed with vacuum state in the oscillator). In Fig.~\ref{fig:action_evolution}(b), the trend in amplitudes $\beta_{S1}$ and $\beta_{S2}$ of the two small conditional displacements in the SBS unitary is particularly insightful, as it helped us identify a limitation of the proposal in Ref.~\cite{Royer2020}, according to which the amplitudes $\beta_{S1}$ and $\beta_{S2}$ should be identical and equal to $i\Delta^2/2$, while the RL agent systematically converges to $|\beta_{S2}|>|\beta_{S1}|$. Using simulations, we verified that in presence of error channels that act during the execution of the SBS unitary, this is indeed a correct inequality. The optimal ratio of  these two amplitudes is found in simulations to be strongly dependent on the error channel. The agent adapts this ratio to the real error channel of our system. 

In Ref.~\cite{Sivak2021}, it was shown that a similar RL agent is able to converge to correct solutions even starting from completely random parameter initializations. Here, we initialize the parameters close to their expected optimal values through various calibrations, but the compounded effect of small errors (at the level of a few percent) in multiple parameters results in a QEC protocol which, although fully functional, is far from optimal. In particular, we were not able to reach break-even with only the independent calibrations and educated guesses, hence model-free RL can be acknowledged as one of the most crucial factors in the success of this project.

\subsection{Syndrome measurement statistics \label{sec:msmt stat}}


\begin{figure}
    \centering
    \includegraphics[width=\figwidthDouble]{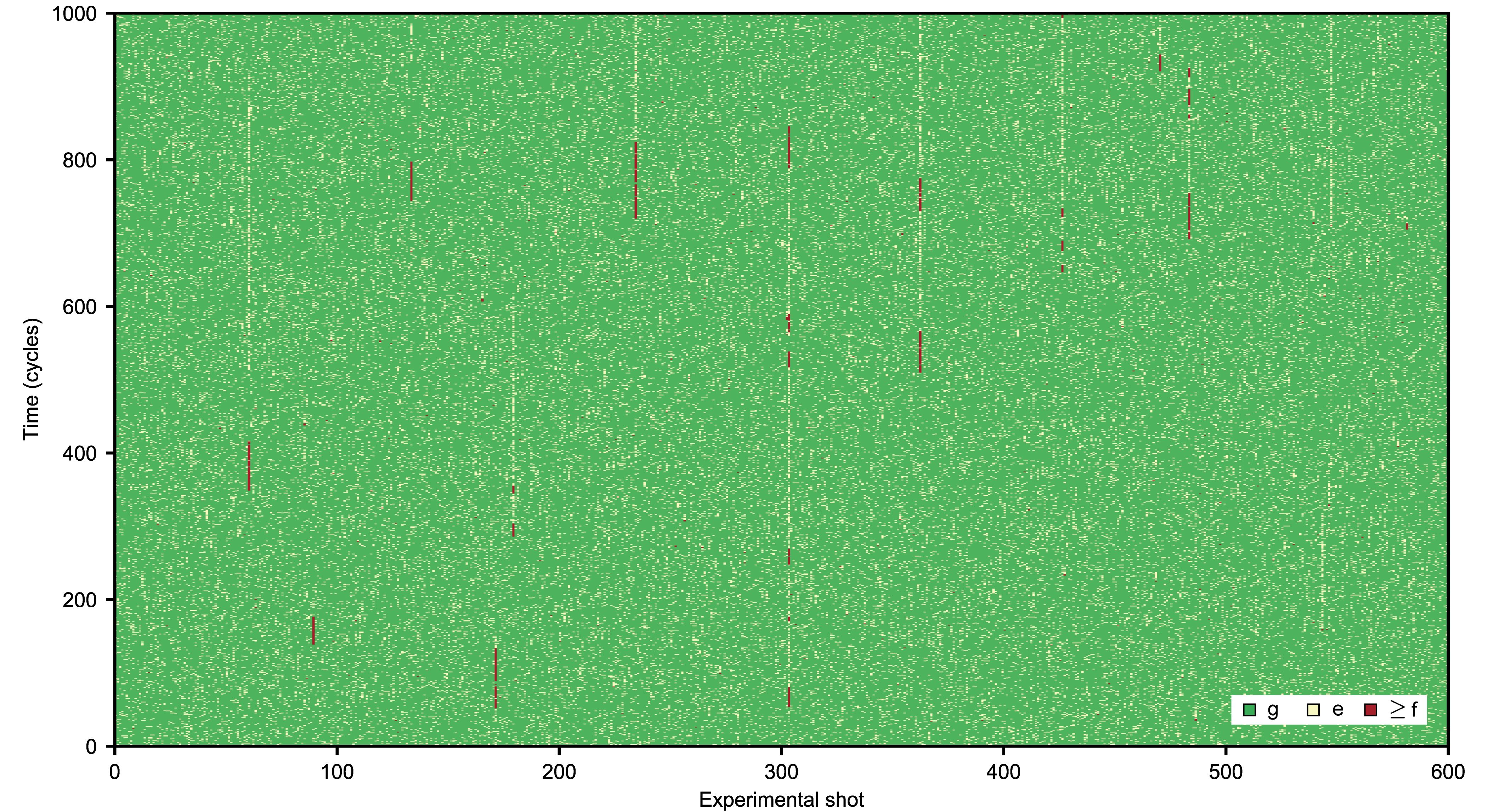}
    \caption[Syndrome measurement outcomes]{
    \textbf{Syndrome measurement outcomes.} A sample of 600 experimental QEC shots of duration $T=1000$ cycles each. The ``$g$'' outcome (green) is prevalent, heralding the no-error process, while occasional ``$e$'' outcomes (yellow) indicate correction of errors, and ``$\geq f$'' outcomes (red) indicate leakage. When transmon escapes to a state higher than $|f\rangle$, which is not addressed by our reset scheme, the leakage outcome persists for multiple cycles (streaks of red). In the readout IQ plane such states occur above the Q threshold, and therefore they are conveniently classified as leakage, but without further identification of the exact leakage state. After transmon stochastically drops back to $|f\rangle$, the controller is able to reset it and return the ancilla to the computational manifold. However, during the cycles when ancilla is effectively inactive, the code is not stabilized. Hence, leakage streaks are often followed by streaks of $e$ outcomes where QEC re-stabilizes the code manifold. 
\label{fig:outcomes}}
\end{figure}


\begin{figure}
    \centering
    \includegraphics[width=\figwidthDouble]{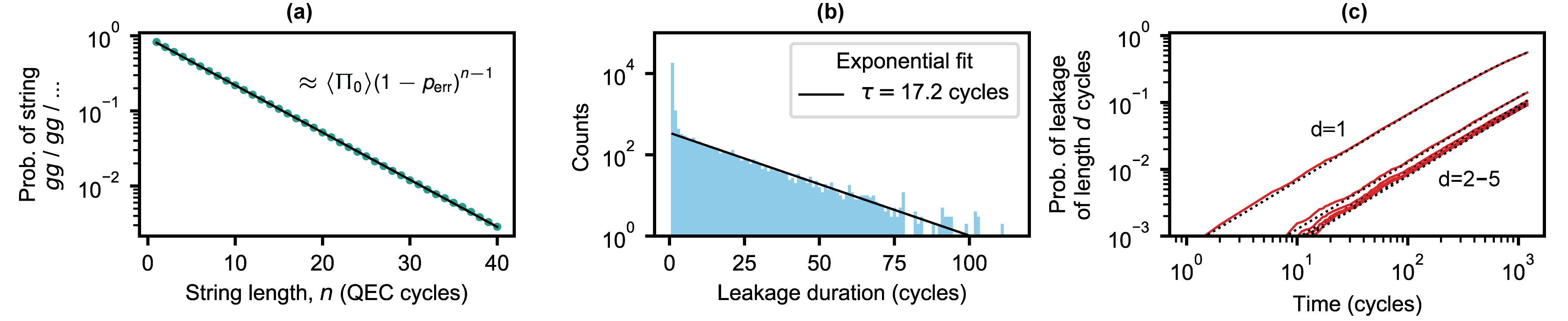}
    \caption[Analysis of syndrome measurement outcomes]{
    \textbf{Analysis of syndrome measurement outcomes.} 
    {\bf (a)} Probability of a string $gg/gg/...$ as a function of its length, together with the fit to a single exponential decay.
    {\bf (b)} Histogram of durations of leakage events. Events of duration 1 or 2 cycles are predominantly  $|f\rangle$ state; longer duration events are likely $|h\rangle$ or higher excited states. The exponential fit gives the effective lifetime of these leakage states. 
    {\bf (c)} Fraction of shots that experienced a leakage event of duration $d$ up to a given time. Dotted lines are fits to a constant-rate model, see Section~\ref{sec:leakage}.
\label{msmts}}
\end{figure}

A sample of $600$ experimental QEC shots is shown in Fig.~\ref{fig:outcomes}, where QEC is ran for $T=1000$ cycles in each shot. Consider a string of measurement outcomes going from any chosen time step $t_i$ to a time step $t_{i+2n}$ (it contains $n$ QEC cycles). The probability $P([gg]^{n})$ that this string contains only $gg$ outcomes is shown in Fig.~\ref{msmts}(a), where it is averaged over the experimental shots and over  initial times $t_i$ (the averaging over $t_i$ is done with a sliding window method which is applicable due to process stationarity). While in the most general case the functional form of $P([gg]^{n})$ is a sum of multiple decaying exponentials, we clearly observe only a single dominant exponential contribution. Hence, we fit this probability to $P([gg]^{n})=a\lambda^n$, obtaining  $a=0.936\pm0.003$ and $\lambda=0.86517\pm0.00013$. By adopting a model for the error process and for the QEC process, we can link the fit parameters $\{a,\lambda\}$ to model parameters. In general, such a model would be quite complex. However, here we are interested in only two characteristic parameters of the process: the probability $\langle\Pi_{0}\rangle$ of occupying the code space in the dynamical equilibrium of the QEC process, and the probability $p_{{\rm err}}$ of having an error that transfers the state out of the code space. These parameters can be extracted with very minimal model assumptions.

Using the transfer matrix approach, it can be shown that $p_{{\rm err}}\approx1-\lambda$ and $\langle \Pi_{0}\rangle \approx a\lambda$ under the following assumptions: 
1) The error probability is small $p_{{\rm err}}\ll1$, which is justified since the cycle duration is small compared to all relevant error rates in the system, and is confirmed by the fit results; 
2) The conditional probability $P_{gg}^{\,{\cal C}_0}$ of emitting $gg$ when the quantum state is in the code space is nearly 1. This is justified, since in the error-free model of the SBS protocol described in Section~\ref{sec:sbs} this probability is $0.999$; 
3) The conditional probability $P_{gg}^{\,\rm err}$ of emitting $gg$ when the quantum state is in any of the error spaces is small $P_{gg}^{\,\rm err}\ll 1$. This assumption is partially justified, since in the error-free model of the SBS protocol this probability is smaller than $0.5$ for the first error space, and then monotonously reduces for the higher levels of the error hierarchy; 
4) The probability $P_{gg}^{\,\rm corr}$ of correcting an error \textit{and} emitting $gg$ is small $P_{gg}^{\,\rm corr}\ll1$. In the error-free model of the SBS protocol this probability is exactly zero, while in practice it is limited to $\sim10^{-2}$ due to the readout infidelity of the $|e\rangle$ state.

This result can be intuitively understood as follows: the two most probable system trajectories that generate the string of all $gg$'s correspond to (i) starting in the code space and remaining there for $n$ steps, which happens with probability $\langle\Pi_{0}\rangle(1-p_{\rm err})^{n}$, and (ii) starting in the code space, remaining there for $n-1$ steps, and transitioning out on the very last step, which happens with probability $\langle\Pi_{0}\rangle(1-p_{\rm err})^{n-1}p_{\rm err}$. The sum of these two contributions equals $\langle\Pi_{0}\rangle(1-p_{\rm err})^{n-1} \equiv a\lambda^n$, leading to $p_{{\rm err}}\approx1-\lambda$ and $\langle \Pi_{0}\rangle \approx a\lambda$. The corrections to these formulas are of the second order in parameters $\{p_{\,\rm err}, 1-P_{gg}^{\,{\cal C}_0}, P_{gg}^{\,\rm err}, P_{gg}^{\,\rm corr}\}$. 
From the fit in Fig.~\ref{msmts}(b), we extract $p_{{\rm err}}=0.13\pm0.02$ and $\langle\Pi_{0}\rangle=0.81\pm0.02$. Note that this result for $\langle\Pi_{0}\rangle$ agrees within the error margin with the result obtained by an independent method based on the reconstruction of the density matrix from the measured Wigner functions in Section~\ref{sec:wigner_tomo}. Here, the error of the fit is negligible compared to the model approximations, hence the quoted error bars are obtained from an estimate of the second-order corrections $\sim p_{{\rm err}}^{2}\approx0.02$. 

Here,  we only considered the string of a special type $gg/gg/...$; an important avenue of future research would include learning the error channel from the full statistics of syndrome outcomes, using the dataset in Fig.~\ref{fig:outcomes}.

\begin{figure}
    \centering
    \includegraphics[width=\figwidthDouble]{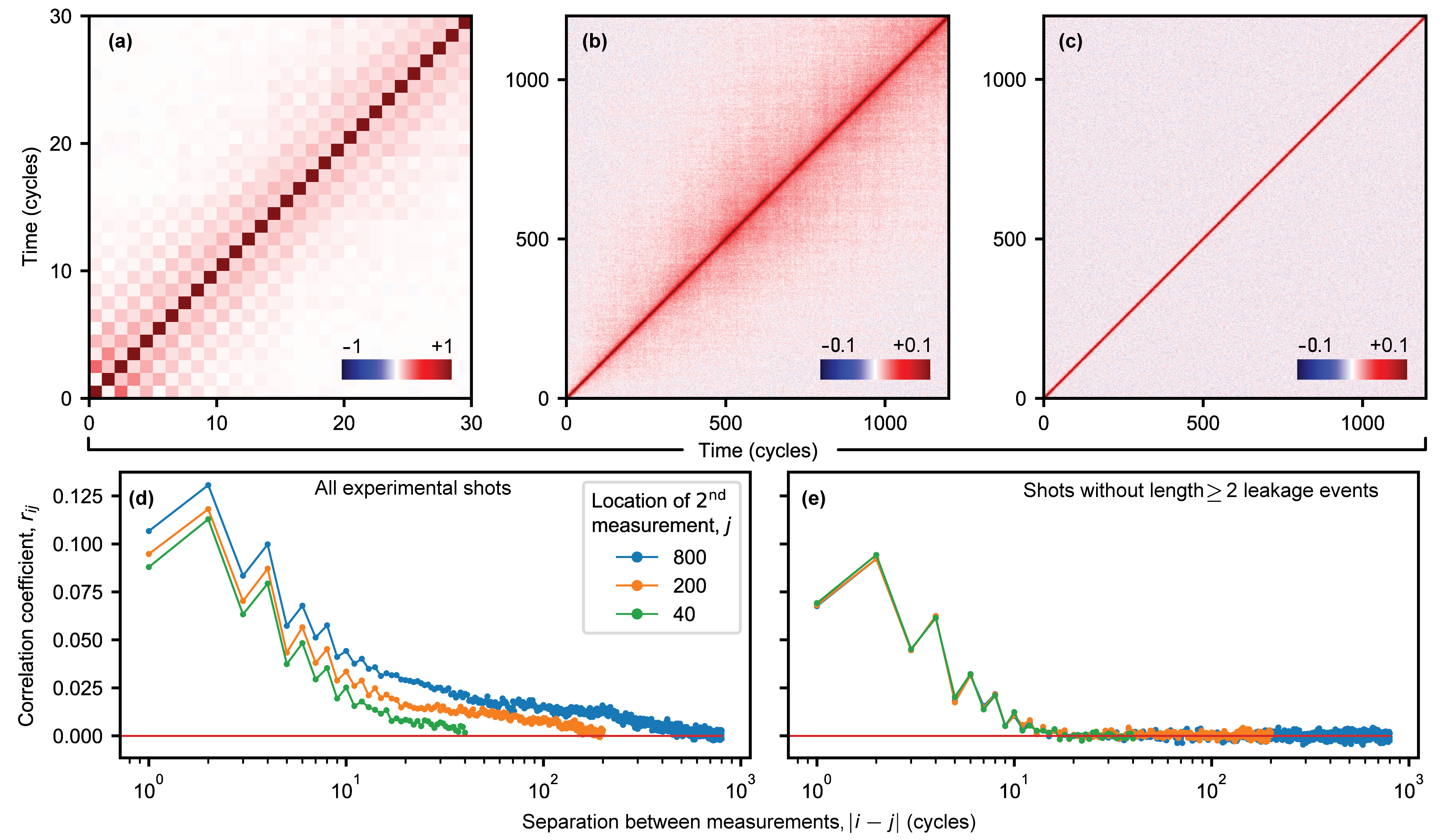}
    \caption[Correlation of syndrome measurement outcomes]{
    \textbf{Correlation of syndrome measurement outcomes.}
    {\bf (a)} Correlation matrix $r_{ij}$ for the first $30$ cycles, computed from the dataset in Fig.~\ref{msmts}(a).
    {\bf (b)} $r_{ij}$ for the full QEC duration of $1200$ cycles with zoomed-in color scale to resolve small numbers.
    {\bf (c)} $r_{ij}$ after removing leakage events of duration $\geq 2$ cycles.
    {\bf (d, e)} Cuts of the correlation matrices from (b) and (c) at different locations in the QEC trajectory. In a stationary process, $r_{ij}$ would only depend on $|i-j|$, which is clearly not satisfied in (d). After removing length $\geq 2$ leakage events, the deviation from stationarity is not resolvable. 
\label{corr}}
\end{figure}

\subsection{Analysis of transmon leakage \label{sec:leakage}}

To quantify transmon leakage, we histogram its duration in Fig.~\ref{msmts}(b). The most likely leakage duration is $1$ cycle, since the controller resets $|f\rangle$ to $|g\rangle$ with high probability. However, because of the finite readout fidelity of $|f\rangle$, shown in Fig.~\ref{fig:readout}(b), the controller sometimes fails to reset this state, resulting in the next most probable leakage duration of $2$ cycles. After that, the histogram follows exponential distribution with decay constant of $17.2$ cycles, corresponding to $85\rm\,\mu s$, which we attribute to the effective lifetime of higher leakage states that are not addressed by our reset scheme. This time scale is consistent with our estimate $\sim280\,{\rm \mu s}/3=93\,{\rm \mu s}$ of the $|h\rangle$ state lifetime, derived from the bosonic statistics and the average measured lifetime of $|e\rangle$ state.

To extract the leakage rate, in Fig.~\ref{msmts}(c) we plot the fraction of shots that experienced a leakage event of a certain duration up to a given cycle.  We fit the data to a constant-rate model $L(t)=1-\exp(-t/\tau_l)$ where $t$ is the cycle index and $p_l=1/\tau_l$ is the leakage rate (i.e. leakage probability per cycle). We find $\tau_l=1480\pm10$ cycles, corresponding to a leakage rate of $p_l=(6.76\pm0.04)\times 10^{-4}$. Similar analysis can be done for leakage events of length $\geq 2$; with the fit to the same model, we find the time scale of $\tau_{l,\geq2}=7820\pm10$ cycles, and the corresponding rate of $p_{l,\geq2}=(1.280\pm0.002)\times 10^{-4}$. Note that this measurement was performed at the time of slightly sub-optimal performance, and therefore the leakage rate at the maximal achieved QEC gain might have been smaller.

Next, we study the correlation of syndrome measurement outcomes across time. 
The correlation matrix is given by
\begin{align}
r_{ij} = \frac{\mathbb{E}[m_i m_j]-\mathbb{E}[m_i] \mathbb{E}[m_j]}{\sqrt{(\mathbb{E}[m_i^2]-\mathbb{E}[m_i]^2)(\mathbb{E}[m_j^2]-\mathbb{E}[m_j]^2)}},
\end{align}
where $m_k$ is the measurement outcome obtained at cycle $k$, and empirical expectation values are obtained by averaging across experimental shots. For the dataset of Fig.~\ref{fig:outcomes} this correlation matrix is shown in Fig.~\ref{corr}(a), where we consider only the first $30$ cycles. Overall, the correlation is weak, and the correlation between ${\cal R}_\Delta^\sX$ and ${\cal R}_\Delta^\sZ$ channels (separated by odd number of cycles) is weaker than the correlation between the same-quadrature channels (separated by even number of cycles). 

By zooming in the color scale to visually resolve small numbers and considering the full duration of the trajectory of $1200$ cycles, as shown in Fig.~\ref{corr}(b), it becomes evident that the process is not perfectly stationary. To emphasize this, we show in Fig.~\ref{corr}(d) the  correlation coefficient $r_{ij}$ as a function of $|i-j|$ for several choices of $j$. Further along the QEC trajectory the process acquires a correlation tail. Although quite weak, this correlation stretches over hundreds of cycles. 

Previously, it was demonstrated that leakage removal helps to reduce correlated errors in the arrays of transmons \cite{McEwen2021}. Our QEC protocol already contains a mechanism for leakage removal from the $|f\rangle$ state through measurement-based feedback in every cycle. However, leakage states higher than $|f\rangle$ are not cleared by our reset. The signature of such leakage events to higher states is two or more consecutive leakage syndrome outcomes. To check the hypothesis that this residual leakage to states higher than $|f\rangle$ is responsible for increase of correlation, we post-select trajectories that do not have any length-two or longer leakage events. In the post-selected dataset, the correlation matrix does not display any detectable non-stationarity, as seen in Fig.~\ref{corr}(c,e), confirming the hypothesis.
By fitting the remaining short-time correlations in Fig.~\ref{corr}(e) to an exponential decay, we conclude that it takes $3.9\pm0.1$ cycles (approximately 2 QEC cycles) to lose the memory of a typical large error. However, the most probable small errors are corrected in a single QEC cycle. 

\begin{table}
\begin{tabular}{|c|c|c|c|c|c|c|c|}
\hline 
 & NP & L & $d\ge5$ & $d\ge4$ & $d\ge3$ & $d\ge2$ & $d\ge1$\tabularnewline
\hline 
\hline 
 \begin{tabular}{@{}c@{}}Survival prob. \\ per cycle\end{tabular} & $1.0000$ & $0.9985$ & $0.9993$ & $0.9986$ & $0.9972$ & $0.9907$ & $0.9396$\tabularnewline
\hline 
\begin{tabular}{@{}c@{}} Improvement \\ of $\Gamma_{{\rm GKP}}$\end{tabular} & $1.00$ & $1.10$ & $1.18$ & $1.36$ & $1.68$ & $2.44$ & $6.31$\tabularnewline
\hline 
\begin{tabular}{@{}c@{}}Lifetime of  \\ $|+Z\rangle$ (ms)\end{tabular} & $1.874\pm0.004$ & $2.11\pm0.01$ & $2.23\pm0.01$ & $2.55\pm0.01$ & $3.13\pm0.02$ & $4.60\pm0.02$ & $10.0\pm0.6$\tabularnewline
\hline 
\begin{tabular}{@{}c@{}}Lifetime of  \\ $|+Y\rangle$ (ms)\end{tabular} & $1.147\pm0.004$ & $1.23\pm0.01$ & $1.36\pm0.01$ & $1.56\pm0.01$ & $1.93\pm0.01$ & $2.80\pm0.02$ & $9.4\pm1.0$\tabularnewline
\hline 
\begin{tabular}{@{}c@{}}Lifetime of  \\ $|+Z\rangle$ (cycles)\end{tabular} & $381\pm1$ & $427\pm1$ & $452\pm2$ & $518\pm2$ & $636\pm4$ & $934\pm4$ & $2000\pm100$\tabularnewline
\hline 
\begin{tabular}{@{}c@{}}Lifetime of  \\ $|+Z\rangle$ (cycles)\end{tabular} & $233\pm1$ & $250\pm1$ & $275\pm2$ & $317\pm2$ & $393\pm3$ & $567\pm5$ & $1900\pm200$\tabularnewline
\hline 
\end{tabular}
\caption{\textbf{Post-selection results.} \label{tab:post-selection} Top row labels the post-selection schemes. NP stands for ``no post-selection''; L stands for ``leakage''; $d\geq N$ means post-selection that discards trajectories containing strings of $N$ or more consecutive $e$ outcomes in the same-quadrature cycles. 
}
\end{table}

\subsection{Post-selection of errors\label{sec:post-selection}}

Here, we provide additional details about the the post-selection experiment that verifies the ability of our QEC scheme to faithfully identify the errors. The post-selection results are summarized in Table~\ref{tab:post-selection}. Note that this experiment was performed at the time of slightly sub-optimal system performance, hence the baseline results with no post-selection are lower than in some other experiments reported here, e.g. in Section~\ref{sec:stability}. 
The main conclusion of this post-selection experiment is that it enables significant improvement of the error probability at a cost of only a modest rejection probability. 

The saturation of lifetimes in the most stringent post-selection scheme (which preserves only the all-$g$ trajectories) can be related to the following mechanisms: (i) Direct logical errors, which are undetectable in any QEC scheme. (ii) Misclassification of $e$ as $g$ (due to ancilla decay during the measurement), which means that some of the all-$g$ trajectories that survived the post-selection actually contained errors, and some of those errors might have been close to a logical operation instead of the identity operation. (iii) The non-orthogonality of logical states, which in our case is not a  limiting factor.

\begin{figure}
    \centering
    \includegraphics[width=\figwidthDouble]{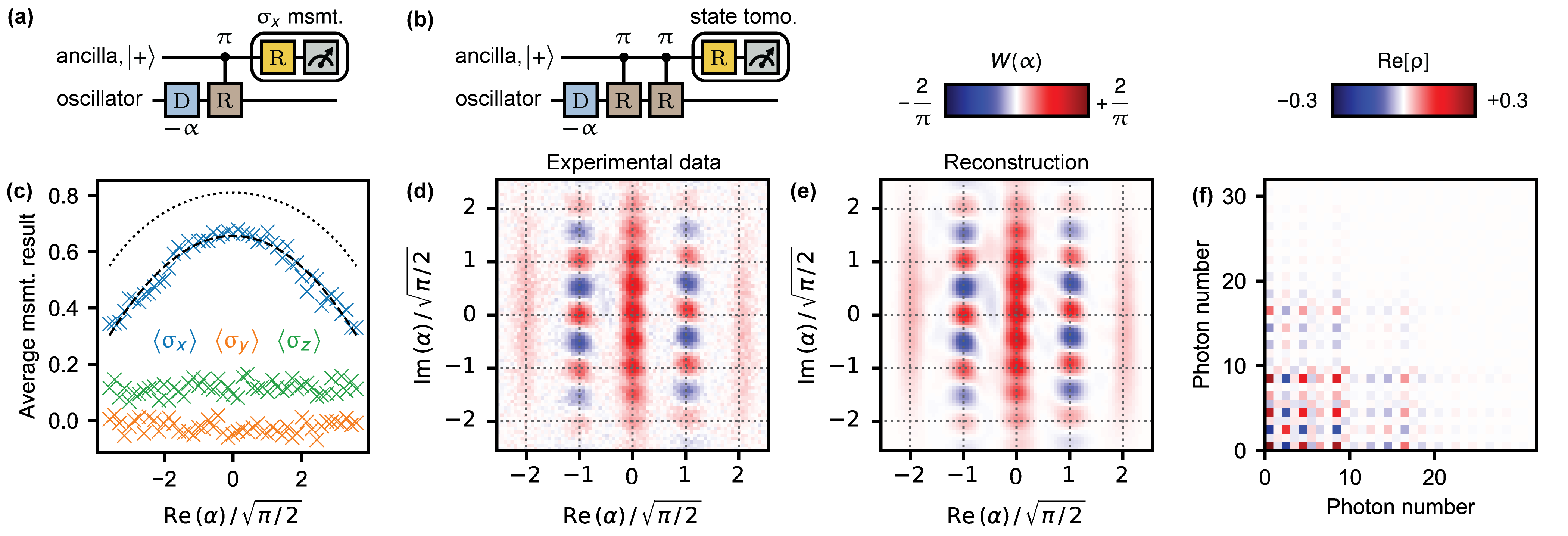}
    \caption[State reconstruction]{
    \textbf{State reconstruction.}
    {\bf (a)} Wigner tomography experiment.
    {\bf (b)} Calibration experiment to extract $\alpha$-dependent contrast of the Wigner tomography. This calibration relies on the assumption that the loss of contrast in tomography is primarily due to incoherent errors during the parity mapping gate ${\rm CR}(\pi)$.
    {\bf (c)} Results of the qubit state tomography in the calibration experiment shown in (a). The measurement of $\langle \sigma_x\rangle$ is fit to a quadratic function of $|\alpha|$ (black dashed line), and its square root (black dotted line) is used as a Wigner function measurement contrast in the state reconstruction. 
    {\bf (d)} Experimental Wigner function of the $|+Z\rangle$ state immediately after initialization. 
    {\bf (e)} Wigner function of the reconstructed state.
    {\bf (f)} Real part of the density matrix of the reconstructed state in the photon number basis. 
\label{fig:state_reconstruction}}
\end{figure}

\subsection{Wigner tomography of logical states \label{sec:wigner_tomo}}

{\bf Tomography and its calibration.}
Wigner tomography is derived from the expression for the Wigner function $W(\alpha)=(2/\pi)\langle \Pi_\alpha\rangle$, where $\Pi_\alpha=D(\alpha)\Pi D^\dagger (\alpha)$ is the displaced parity operator, and $\Pi=\exp(i\pi a^\dagger a)$ is the photon number parity. The displaced parity operator is unitary and can be measured with phase estimation. It is also hermitian, and hence its eigenvalues are constrained to be $\pm 1$. Therefore, it is particularly convenient to measure displaced parity by mapping it onto the qubit observable \cite{Vlastakis2013}, which is achieved in our system with a circuit shown in Fig.~\ref{fig:state_reconstruction}(a). 
The conditional rotation gate ${\rm CR}(\pi)$ is realized with a delay of duration $\pi/\chi$ under the dispersive coupling Hamiltonian, which amounts to approximately $10\,\rm \mu s$. Because of such long duration, previous GKP experiments with similarly small $\chi$ chose to perform state tomography using the characteristic function instead \cite{Campagne-Ibarcq2020,Eickbusch2021}. However, the long coherence of our system allows to measure Wigner function with reasonably high fidelity. 

We use several calibration techniques to improve the quality of the subsequent state reconstruction from the tomographic data. First, to symmetrize the effect of ancilla relaxation during the readout, in half of the phase estimation runs we map $+1$ eigenvalue of $\Pi_\alpha$ to the $g$ outcome, and in another half to the $e$ outcome. This technique eliminates any finite offset in $W(\alpha)$, but maintains reduced contrast due to ancilla relaxation and decoherence. Next, to  calibrate the contrast reduction, we perform an experiment with a similar circuit in which ${\rm CR}(\pi)$ is replaced with $[{\rm CR}(\pi)]^2=I$, see Fig.~\ref{fig:state_reconstruction}(b). We fit the result of this experiment, shown in Fig.~\ref{fig:state_reconstruction}(c), to $\langle\sigma_x\rangle=1-p[\alpha]$, where $p[\alpha]=\eta_0+\eta_2 |\alpha|^2$ is the purity loss per ${\rm CR}(\pi)$ gate. Under the assumption that contrast reduction in tomography is primarily due to incoherent processes (ancilla relaxation and dephasing, and oscillator photon loss), the inferred tomography contrast is $P(\alpha)=\sqrt{1-p[\alpha]}$. At $\alpha=0$, this inferred contrast is equal to $0.8$, which matches the measured contrast of the Wigner function of vacuum in Fig.~\ref{fig:calibration of primitive pulses}(c), justifying the assumptions of this calibration method.

The phase space points $\alpha_i$ for Wigner tomography are chosen on a square $81\times81$ grid in a complex plane restricted to $|{\rm Re}[\alpha_i]|, |{\rm Im}[\alpha_i]|\leq 3.2$. We acquire $2400$ shots per point in $6$ separate acquisition time frames. Between the time frames we perform system performance checks; data acquisition is put on hold if the spurious resonance in $T_1^{\, t}(\overline{n})$ reappears [see Fig.~\ref{fig:readout}(d)]. A single state tomography dataset consists of $15.7$ million shots, and takes a long time to acquire -- from $6$ hours in the case of $T=0$ cycles, to $26$ hours in the case of $T=800$ cycles. Therefore, conclusions derived from the analysis of tomography data apply to long-time average system performance.

{\bf State reconstruction.} Tomographic data is used to produce a best guess for the density matrix of the state. We parametrize the density matrix as $\rho = C^\dagger C/{\rm{Tr}}[C^\dagger C]$, where $C=A+iB$, and $A$ and $B$ are real-valued matrices. Such parametrization ensures that $\rho$ is positive semi-definite with trace 1. We truncate the density matrix to dimension $N=32$ in photon number basis. Coefficients of matrices $A$ and $B$ are optimized using the least squares fit of the Wigner tomography and contrast data, with the cost function given by
\begin{align}
{\rm cost} = \sum_{i=1}^{N_\alpha} \left( \frac{2}{\pi} {\rm Tr}[\rho\, D(\alpha_i)\Pi D^\dagger(\alpha_i)]\cdot  P(\alpha_i) - W(\alpha_i)\right)^2.
\end{align}

An example Wigner tomography of the $|+Z\rangle$ grid state together with its reconstruction is shown in Fig.~\ref{fig:state_reconstruction}(d-f). As will be described shortly, using the reconstructed density matrix we extract various parameters of the state: its purity, mean photon number, and envelope size.

\begin{figure}
    \centering
    \includegraphics[width=\figwidthDouble]{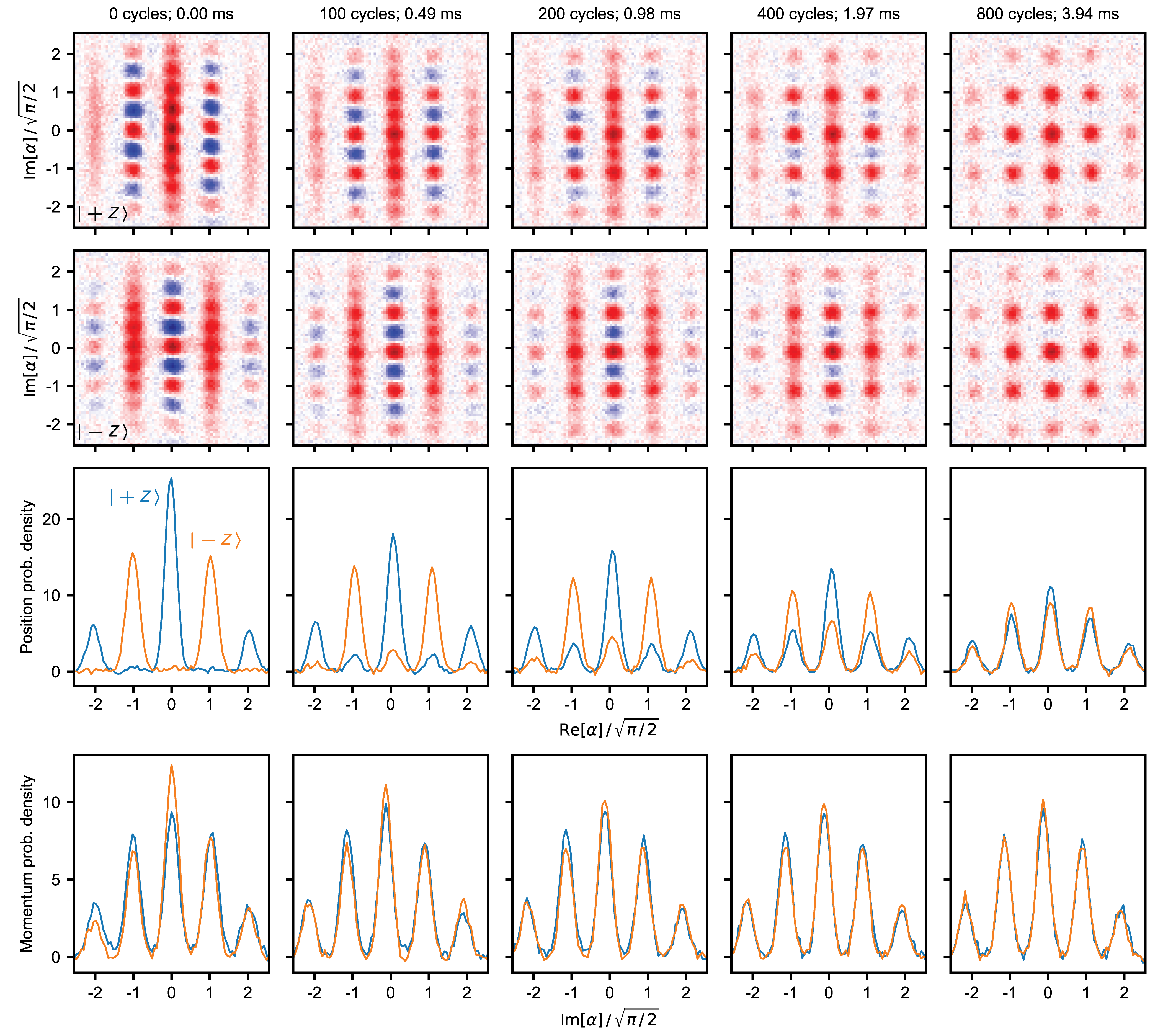}
    \caption[Wigner tomography after QEC]{
    \textbf{Wigner tomography after QEC.} Evolution of $|+Z\rangle$ state (1st row) and  $|-Z\rangle$ state (2nd row) is followed for $800$ cycles. Color scheme is the same as in Fig.~\ref{fig:state_reconstruction}, with the range scaled to $[-0.63, 0.63]$. Marginal of the Wigner function along momentum (position) quadrature, which gives probability density of the oscillator position (momentum), is shown in the 3rd (4th) row in blue for $|+Z\rangle$ state, and orange for $|-Z\rangle$ state. The probability density is not normalized.
\label{fig:wigner_tomo_after_qec}}
\end{figure}

{\bf Evolution of logical states.} We visualize the evolution of logical $|+Z\rangle$ and $|-Z\rangle$ grid states during the QEC by taking Wigner tomography snapshots after $0$, $100$, $200$, $400$, and $800$ cycles, with results shown in Fig.~\ref{fig:wigner_tomo_after_qec}. 

The marginal of the Wigner function along momentum quadrature gives the probability density of the oscillator position, shown in the third row of Fig.~\ref{fig:wigner_tomo_after_qec}. The $|+Z\rangle$ and $|-Z\rangle$ states have non-overlapping support in position representation, clearly observed in the data at $T=0$ cycles. During the QEC process these basis states mix under the logical Pauli channel, which is manifested in the appearance of position peaks of the opposite state, until finally they become almost (but not completely) indistinguishable after $T=800$ cycles.

On the other hand, the marginal of the Wigner function along position quadrature gives the probability density of the oscillator momentum, shown in the last row of Fig.~\ref{fig:wigner_tomo_after_qec}. In the momentum representation, $|+Z\rangle$ and $|-Z\rangle$ states share the same support, but have a different  pattern of phases associated with the peaks of the wavefunction. The phase information is discarded in the probability density function, which looks identical for both states. 


\begin{figure}
  \begin{minipage}[c]{0.62\textwidth}
    \includegraphics[width=4in]{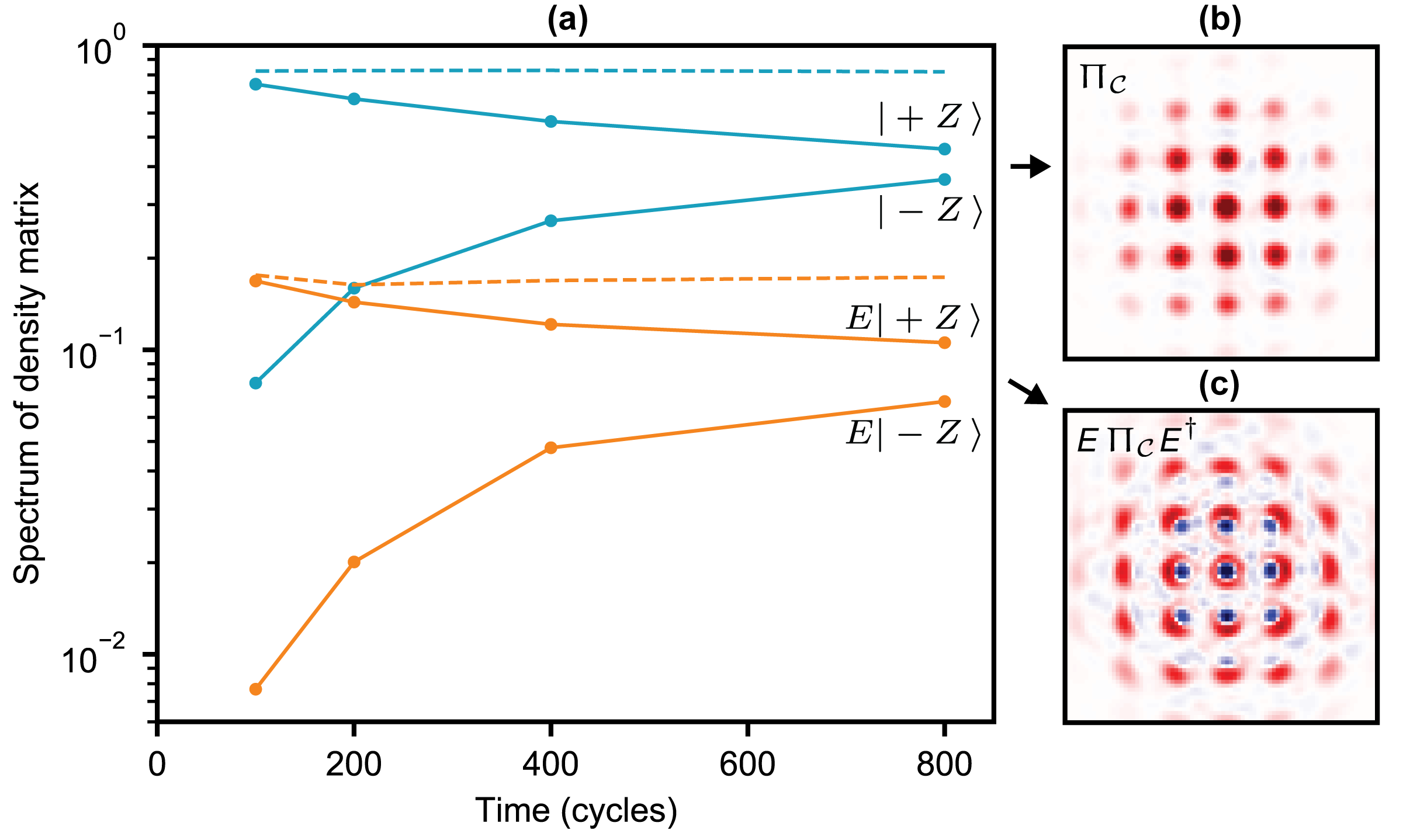}
  \end{minipage}\hfill
  \begin{minipage}[c]{0.38\textwidth}
	\caption{\textbf{QEC subspaces.}
{\bf (a)} Spectrum of the reconstructed density matrices of the $|+Z\rangle$ state evolving under the QEC process. The spectrum separates into two pairs of eigenvalues: one pair corresponds to the code space ${\cal C}_0$, and another pair corresponds to an error space obtained from the code space by application of an error operator $E$. Dashed lines show the sum of the eigenvalues within each pair, which gives the probability of occupying the code space and the error space.
{\bf (b)} Wigner function of the projector $\Pi_{0}$ onto the first pair of eigenvectors (taken at $T=800$), defining the code space. 
{\bf (c)} Wigner function of the projector $E\,\Pi_{0}E^\dagger$ onto the second pair of eigenvectors, defining the error space. This error space corresponds to ${\cal C}_2$ subspace in Section~\ref{sec:sbs}.
\label{fig:spectrum_of_dm}}
  \end{minipage}
\end{figure}

{\bf Spectral analysis of reconstructed states.} Focusing on the time evolution of the $|+Z\rangle$ state, we perform spectral decomposition of its reconstructed density matrices at $T=100,200,400,800$.  We find that the eigenvalues of the density matrix are arranged in pairs corresponding to the images of this state and of its complement $|-Z\rangle$ in different subspaces of the QEC, see Fig.~\ref{fig:spectrum_of_dm}(a). In particular, we identify only two subspaces with a substantial presence of the state during the QEC process: the code space ${\cal C}_0$, shown in Fig.~\ref{fig:spectrum_of_dm}(b), and the error space ${\cal C}_2$ corresponding to an error $E$ that most closely resembles $a^\dagger$,  see Fig.~\ref{fig:spectrum_of_dm}(c) and Section~\ref{sec:sbs}. 

While the QEC circuit imposes the structure of the error subspaces, as described in Section~\ref{sec:sbs}, the properties of the ``thermal'' distribution across these subspaces in the dynamical equilibrium is defined by the strength of the various error mechanisms in our system as well as the rate at which these errors are corrected. 
The probability of occupying the code space, given by the sum of the first two eigenvalues, remains constant over time and equal to $\langle \Pi_{0}\rangle=0.825\pm0.003$, where error bar represents the standard deviation with respect to different durations of the QEC process. This value agrees well with an independent analysis in Section~\ref{sec:msmt stat}. 
Having only one relevant error subspace in the steady-state distribution also qualitatively agrees with an observation in Section~\ref{sec:msmt stat} that errors are rare. 
We believe that other error subspaces are populated with probability $<1\%$, which is beyond the resolution power of this method. 
Developing a more accurate and sample-efficient reconstruction technique for characterizing the  distribution across the error spaces is an important direction left for the future.

\begin{figure}
    \centering
    \includegraphics[width=\figwidthDouble]{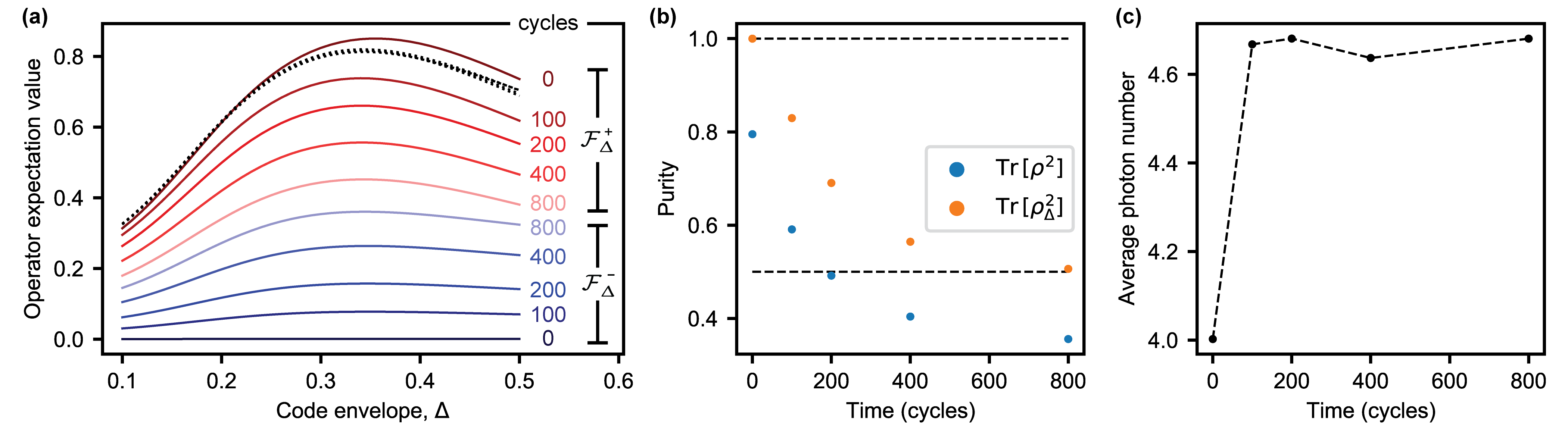}
    \caption[Analysis of state evolution]{
    \textbf{Analysis of $\boldsymbol{|+Z\rangle}$ evolution.}
    {\bf (a)} Expectation value ${\cal F}_\Delta^\pm$ of the state projectors $|\pm Z_\Delta\rangle\langle \pm Z_\Delta|$ for a range of values of $\Delta$ and for different durations of the QEC process. The fidelity ${\cal F}_\Delta^+$ decreases, while ${\cal F}_\Delta^-$ increases as a function of time, as expected for a logical Pauli channel. The expectation value of the code projector $\langle \Pi_\Delta\rangle = {\cal F}_\Delta^+ + {\cal F}_\Delta^-$ (black dotted lines) remains nearly time-independent for $T>0$. 
    {\bf (b)} Purity of the reconstructed state $\rho$ and of its projection onto the code space $\rho_{\Delta}$ as a function of time. 
    {\bf (c)} Average photon number as a function of time. 
\label{fig:wigner_analysis}}
\end{figure}

{\bf Extracting code envelope size.}
For each QEC duration $T$, the reconstructed density matrix is used to find the fidelity of the experimental states to the family of finite-energy codewords $\{|\pm Z_\Delta\rangle\}$ parametrized by the envelope size $\Delta$. For $T>0$, we additionally displace the target codewords by $(0.08-0.12i)\sqrt{\pi/2}$ to account for a small shift visible in the tomography. Since these target states are pure, the fidelity is given by ${\cal F}_\Delta^{+}={\rm Tr}\left[|+ Z_\Delta\rangle\langle + Z_\Delta|\, \rho\right]$ and ${\cal F}_\Delta^{-}={\rm Tr}\left[|- Z_\Delta\rangle\langle - Z_\Delta|\, \rho\right]$. For experiments that start with a preparation of $|+Z\rangle$, these fidelities are shown in Fig.~\ref{fig:wigner_analysis}(a) for each QEC duration from the dataset in Fig.~\ref{fig:wigner_tomo_after_qec}. Immediately after the initialization, the fidelity ${\cal F}_\Delta^{+}$ is maximized for $\Delta=0.36$, where it reaches $0.85$. During the QEC process, ${\cal F}_\Delta^{+}$ gradually reduces while ${\cal F}_\Delta^{-}$ increases, consistent with the logical Pauli channel.  The sum $\langle \Pi_\Delta\rangle = {\cal F}_\Delta^{-}+{\cal F}_\Delta^{+}$, which is equal to the expectation value of the code projector, remains nearly constant for $T>0$. It is maximized at $\Delta=0.34$ ($9.4\,\rm dB$), where it is equal to $\langle \Pi_\Delta\rangle=0.817\pm0.003$. This value is close to $\langle \Pi_{0}\rangle=0.825\pm0.003$ extracted from the density matrix spectrum, which indicates that the code ${\cal C}_0$ stabilized in the experiment is indeed the one defined by the envelope operator $\exp(-\Delta^2 n)$ with $\Delta=0.34$. 


The purity of $|+Z\rangle$ state is shown in Fig.~\ref{fig:wigner_analysis}(b). During the QEC process, it reduces below $0.5$, since the steady state  contains a mixture of the codewords with their images in the error spaces. However, the part of this mixed state that resides within the code space should approach a purity of $0.5$. To confirm that this is the case, for every state $\rho$ we define its projection onto the code space as $\rho_{\Delta}=\Pi_{\Delta}\, \rho\, \Pi_{\Delta} /\, {\rm Tr} [\Pi_{\Delta}\, \rho\, \Pi_{\Delta}]$, where we only consider the code with the optimal envelope $\Delta=0.34$. As seen in Fig.~\ref{fig:wigner_analysis}(b), the purity of $\rho_{\Delta}$ after initialization is close to $1$, and after hundreds of cycles it approaches $0.5$, as expected for the logical Pauli channel. 

Lastly, in Fig.~\ref{fig:wigner_analysis}(c) we plot the evolution of the average photon number $\langle n\rangle={\rm Tr}[a^\dagger a\,\rho]$ during the QEC process. The extracted steady-state photon number is $\langle n \rangle=4.67\pm 0.02$. 


\subsection{Sensitivity to ancilla errors\label{sec:noise injection}}

{\bf Ancilla phase flips.}
Our QEC circuit is fault-tolerant with respect to ancilla phase flips by design \cite{Royer2020}. To see this, consider that if $\sigma_z$ error happens during the ancilla readout, it would have no effect because the readout projects the ancilla onto an eigenstate of $\sigma_z$. Likewise, during the virtual rotation gate or the idle time the ancilla is nominally in the $|g\rangle$ state, which is an eigenstate of $\sigma_z$. Finally, the effect of $\sigma_z$ errors on the SBS unitary can be understood by propagating them through the circuit layers. For example, if such an error happens during the big conditional displacement, it is equivalent to changing the circuit parameters from $\vec{\beta}=l_S\times(i\Delta^2/2,1,i\Delta^2/2)$ to $\vec{\beta}=l_S\times(i\Delta^2/2,1,-i\Delta^2/2)$. Since $\Delta^2\ll 1$, this change is equivalent to a small error that will be corrected in the following QEC cycles. 

{\bf Ancilla bit flips.} In contrast, ancilla bit flips can detrimentally affect the logical qubit in several ways. If such an error happens during the middle half of the big conditional displacement of amplitude $l_S$, it will with high probability generate a logical error. This mechanism accounts for a significant fraction of logical errors in the experiment. For example, its contribution to the Pauli error probability $p_X=1.8\times10^{-3}$ (per QEC cycle), is estimated to be $\sim 0.5\times0.5\times0.5\times(700{\,\rm ns} / 280{\rm\, \mu s})\approx 0.3\times 10^{-3}$, where the factors of $0.5$ account for (i) half of the superposition state being sensitive to ancilla decay, (ii) half of the QEC cycle is devoted to position quadrature, (iii) half of the big conditional displacement gate. In practice, the relaxation time of the ancilla is likely degraded during the execution of the conditional displacement due to the large number of intermediate photons in the oscillator, see \cite{Eickbusch2021} and evidence in Section~\ref{sec:stability}. Hence, this estimate provides an optimistic lower bound. Ancilla bit flips can also create detrimental back-action on the oscillator if they happen during the readout time. Since readout outcome is used in a feedback loop to implement a virtual rotation gate, misclassification of the ancilla state generates rotational errors that the GKP code is not well suited to correct, with erroneous rotation angle distributed in the range $0.0-0.6$ radians. To estimate the contribution of such errors to the logical error rate, consider that a rotation by $\sim\Delta/(l_S/2)\approx 0.3$ radians would diminish the overlap of the blobs in the Wigner function; therefore, a significant  fraction of misclassification-induced rotation errors cause large disturbance of the stabilized code space. The transmon $|e\rangle$ state is the most prone to misclassification. Since readout fidelity of the $|e\rangle$ state is close to $99\%$, and this outcome is generated $10\%$ of the time, we estimate an additional $\sim 0.5\times 10^{-3}$ contribution to logical error probability per QEC cycle from this mechanism.  The two contributions described here account for half of the logical error probability $p_X$, and the remaining half is not yet well understood. 

\begin{figure}
  \begin{minipage}[c]{0.5\textwidth}
    \includegraphics[width=\figwidth]{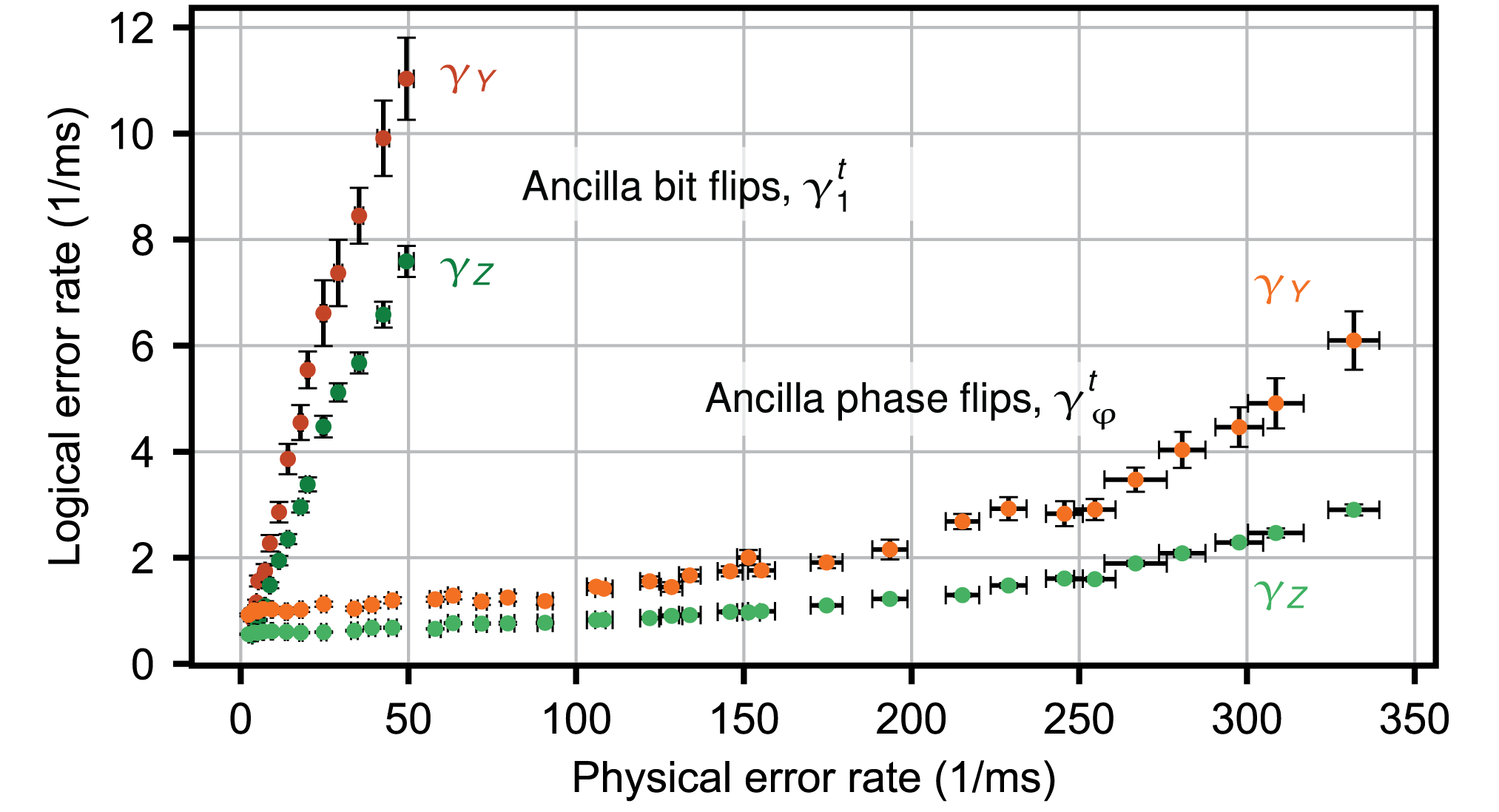}
  \end{minipage}\hfill
  \begin{minipage}[c]{0.5\textwidth}
	\caption{\textbf{Effect of ancilla errors.} Logical error rates $\gamma_Z$ and $\gamma_Y$ as a function of physical error rates $\gamma_\varphi^{t}$ and $\gamma_1^t$ of the ancilla transmon ($\gamma_X$ is expected to behave identically to $\gamma_Z$). Physical error rates are varied with noise injection. The error sensitivities $d\gamma_Z/d\gamma_1^t=0.17$, $d\gamma_Y/d\gamma_1^t=0.25$, $d\gamma_Z/d\gamma_\varphi^t=0.0027$, and $d\gamma_Y/d\gamma_\varphi^t=0.0050$ are extracted by linear fits in the low-error region. The derived sensitivity of $\Gamma_{\rm GKP}=(\gamma_X+\gamma_Y+\gamma_Z)/3$ to phase flips is 65 times smaller than to bit flips.
	\label{fig:error_injection}}
  \end{minipage}
\end{figure}

{\bf Transmon noise injection.}
To check the effect of ancilla errors on the logical performance in a controllable way, we perform noise injection experiments that selectively increase the transmon phase-flip rate $\gamma_\varphi^{\,t}$ or bit-flip rate $\gamma_1^{\,t}$, with the results shown in Fig.~\ref{fig:error_injection}. With noise injection, we are able to increase $\gamma_1^{\,t}$ by a factor of $14$ (spoiling $T_1^{\,t}$ from $290\,\rm \mu s$ to $20\,\rm \mu s$), and $\gamma_\varphi^{\,t}$ by a factor of $140$ (spoiling $T_{\varphi}^{\,t}$ from $430\,\rm \mu s$ to $3\,\rm \mu s$).
Using linear fits in the low-error region, we extract the error sensitivities $d\gamma_Z/d\gamma_1^t=0.17$, $d\gamma_Y/d\gamma_1^t=0.25$, $d\gamma_Z/d\gamma_\varphi^t=0.0027$, and $d\gamma_Y/d\gamma_\varphi^t=0.0050$. The derived sensitivity of $\Gamma_{\rm GKP}=(\gamma_X+\gamma_Y+\gamma_Z)/3$ to ancilla phase flips is 65 times smaller than the sensitivity to ancilla bit flips, confirming the qualitative arguments provided above.  Mitigating the effect of ancilla bit flips on the logical performance is one of the most important future directions in grid-code QEC. 

{\bf Verifying noise injection.}
In the following, we explain how the noise injection experiments were conducted and how we verified that the noise affects the system as intended, i.e. selectively tunes $\gamma_\varphi^{\,t}$ or $\gamma_1^{\,t}$. We are able to achieve high degree of  selectivity, with negligibly small spurious effects. 
To spoil $\gamma_1^{\,t}$, we inject noise at the transmon frequency, and to spoil $\gamma_\varphi^{\,t}$, we inject noise at low frequency \cite{Clerk2010}.
The baseband white noise with flat spectral density up to $80\,\rm MHz$ is sourced from an Agilent 33250A arbitrary waveform generator. In the $\gamma_1^{\,t}$ tuning experiment, it is upconverted to the qubit frequency using a double-balanced mixer with an LO blue-detuned by $30\,\rm MHz$ from the qubit frequency. 
After this pre-processing, the noise is filtered and combined with the qubit control line after the switch (in contrast to all control pulses, the noise is not gated). 

\begin{figure}
    \centering
    \includegraphics[width=\figwidthDouble]{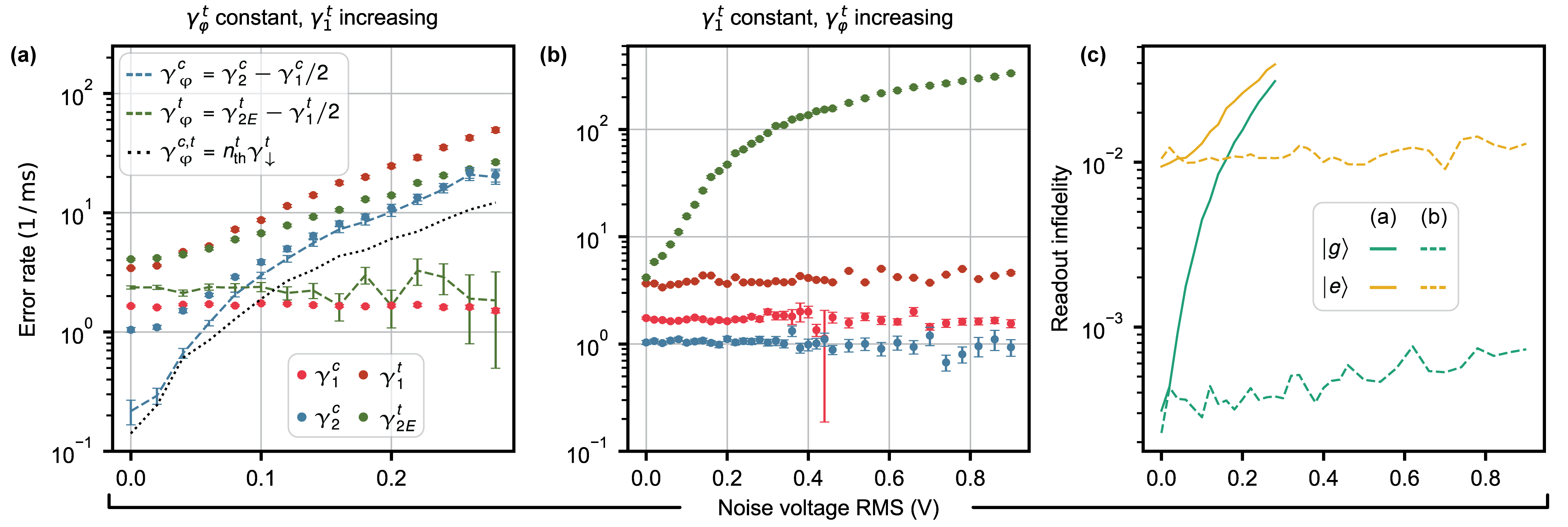}
    \caption[Noise verification experiments]{
    \textbf{Noise verification experiments.}
    {\bf (a)} Component error rates as a function of the root mean square (RMS) voltage of the injected noise (at the generator plane). Here, noise is up-converted to the qubit frequency to increase $\gamma_1^{\,t}$.
    {\bf (b)} Same as in (a), but with the baseband noise that increases $\gamma_\varphi^{\,t}$.
    {\bf (c)} Readout infidelity of the transmon $|g\rangle$ and $|e\rangle$ states in these two noise injection settings. 
\label{fig:noise_verification}}
\end{figure}

In Fig.~\ref{fig:noise_verification}(a), we inject resonant noise to tune $\gamma_1^{\,t}$. 
This noise couples to the $\sigma_x$ operator and therefore changes $\gamma_\downarrow^{\,t}$ and $\gamma_\uparrow^{\,t}$ symmetrically, which results in increased steady-state population of the qubit, approaching $0.5$ at the largest applied noise power. Note that this noise also affects the dephasing rate $\gamma_{2E}^{\,t}$, but the changes in $\gamma_{2E}^{\,t}$ are explained by changes in $\gamma_1^{\,t}$: the extracted pure dephasing rate $\gamma_\varphi^{\,t}=\gamma_{2E}^{\,t}-\gamma_1^{\,t}/2$ remains independent of the noise power, as intended in this experiment. The error bars on $\gamma_\varphi^{\,t}$ increase at large noise power, because this small rate is extracted as a difference of two large rates. The oscillator dephasing rate $\gamma_2^{\,c}$ is also affected by the noise, which is explained by the increased rate of qubit up- and down-transitions that dephase the oscillator through the dispersive coupling \cite{Reagor2016}. The pure dephasing rate $\gamma_\varphi^{\,c}=\gamma_2^{\,c}-\gamma_1^{\,c}/2$ of the oscillator agrees reasonably well with the prediction $\gamma_\varphi^{\,c,t}=n_{\rm th}^{\,t}\gamma_\downarrow^{\,t}$ derived from this mechanism (black dotted line). The disagreement at high noise power is under investigation; it likely comes from the breakdown of the simple formula for $\gamma_\varphi^{\,c,t}$ in the limit where $n_{\rm th}^{\,t}$ is not small.

In Fig.~\ref{fig:noise_verification}(b), we inject baseband noise to tune $\gamma_\varphi^{\,t}$.
In addition to this desired effect, within the same dynamic range of the noise we observe an undesired increase of the qubit excited state population by a factor of $2$ (data not shown), likely due to the heating of the attenuators by the dissipated noise power. Since $\gamma_\uparrow^{\,t}/\gamma_1^{\,t}\ll 1$, the qubit lifetime is not significantly affected by this heating. The lifetime and coherence of the oscillator also remain independent of the noise power. The increase of the error bars on $\gamma_1^{\,c}$ and $\gamma_2^{\,c}$ with the noise power is related to strong degradation of the fidelity of the transmon selective pulse used to read out the population of the oscillator $|0\rangle$ and $|1\rangle$ states as described in Section~\ref{sec:Fock encoding characterization}. This pulse has a duration of $\sim 20\,\rm \mu s$ and it is directly sensitive to the transmon coherence; at the highest injected noise power, where coherence time is spoiled down to $3\,\rm \mu s$, the fidelity of this selective pulse is only a few percent. 

In Fig.~\ref{fig:noise_verification}(c), we show the effect of the noise on the readout fidelity of the transmon $|g\rangle$ and $|e\rangle$ states. In principle, the noise that induces phase flips ($\sigma_z$ errors) should not affect the readout of $\sigma_z$. However, due to the aforementioned heating of the qubit, we observe a weak degradation of ${\cal F}_r^{(g)}$. On the other hand, noise at the qubit frequency couples to $\sigma_x$ and results in significant degradation of both ${\cal F}_r^{(g)}$ and ${\cal F}_r^{(e)}$. 

\subsection{Long-time system stability \label{sec:stability}}

\begin{figure}
    \centering
    \includegraphics[width=\figwidthDouble]{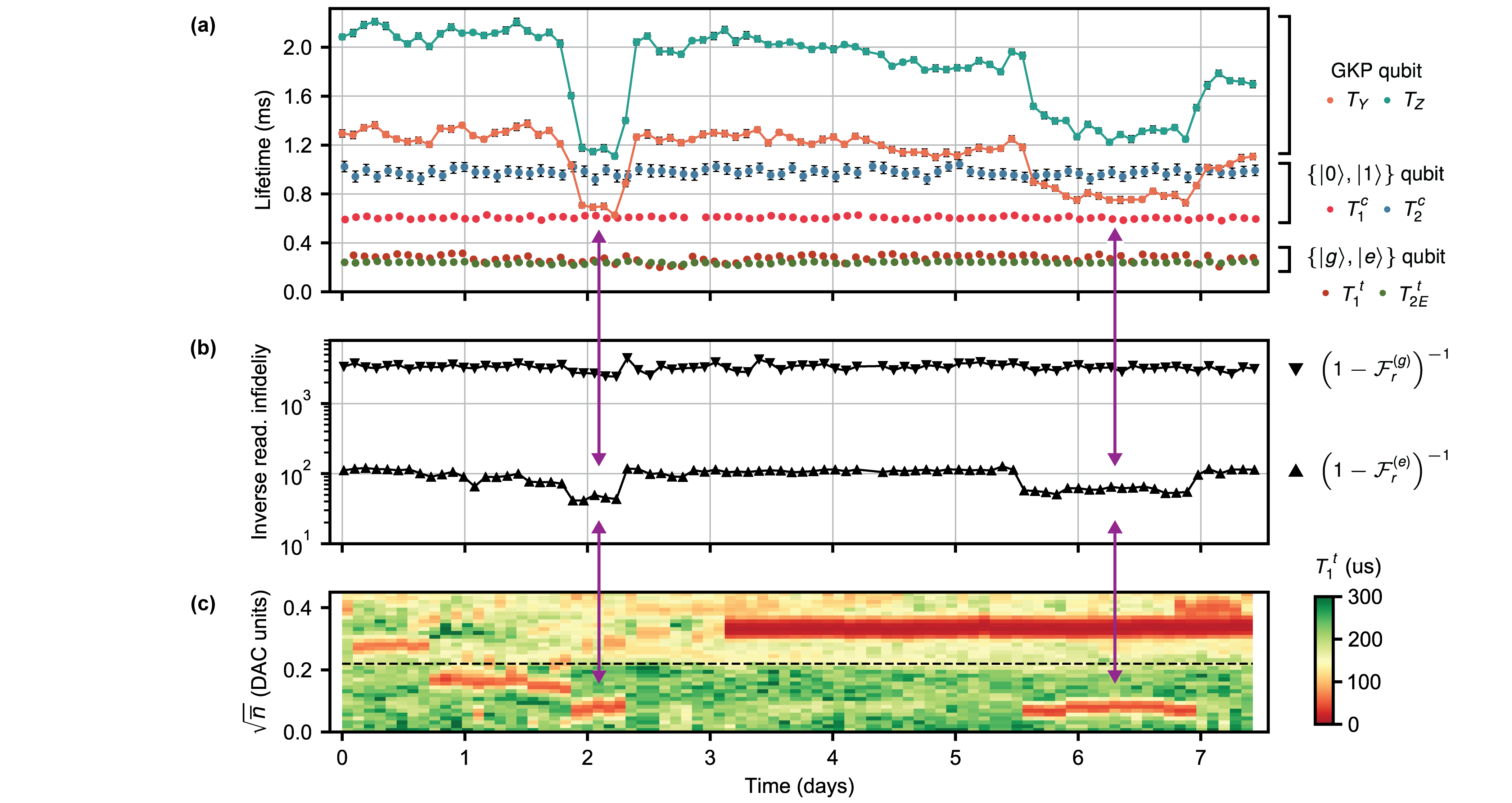}
    \caption[Quantum system stability]{
    \textbf{Quantum system stability.}
    {\bf (a)} Lifetimes of Pauli $Y$ and $Z$ eigenstates of $\{|0\rangle, |1\rangle\}$ qubit, $\{|g\rangle, |e\rangle\}$ qubit, and an error-corrected GKP qubit. 
    {\bf (b)} Inverse readout infidelity of the transmon $|g\rangle$ and $|e\rangle$ states. Logical lifetime is strongly correlated with $(1-{\cal F}_r^{(e)})^{-1}$.
    {\bf (c)} Transmon lifetime $T_1^{\,t}$ as a function of the number $\overline{n}$ of the steady-state photons in the readout resonator.  The dashed line denotes the DAC amplitude used for the actual readout. The correlated degradation of the system performance and appearance of a spurious resonance that degrades $T_1^{\,t}(\overline{n})$ around $\sqrt{\overline{n}}=0.1$ is indicated with purple arrows. 
\label{fig:lifetimes_stability}}
\end{figure}

With repetitive measurements of the lifetimes of $\{|0\rangle, |1\rangle\}$ qubit, $\{|g\rangle, |e\rangle\}$ qubit, and error-corrected GKP qubit, we investigate the stability of our quantum system over time, with results of a week-long scan shown in Fig.~\ref{fig:lifetimes_stability}. 
We find that the $\{|0\rangle, |1\rangle\}$ qubit is the most stable, which we attribute to the fact that most of the electromagnetic field is stored in the vacuum of the cavity. In contrast, the $\{|g\rangle, |e\rangle\}$ qubit exhibits notable fluctuations of the $|e\rangle$ state lifetime; such fluctuations are often observed in transmons \cite{Klimov2018,Carroll2021}, and are typically attributed to two-level defects in the amorphous dielectric, although there are other mechanisms that could lead to such fluctuations, and their source in our system is not yet understood.

We also find significant fluctuations of the lifetime of an error-corrected GKP qubit. Periods of relative stability are regularly interrupted with sudden drops and resurgences of performance, correlated with the appearance and disappearance of a resonant feature in the $T_1^{\, t}(\overline{n})$ dependence, see Fig.~\ref{fig:lifetimes_stability}(c). The behavior of readout infidelity of the $|e\rangle$ state is also correlated with this feature, see Fig.~\ref{fig:lifetimes_stability}(b). We find that the correlation coefficient between the logical error rate and the readout infidelity is $r=0.81$. However, preliminary simulations indicate that degradation of the readout fidelity alone is not sufficient to explain the collapses of the logical performance. We therefore believe that the presence of the spurious resonance affects not only the readout fidelity, but also the fidelity of the SBS unitary. 

A plausible causal chain is the following: 1) for unknown reason, the spurious degrees of freedom (defects) appear and disappear; 2) when the transmon $|g\rangle\leftrightarrow|e\rangle$ transition frequency is resonant with the defect, their interaction strength is enhanced, which reduces the lifetime of the $|e\rangle$ state; 3) during the readout, the transmon is Stark-shifted by the readout photons into resonance with the defects, which increases the probability of readout errors; 4) when the transmon state is misclassified, the virtual rotation gate is executed with an incorrect angle, inducing a phase-space rotation error on the oscillator; 5) during the conditional displacement gates, the transmon is also Stark-shifted into resonance with the defects by the intermediate photons of the cavity mode; 6) when transmon decay happens during the big conditional displacement gate, it has a significant chance of inducing a logical error. 
This proposed connection between spurious defects and fluctuations of the logical performance could be verified with detailed system-level simulations that take into account time-dependent Stark shift of the transmon and Stark-shift-dependent degradation of $T_1^{\,t}$, which is left for the  future analysis.

Apart from the stochastic fluctuations, we observe a systematic drift that warrants periodic retraining of the QEC circuit. This drift can be seen by comparing the initial and final data points of the scan in Fig.~\ref{fig:lifetimes_stability}, where all the monitored physical error sources are similar in magnitude, but the logical lifetime is reduced in the final point as compared to the initial point. Due to this effect, the various datasets reported in our work were acquired with different version of the QEC circuit that were retrained every 1-2 weeks.


\subsection{Average channel fidelity \label{sec:channel fidelity}}

The {\sl average channel fidelity} of a quantum channel ${\cal E}:\rho\to{\cal E}(\rho)$ to a target unitary channel ${\cal U}: \rho\to U\rho\, U^\dagger$ is
\begin{align}
\overline{{\cal F}}({\cal E}, {\cal U}) = \int d\psi \langle \psi | U^\dagger {\cal E}(|\psi\rangle\langle \psi|)U|\psi\rangle, \label{fidelity_Haar}
\end{align}
where the integral is over the uniform measure on the state space, normalized so that $\int d\psi=1$. Henceforth, we refer to this metric simply as {\sl fidelity}.

To derive an equivalent but experimentally-compatible expression, we make use of the Pauli transfer matrix (PTM) representation of a channel ${R}_{ij}[{\cal E}]=\frac{1}{2}{\rm Tr}(\sigma_{i}{\cal E}[\sigma_{j}])$, where $\{\sigma_k,\, k=I,X,Y,Z\}$ are Pauli matrices. This representation has several useful properties, e.g. that composition of channels corresponds to a product of their PTMs \cite{Greenbaum2015}. In terms of the PTM, we have the following expression for fidelity:
\begin{align}
\overline{{\cal F}} = \frac{2{\cal F}_e+1}{3}, \quad
{\cal F}_e= \frac{1}{4}{\rm Tr}\big(R^{T}[{\cal U}]\,R[{\cal E}]\big),\label{fidelity definition}
\end{align}
where ${\cal F}_e$ is often called the {\sl entanglement fidelity}. 

\begin{figure}
  \begin{minipage}[c]{0.7\textwidth}
    \includegraphics[width=4.5in]{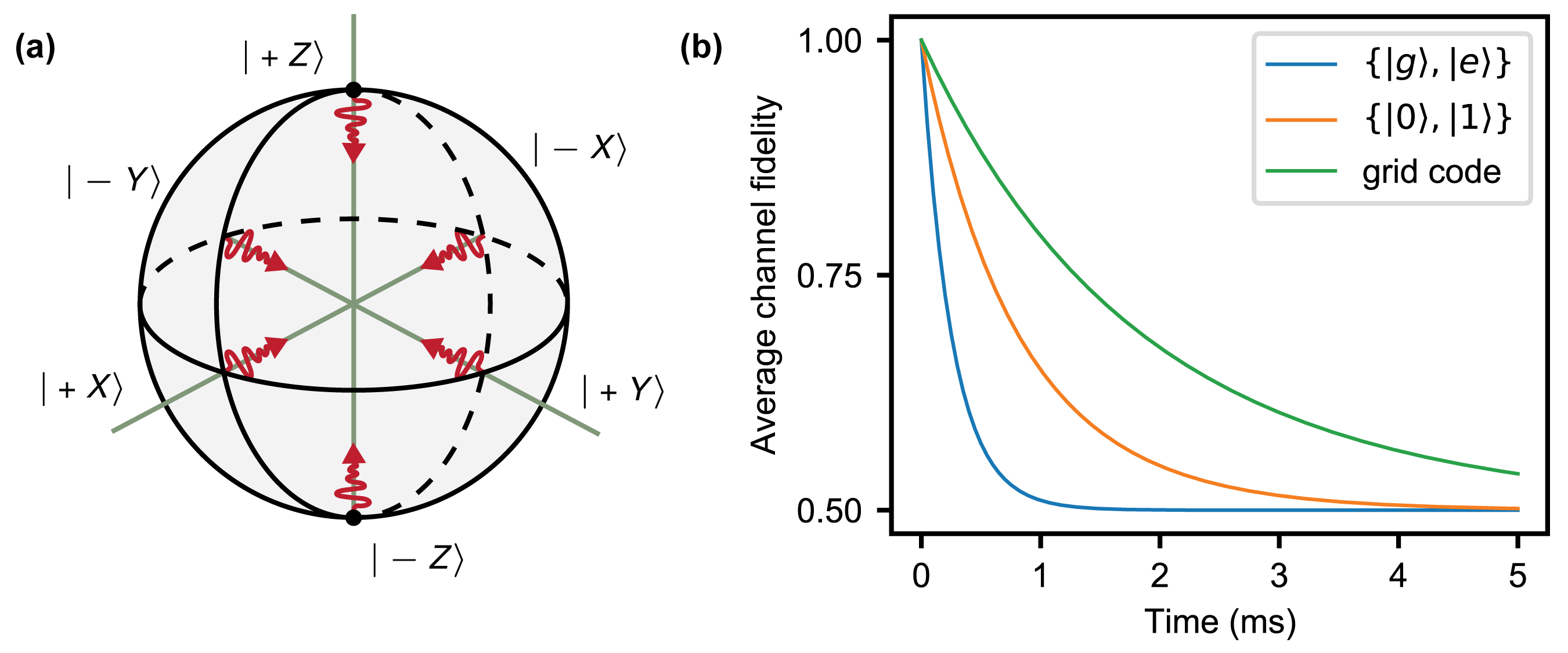}
  \end{minipage}\hfill
  \begin{minipage}[c]{0.3\textwidth}
	\caption{\textbf{Average channel fidelity.}
{\bf (a)} Illustration of channel action on the Pauli eigenstates. By linearity, the evolution of these six cardinal points is sufficient to predict the average effect across the whole Bloch sphere. 
{\bf (b)} Expected time evolution of average channel fidelity for three different qubits, calculated using experimentally extracted lifetimes of Pauli eigenstates.
\label{fig:channel_fidelity}}
  \end{minipage}
\end{figure}

To benchmark a quantum error correction channel, we compare it to an identity channel ${\cal I}: \rho\to \rho$ with $R_{ij}[{\cal I}]=\delta_{ij}$. In this case, Eq.~\eqref{fidelity definition} can be further simplified to 
\begin{align}
\overline{{\cal F}}=\frac{1}{12}\sum_{{P}=X,Y,Z}\big({\rm Tr}[{P}\,{\cal E}(|+{P}\rangle\langle+{P}|)]-{\rm Tr}[{P}\,{\cal E}(|-{P}\rangle\langle-{P}|)]\big)
+\frac{1}{2}, \label{eq:process fidelity}
\end{align}
where we made use of the identities $\text{{\rm Tr}}[\sigma_{P}{\cal E}(\sigma_{P})]={\rm Tr}[P\,{\cal E}(|+P\rangle\langle+P|)]-{\rm Tr}[P\,{\cal E}(|-P\rangle\langle-P|)]$ for $P\in\{X,Y,Z\}$, and ${\rm Tr}({\cal E}[I])=2$. The complete derivation of this formula starting from Eq.~\eqref{fidelity_Haar} can be found in Ref.~\cite{Nielsen2002}.
In experiment, the right-hand side of Eq.~\eqref{eq:process fidelity} is measured by preparing $\pm1$ Pauli eigenstates, passing them through the channel ${\cal E}$, and then measuring the corresponding Pauli operator, as cartooned in Fig.~\ref{fig:channel_fidelity}(a). Such a procedure is applicable to an arbitrary duration $t$ of the channel ${\cal E}(t)$. 

We focus on the comparison of three different qubits in our system: $\{|0\rangle, |1\rangle\}$, $\{|g\rangle, |e\rangle\}$, and an error-corrected GKP qubit. The free evolution of the two passive qubits is modeled using a composite amplitude damping and white-noise dephasing channel, while the evolution of an error-corrected GKP qubit is modeled using a logical Pauli channel. Given these well-justified assumptions on the error channels, from Eq.~(\ref{eq:process fidelity}) we find:
\begin{align}
\overline{{\cal F}}_{\{01\}}(t)&=\frac{1}{6}e^{-\gamma_{1}^{c}t}+\frac{1}{3}e^{-\gamma_{2}^c t}+\frac{1}{2},\label{eq:process fidelity Fock} \\
\overline{{\cal F}}_{\{ge\}}(t)&=\frac{1}{6}e^{-\gamma_{1}^{t}t}+\frac{1}{3}e^{-\gamma_{2E}^t t}+\frac{1}{2},\label{eq:process fidelity transmon} \\
\overline{{\cal F}}_{{\rm GKP}}(t)&=\frac{1}{6}e^{-\gamma_{X}t}+\frac{1}{6}e^{-\gamma_{Y}t}+\frac{1}{6}e^{-\gamma_{Z}t}+\frac{1}{2}. \label{eq:process fidelity GKP}
\end{align}

We show the time evolution of the fidelity given by Eqs.~(\ref{eq:process fidelity Fock}$-$\ref{eq:process fidelity GKP}) in Fig.~\ref{fig:channel_fidelity}(b), using experimentally extracted decay rates at the highest QEC gain measured in our experiment. 

As seen above, in general the fidelity decays to its steady-state value in a way that cannot be characterized by a single time constant even in the simplest error models such as Pauli noise or amplitude damping. Therefore, fitting the fidelity decay to a single exponential is not strictly valid, although this heuristic approach was adopted in the previous works on bosonic QEC \cite{Ofek2016a, Hu2019a, Gertler2020}. 
To avoid such an inconsistency, we consider the channel ${\cal E}$ acting for only a short time $\delta t$. 
Any time dependence of the fidelity, even if it contains multiple exponentially decaying contributions, at short times is equivalent to a linear decay:
\begin{align}
\overline{{\cal F}}(\delta t)=1-\frac{1}{2}\Gamma\,\delta t, \label{short time expansion}
\end{align}
where $\Gamma$ is an effective depolarization rate, and $1/\Gamma$ is the fidelity lifetime. For a depolarizing channel ${\cal E}_{\rm dep}(\rho)=(1-p)\rho + p \frac{I}{2}$ with a depolarization probability $p=1-e^{-\gamma t}$, we have $\Gamma = \gamma$, motivating the name and the coefficient $1/2$ in Eq. \eqref{short time expansion}. For qubits considered here, the effective depolarization rates are:
\begin{align}
\Gamma_{\{01\}} =\frac{2\gamma_{2}^{\,c} + \gamma_{1}^{\,c}}{3}, \quad
\Gamma_{\{ge\}} =\frac{2\gamma_{2E}^{\,t} + \gamma_{1}^{\,t}}{3}, \quad
\Gamma_{{\rm GKP}} =\frac{\gamma_{X} + \gamma_{Y} + \gamma_{Z}}{3}.
\end{align}

We define the coherence gain $G$ as an improvement of the effective depolarization rate of an error-corrected logical qubit over the best physical qubit in the same system (with the break-even point corresponding to $G=1$). In a bosonic circuit QED system, the latter is typically the $\{|0\rangle, |1\rangle\}$ qubit, hence $G=\Gamma_{\{01\}}/\Gamma_{\rm GKP}$. 
The highest gain achieved in our experiment is $G_{\rm max}=2.27\pm0.07$. During a scan discussed in Section~\ref{sec:stability}, gain remained above break-even $100\%$ of this week-long time window, with a median of $\overline{G}=2.0$.

Lastly,  we acknowledge that the decay constant of the average channel fidelity is not the only relevant metric that we expect to be correlated with the future ability of such systems to participate in quantum computations. Other metrics, such as the SPAM fidelity and the fidelity of gates, are important as well, and we leave their optimization and detailed characterization for future work.




\putbib[lib_clean]
\end{bibunit}
\makeatother


\begin{thebibliography}{55}
\expandafter\ifx\csname natexlab\endcsname\relax\def\natexlab#1{#1}\fi
\expandafter\ifx\csname bibnamefont\endcsname\relax
  \def\bibnamefont#1{#1}\fi
\expandafter\ifx\csname bibfnamefont\endcsname\relax
  \def\bibfnamefont#1{#1}\fi
\expandafter\ifx\csname citenamefont\endcsname\relax
  \def\citenamefont#1{#1}\fi
\expandafter\ifx\csname url\endcsname\relax
  \def\url#1{\texttt{#1}}\fi
\expandafter\ifx\csname urlprefix\endcsname\relax\def\urlprefix{URL }\fi
\providecommand{\bibinfo}[2]{#2}
\providecommand{\eprint}[2][]{\url{#2}}

\bibitem[{\citenamefont{Knill and Laflamme}(1997)}]{Knill1997}
\bibinfo{author}{\bibfnamefont{E.}~\bibnamefont{Knill}} \bibnamefont{and}
  \bibinfo{author}{\bibfnamefont{R.}~\bibnamefont{Laflamme}},
  \emph{\bibinfo{title}{Theory of quantum error-correcting codes}},
  \bibinfo{journal}{Phys. Rev. A} \textbf{\bibinfo{volume}{55}},
  \bibinfo{pages}{900} (\bibinfo{year}{1997}).

\bibitem[{\citenamefont{Gottesman et~al.}(2001)\citenamefont{Gottesman, Kitaev,
  and Preskill}}]{Gottesman2001}
\bibinfo{author}{\bibfnamefont{D.}~\bibnamefont{Gottesman}},
  \bibinfo{author}{\bibfnamefont{A.}~\bibnamefont{Kitaev}}, \bibnamefont{and}
  \bibinfo{author}{\bibfnamefont{J.}~\bibnamefont{Preskill}},
  \emph{\bibinfo{title}{{Encoding a qubit in an oscillator}}},
  \bibinfo{journal}{Physical Review A} \textbf{\bibinfo{volume}{64}},
  \bibinfo{pages}{012310} (\bibinfo{year}{2001}).

\bibitem[{\citenamefont{Mirrahimi et~al.}(2014)\citenamefont{Mirrahimi,
  Leghtas, Albert, Touzard, Schoelkopf, Jiang, and Devoret}}]{Mirrahimi2014}
\bibinfo{author}{\bibfnamefont{M.}~\bibnamefont{Mirrahimi}},
  \bibinfo{author}{\bibfnamefont{Z.}~\bibnamefont{Leghtas}},
  \bibinfo{author}{\bibfnamefont{V.~V.} \bibnamefont{Albert}},
  \bibinfo{author}{\bibfnamefont{S.}~\bibnamefont{Touzard}},
  \bibinfo{author}{\bibfnamefont{R.~J.} \bibnamefont{Schoelkopf}},
  \bibinfo{author}{\bibfnamefont{L.}~\bibnamefont{Jiang}}, \bibnamefont{and}
  \bibinfo{author}{\bibfnamefont{M.~H.} \bibnamefont{Devoret}},
  \emph{\bibinfo{title}{{Dynamically protected cat-qubits: a new paradigm for
  universal quantum computation}}}, \bibinfo{journal}{New Journal of Physics}
  \textbf{\bibinfo{volume}{16}}, \bibinfo{pages}{045014}
  (\bibinfo{year}{2014}).

\bibitem[{\citenamefont{Michael et~al.}(2016)\citenamefont{Michael, Silveri,
  Brierley, Albert, Salmilehto, Jiang, and Girvin}}]{Michael2016}
\bibinfo{author}{\bibfnamefont{M.~H.} \bibnamefont{Michael}},
  \bibinfo{author}{\bibfnamefont{M.}~\bibnamefont{Silveri}},
  \bibinfo{author}{\bibfnamefont{R.~T.} \bibnamefont{Brierley}},
  \bibinfo{author}{\bibfnamefont{V.~V.} \bibnamefont{Albert}},
  \bibinfo{author}{\bibfnamefont{J.}~\bibnamefont{Salmilehto}},
  \bibinfo{author}{\bibfnamefont{L.}~\bibnamefont{Jiang}}, \bibnamefont{and}
  \bibinfo{author}{\bibfnamefont{S.~M.} \bibnamefont{Girvin}},
  \emph{\bibinfo{title}{New class of quantum error-correcting codes for a
  bosonic mode}}, \bibinfo{journal}{Physical Review X}
  \textbf{\bibinfo{volume}{6}}, \bibinfo{pages}{031006} (\bibinfo{year}{2016}).

\bibitem[{\citenamefont{Grimsmo et~al.}(2020)\citenamefont{Grimsmo, Combes, and
  Baragiola}}]{Grimsmo2020}
\bibinfo{author}{\bibfnamefont{A.~L.} \bibnamefont{Grimsmo}},
  \bibinfo{author}{\bibfnamefont{J.}~\bibnamefont{Combes}}, \bibnamefont{and}
  \bibinfo{author}{\bibfnamefont{B.~Q.} \bibnamefont{Baragiola}},
  \emph{\bibinfo{title}{Quantum computing with rotation-symmetric bosonic
  codes}}, \bibinfo{journal}{Physical Review X} \textbf{\bibinfo{volume}{10}},
  \bibinfo{pages}{011058} (\bibinfo{year}{2020}).

\bibitem[{\citenamefont{Shor}(1995)}]{Shor1995}
\bibinfo{author}{\bibfnamefont{P.~W.} \bibnamefont{Shor}},
  \emph{\bibinfo{title}{Scheme for reducing decoherence in quantum computer
  memory}}, \bibinfo{journal}{Phys. Rev. A} \textbf{\bibinfo{volume}{52}},
  \bibinfo{pages}{R2493} (\bibinfo{year}{1995}).

\bibitem[{\citenamefont{Steane}(1996)}]{Steane1996}
\bibinfo{author}{\bibfnamefont{A.~M.} \bibnamefont{Steane}},
  \emph{\bibinfo{title}{Error correcting codes in quantum theory}},
  \bibinfo{journal}{Phys. Rev. Lett.} \textbf{\bibinfo{volume}{77}},
  \bibinfo{pages}{793} (\bibinfo{year}{1996}).

\bibitem[{\citenamefont{Fowler et~al.}(2012)\citenamefont{Fowler, Mariantoni,
  Martinis, and Cleland}}]{Fowler2012}
\bibinfo{author}{\bibfnamefont{A.~G.} \bibnamefont{Fowler}},
  \bibinfo{author}{\bibfnamefont{M.}~\bibnamefont{Mariantoni}},
  \bibinfo{author}{\bibfnamefont{J.~M.} \bibnamefont{Martinis}},
  \bibnamefont{and} \bibinfo{author}{\bibfnamefont{A.~N.}
  \bibnamefont{Cleland}}, \emph{\bibinfo{title}{{Surface codes: Towards
  practical large-scale quantum computation}}}, \bibinfo{journal}{Physical
  Review A} \textbf{\bibinfo{volume}{86}}, \bibinfo{pages}{032324}
  (\bibinfo{year}{2012}).

\bibitem[{\citenamefont{Noh and Chamberland}(2020)}]{Noh2020}
\bibinfo{author}{\bibfnamefont{K.}~\bibnamefont{Noh}} \bibnamefont{and}
  \bibinfo{author}{\bibfnamefont{C.}~\bibnamefont{Chamberland}},
  \emph{\bibinfo{title}{{Fault-tolerant bosonic quantum error correction with
  the surface--Gottesman-Kitaev-Preskill code}}}, \bibinfo{journal}{Phys. Rev.
  A} \textbf{\bibinfo{volume}{101}}, \bibinfo{pages}{012316}
  (\bibinfo{year}{2020}).

\bibitem[{\citenamefont{Darmawan et~al.}(2021)\citenamefont{Darmawan, Brown,
  Grimsmo, Tuckett, and Puri}}]{Darmawan2021}
\bibinfo{author}{\bibfnamefont{A.~S.} \bibnamefont{Darmawan}},
  \bibinfo{author}{\bibfnamefont{B.~J.} \bibnamefont{Brown}},
  \bibinfo{author}{\bibfnamefont{A.~L.} \bibnamefont{Grimsmo}},
  \bibinfo{author}{\bibfnamefont{D.~K.} \bibnamefont{Tuckett}},
  \bibnamefont{and} \bibinfo{author}{\bibfnamefont{S.}~\bibnamefont{Puri}},
  \emph{\bibinfo{title}{{Practical quantum error correction with the XZZX code
  and Kerr-cat qubits}}}, \bibinfo{journal}{PRX Quantum}
  \textbf{\bibinfo{volume}{2}}, \bibinfo{pages}{030345} (\bibinfo{year}{2021}).

\bibitem[{\citenamefont{Terhal et~al.}(2020)\citenamefont{Terhal, Conrad, and
  Vuillot}}]{Terhal2020}
\bibinfo{author}{\bibfnamefont{B.~M.} \bibnamefont{Terhal}},
  \bibinfo{author}{\bibfnamefont{J.}~\bibnamefont{Conrad}}, \bibnamefont{and}
  \bibinfo{author}{\bibfnamefont{C.}~\bibnamefont{Vuillot}},
  \emph{\bibinfo{title}{Towards scalable bosonic quantum error correction}},
  \bibinfo{journal}{Quantum Science and Technology}
  \textbf{\bibinfo{volume}{5}}, \bibinfo{pages}{043001} (\bibinfo{year}{2020}).

\bibitem[{\citenamefont{Ofek et~al.}(2016)\citenamefont{Ofek, Petrenko, Heeres,
  Reinhold, Leghtas, Vlastakis, Liu, Frunzio, Girvin, Jiang
  et~al.}}]{Ofek2016a}
\bibinfo{author}{\bibfnamefont{N.}~\bibnamefont{Ofek}},
  \bibinfo{author}{\bibfnamefont{A.}~\bibnamefont{Petrenko}},
  \bibinfo{author}{\bibfnamefont{R.}~\bibnamefont{Heeres}},
  \bibinfo{author}{\bibfnamefont{P.}~\bibnamefont{Reinhold}},
  \bibinfo{author}{\bibfnamefont{Z.}~\bibnamefont{Leghtas}},
  \bibinfo{author}{\bibfnamefont{B.}~\bibnamefont{Vlastakis}},
  \bibinfo{author}{\bibfnamefont{Y.}~\bibnamefont{Liu}},
  \bibinfo{author}{\bibfnamefont{L.}~\bibnamefont{Frunzio}},
  \bibinfo{author}{\bibfnamefont{S.~M.} \bibnamefont{Girvin}},
  \bibinfo{author}{\bibfnamefont{L.}~\bibnamefont{Jiang}},
  \bibnamefont{et~al.}, \emph{\bibinfo{title}{{Extending the lifetime of a
  quantum bit with error correction in superconducting circuits}}},
  \bibinfo{journal}{Nature} \textbf{\bibinfo{volume}{536}},
  \bibinfo{pages}{441} (\bibinfo{year}{2016}).

\bibitem[{\citenamefont{Hu et~al.}(2019)\citenamefont{Hu, Ma, Cai, Mu, Xu,
  Wang, Wu, Wang, Song, Zou et~al.}}]{Hu2019a}
\bibinfo{author}{\bibfnamefont{L.}~\bibnamefont{Hu}},
  \bibinfo{author}{\bibfnamefont{Y.}~\bibnamefont{Ma}},
  \bibinfo{author}{\bibfnamefont{W.}~\bibnamefont{Cai}},
  \bibinfo{author}{\bibfnamefont{X.}~\bibnamefont{Mu}},
  \bibinfo{author}{\bibfnamefont{Y.}~\bibnamefont{Xu}},
  \bibinfo{author}{\bibfnamefont{W.}~\bibnamefont{Wang}},
  \bibinfo{author}{\bibfnamefont{Y.}~\bibnamefont{Wu}},
  \bibinfo{author}{\bibfnamefont{H.}~\bibnamefont{Wang}},
  \bibinfo{author}{\bibfnamefont{Y.~P.} \bibnamefont{Song}},
  \bibinfo{author}{\bibfnamefont{C.-L.} \bibnamefont{Zou}},
  \bibnamefont{et~al.}, \emph{\bibinfo{title}{{Quantum error correction and
  universal gate set operation on a binomial bosonic logical qubit}}},
  \bibinfo{journal}{Nature Physics} \textbf{\bibinfo{volume}{15}},
  \bibinfo{pages}{503} (\bibinfo{year}{2019}).

\bibitem[{\citenamefont{Campagne-Ibarcq
  et~al.}(2020)\citenamefont{Campagne-Ibarcq, Eickbusch, Touzard, Zalys-Geller,
  Frattini, Sivak, Reinhold, Puri, Shankar, Schoelkopf
  et~al.}}]{Campagne-Ibarcq2020}
\bibinfo{author}{\bibfnamefont{P.}~\bibnamefont{Campagne-Ibarcq}},
  \bibinfo{author}{\bibfnamefont{A.}~\bibnamefont{Eickbusch}},
  \bibinfo{author}{\bibfnamefont{S.}~\bibnamefont{Touzard}},
  \bibinfo{author}{\bibfnamefont{E.}~\bibnamefont{Zalys-Geller}},
  \bibinfo{author}{\bibfnamefont{N.~E.} \bibnamefont{Frattini}},
  \bibinfo{author}{\bibfnamefont{V.~V.} \bibnamefont{Sivak}},
  \bibinfo{author}{\bibfnamefont{P.}~\bibnamefont{Reinhold}},
  \bibinfo{author}{\bibfnamefont{S.}~\bibnamefont{Puri}},
  \bibinfo{author}{\bibfnamefont{S.}~\bibnamefont{Shankar}},
  \bibinfo{author}{\bibfnamefont{R.~J.} \bibnamefont{Schoelkopf}},
  \bibnamefont{et~al.}, \emph{\bibinfo{title}{{Quantum error correction of a
  qubit encoded in grid states of an oscillator}}}, \bibinfo{journal}{Nature}
  \textbf{\bibinfo{volume}{584}}, \bibinfo{pages}{368} (\bibinfo{year}{2020}).

\bibitem[{\citenamefont{Gertler et~al.}(2021)\citenamefont{Gertler, Baker, Li,
  Shirol, Koch, and Wang}}]{Gertler2020}
\bibinfo{author}{\bibfnamefont{J.~M.} \bibnamefont{Gertler}},
  \bibinfo{author}{\bibfnamefont{B.}~\bibnamefont{Baker}},
  \bibinfo{author}{\bibfnamefont{J.}~\bibnamefont{Li}},
  \bibinfo{author}{\bibfnamefont{S.}~\bibnamefont{Shirol}},
  \bibinfo{author}{\bibfnamefont{J.}~\bibnamefont{Koch}}, \bibnamefont{and}
  \bibinfo{author}{\bibfnamefont{C.}~\bibnamefont{Wang}},
  \emph{\bibinfo{title}{{Protecting a bosonic qubit with autonomous quantum
  error correction}}}, \bibinfo{journal}{Nature}
  \textbf{\bibinfo{volume}{590}}, \bibinfo{pages}{243} (\bibinfo{year}{2021}).

\bibitem[{\citenamefont{Krinner et~al.}(2022)\citenamefont{Krinner, Lacroix,
  Remm, Di~Paolo, Genois, Leroux, Hellings, Lazar, Swiadek, Herrmann
  et~al.}}]{Krinner2021}
\bibinfo{author}{\bibfnamefont{S.}~\bibnamefont{Krinner}},
  \bibinfo{author}{\bibfnamefont{N.}~\bibnamefont{Lacroix}},
  \bibinfo{author}{\bibfnamefont{A.}~\bibnamefont{Remm}},
  \bibinfo{author}{\bibfnamefont{A.}~\bibnamefont{Di~Paolo}},
  \bibinfo{author}{\bibfnamefont{E.}~\bibnamefont{Genois}},
  \bibinfo{author}{\bibfnamefont{C.}~\bibnamefont{Leroux}},
  \bibinfo{author}{\bibfnamefont{C.}~\bibnamefont{Hellings}},
  \bibinfo{author}{\bibfnamefont{S.}~\bibnamefont{Lazar}},
  \bibinfo{author}{\bibfnamefont{F.}~\bibnamefont{Swiadek}},
  \bibinfo{author}{\bibfnamefont{J.}~\bibnamefont{Herrmann}},
  \bibnamefont{et~al.}, \emph{\bibinfo{title}{Realizing repeated quantum error
  correction in a distance-three surface code}}, \bibinfo{journal}{Nature}
  \textbf{\bibinfo{volume}{605}}, \bibinfo{pages}{669} (\bibinfo{year}{2022}).

\bibitem[{\citenamefont{Zhao et~al.}(2022)\citenamefont{Zhao, Ye, Huang, Zhang,
  Wu, Guan, Zhu, Wei, He, Cao et~al.}}]{Zhao2021}
\bibinfo{author}{\bibfnamefont{Y.}~\bibnamefont{Zhao}},
  \bibinfo{author}{\bibfnamefont{Y.}~\bibnamefont{Ye}},
  \bibinfo{author}{\bibfnamefont{H.-L.} \bibnamefont{Huang}},
  \bibinfo{author}{\bibfnamefont{Y.}~\bibnamefont{Zhang}},
  \bibinfo{author}{\bibfnamefont{D.}~\bibnamefont{Wu}},
  \bibinfo{author}{\bibfnamefont{H.}~\bibnamefont{Guan}},
  \bibinfo{author}{\bibfnamefont{Q.}~\bibnamefont{Zhu}},
  \bibinfo{author}{\bibfnamefont{Z.}~\bibnamefont{Wei}},
  \bibinfo{author}{\bibfnamefont{T.}~\bibnamefont{He}},
  \bibinfo{author}{\bibfnamefont{S.}~\bibnamefont{Cao}}, \bibnamefont{et~al.},
  \emph{\bibinfo{title}{Realization of an error-correcting surface code with
  superconducting qubits}}, \bibinfo{journal}{Physical Review Letters}
  \textbf{\bibinfo{volume}{129}}, \bibinfo{pages}{030501}
  (\bibinfo{year}{2022}).

\bibitem[{\citenamefont{Sundaresan et~al.}(2022)\citenamefont{Sundaresan,
  Yoder, Kim, Li, Chen, Harper, Thorbeck, Cross, C{\'o}rcoles, and
  Takita}}]{Sundaresan2022}
\bibinfo{author}{\bibfnamefont{N.}~\bibnamefont{Sundaresan}},
  \bibinfo{author}{\bibfnamefont{T.~J.} \bibnamefont{Yoder}},
  \bibinfo{author}{\bibfnamefont{Y.}~\bibnamefont{Kim}},
  \bibinfo{author}{\bibfnamefont{M.}~\bibnamefont{Li}},
  \bibinfo{author}{\bibfnamefont{E.~H.} \bibnamefont{Chen}},
  \bibinfo{author}{\bibfnamefont{G.}~\bibnamefont{Harper}},
  \bibinfo{author}{\bibfnamefont{T.}~\bibnamefont{Thorbeck}},
  \bibinfo{author}{\bibfnamefont{A.~W.} \bibnamefont{Cross}},
  \bibinfo{author}{\bibfnamefont{A.~D.} \bibnamefont{C{\'o}rcoles}},
  \bibnamefont{and} \bibinfo{author}{\bibfnamefont{M.}~\bibnamefont{Takita}},
  \emph{\bibinfo{title}{Matching and maximum likelihood decoding of a
  multi-round subsystem quantum error correction experiment}},
  \bibinfo{journal}{arXiv:2203.07205}  (\bibinfo{year}{2022}).

\bibitem[{\citenamefont{Acharya et~al.}(2022)\citenamefont{Acharya, Aleiner,
  Allen, Andersen, Ansmann, Arute, Arya, Asfaw, Atalaya, Babbush
  et~al.}}]{Acharya2022}
\bibinfo{author}{\bibfnamefont{R.}~\bibnamefont{Acharya}},
  \bibinfo{author}{\bibfnamefont{I.}~\bibnamefont{Aleiner}},
  \bibinfo{author}{\bibfnamefont{R.}~\bibnamefont{Allen}},
  \bibinfo{author}{\bibfnamefont{T.~I.} \bibnamefont{Andersen}},
  \bibinfo{author}{\bibfnamefont{M.}~\bibnamefont{Ansmann}},
  \bibinfo{author}{\bibfnamefont{F.}~\bibnamefont{Arute}},
  \bibinfo{author}{\bibfnamefont{K.}~\bibnamefont{Arya}},
  \bibinfo{author}{\bibfnamefont{A.}~\bibnamefont{Asfaw}},
  \bibinfo{author}{\bibfnamefont{J.}~\bibnamefont{Atalaya}},
  \bibinfo{author}{\bibfnamefont{R.}~\bibnamefont{Babbush}},
  \bibnamefont{et~al.}, \emph{\bibinfo{title}{{Suppressing quantum errors by
  scaling a surface code logical qubit}}}, \bibinfo{journal}{arXiv:2207.06431}
  (\bibinfo{year}{2022}).

\bibitem[{\citenamefont{de~Neeve et~al.}(2022)\citenamefont{de~Neeve, Nguyen,
  Behrle, and Home}}]{DeNeeve2020}
\bibinfo{author}{\bibfnamefont{B.}~\bibnamefont{de~Neeve}},
  \bibinfo{author}{\bibfnamefont{T.-L.} \bibnamefont{Nguyen}},
  \bibinfo{author}{\bibfnamefont{T.}~\bibnamefont{Behrle}}, \bibnamefont{and}
  \bibinfo{author}{\bibfnamefont{J.~P.} \bibnamefont{Home}},
  \emph{\bibinfo{title}{Error correction of a logical grid state qubit by
  dissipative pumping}}, \bibinfo{journal}{Nature Physics}
  \textbf{\bibinfo{volume}{18}}, \bibinfo{pages}{296} (\bibinfo{year}{2022}).

\bibitem[{\citenamefont{Ryan-Anderson et~al.}(2021)\citenamefont{Ryan-Anderson,
  Bohnet, Lee, Gresh, Hankin, Gaebler, Francois, Chernoguzov, Lucchetti, Brown
  et~al.}}]{Ryan2021}
\bibinfo{author}{\bibfnamefont{C.}~\bibnamefont{Ryan-Anderson}},
  \bibinfo{author}{\bibfnamefont{J.}~\bibnamefont{Bohnet}},
  \bibinfo{author}{\bibfnamefont{K.}~\bibnamefont{Lee}},
  \bibinfo{author}{\bibfnamefont{D.}~\bibnamefont{Gresh}},
  \bibinfo{author}{\bibfnamefont{A.}~\bibnamefont{Hankin}},
  \bibinfo{author}{\bibfnamefont{J.}~\bibnamefont{Gaebler}},
  \bibinfo{author}{\bibfnamefont{D.}~\bibnamefont{Francois}},
  \bibinfo{author}{\bibfnamefont{A.}~\bibnamefont{Chernoguzov}},
  \bibinfo{author}{\bibfnamefont{D.}~\bibnamefont{Lucchetti}},
  \bibinfo{author}{\bibfnamefont{N.}~\bibnamefont{Brown}},
  \bibnamefont{et~al.}, \emph{\bibinfo{title}{Realization of real-time
  fault-tolerant quantum error correction}}, \bibinfo{journal}{Physical Review
  X} \textbf{\bibinfo{volume}{11}}, \bibinfo{pages}{041058}
  (\bibinfo{year}{2021}).

\bibitem[{\citenamefont{Egan et~al.}(2021)\citenamefont{Egan, Debroy, Noel,
  Risinger, Zhu, Biswas, Newman, Li, Brown, Cetina et~al.}}]{Egan2021}
\bibinfo{author}{\bibfnamefont{L.}~\bibnamefont{Egan}},
  \bibinfo{author}{\bibfnamefont{D.~M.} \bibnamefont{Debroy}},
  \bibinfo{author}{\bibfnamefont{C.}~\bibnamefont{Noel}},
  \bibinfo{author}{\bibfnamefont{A.}~\bibnamefont{Risinger}},
  \bibinfo{author}{\bibfnamefont{D.}~\bibnamefont{Zhu}},
  \bibinfo{author}{\bibfnamefont{D.}~\bibnamefont{Biswas}},
  \bibinfo{author}{\bibfnamefont{M.}~\bibnamefont{Newman}},
  \bibinfo{author}{\bibfnamefont{M.}~\bibnamefont{Li}},
  \bibinfo{author}{\bibfnamefont{K.~R.} \bibnamefont{Brown}},
  \bibinfo{author}{\bibfnamefont{M.}~\bibnamefont{Cetina}},
  \bibnamefont{et~al.}, \emph{\bibinfo{title}{Fault-tolerant control of an
  error-corrected qubit}}, \bibinfo{journal}{Nature}
  \textbf{\bibinfo{volume}{598}}, \bibinfo{pages}{281} (\bibinfo{year}{2021}).

\bibitem[{\citenamefont{Waldherr et~al.}(2014)\citenamefont{Waldherr, Wang,
  Zaiser, Jamali, Schulte-Herbr{\"u}ggen, Abe, Ohshima, Isoya, Du, Neumann
  et~al.}}]{Waldherr2014}
\bibinfo{author}{\bibfnamefont{G.}~\bibnamefont{Waldherr}},
  \bibinfo{author}{\bibfnamefont{Y.}~\bibnamefont{Wang}},
  \bibinfo{author}{\bibfnamefont{S.}~\bibnamefont{Zaiser}},
  \bibinfo{author}{\bibfnamefont{M.}~\bibnamefont{Jamali}},
  \bibinfo{author}{\bibfnamefont{T.}~\bibnamefont{Schulte-Herbr{\"u}ggen}},
  \bibinfo{author}{\bibfnamefont{H.}~\bibnamefont{Abe}},
  \bibinfo{author}{\bibfnamefont{T.}~\bibnamefont{Ohshima}},
  \bibinfo{author}{\bibfnamefont{J.}~\bibnamefont{Isoya}},
  \bibinfo{author}{\bibfnamefont{J.}~\bibnamefont{Du}},
  \bibinfo{author}{\bibfnamefont{P.}~\bibnamefont{Neumann}},
  \bibnamefont{et~al.}, \emph{\bibinfo{title}{Quantum error correction in a
  solid-state hybrid spin register}}, \bibinfo{journal}{Nature}
  \textbf{\bibinfo{volume}{506}}, \bibinfo{pages}{204} (\bibinfo{year}{2014}).

\bibitem[{\citenamefont{Abobeih et~al.}(2022)\citenamefont{Abobeih, Wang,
  Randall, Loenen, Bradley, Markham, Twitchen, Terhal, and
  Taminiau}}]{Abobeih2022}
\bibinfo{author}{\bibfnamefont{M.}~\bibnamefont{Abobeih}},
  \bibinfo{author}{\bibfnamefont{Y.}~\bibnamefont{Wang}},
  \bibinfo{author}{\bibfnamefont{J.}~\bibnamefont{Randall}},
  \bibinfo{author}{\bibfnamefont{S.}~\bibnamefont{Loenen}},
  \bibinfo{author}{\bibfnamefont{C.}~\bibnamefont{Bradley}},
  \bibinfo{author}{\bibfnamefont{M.}~\bibnamefont{Markham}},
  \bibinfo{author}{\bibfnamefont{D.}~\bibnamefont{Twitchen}},
  \bibinfo{author}{\bibfnamefont{B.}~\bibnamefont{Terhal}}, \bibnamefont{and}
  \bibinfo{author}{\bibfnamefont{T.}~\bibnamefont{Taminiau}},
  \emph{\bibinfo{title}{Fault-tolerant operation of a logical qubit in a
  diamond quantum processor}}, \bibinfo{journal}{Nature}
  \textbf{\bibinfo{volume}{606}}, \bibinfo{pages}{884} (\bibinfo{year}{2022}).

\bibitem[{\citenamefont{Xue et~al.}(2022)\citenamefont{Xue, Russ, Samkharadze,
  Undseth, Sammak, Scappucci, and Vandersypen}}]{Xue2022}
\bibinfo{author}{\bibfnamefont{X.}~\bibnamefont{Xue}},
  \bibinfo{author}{\bibfnamefont{M.}~\bibnamefont{Russ}},
  \bibinfo{author}{\bibfnamefont{N.}~\bibnamefont{Samkharadze}},
  \bibinfo{author}{\bibfnamefont{B.}~\bibnamefont{Undseth}},
  \bibinfo{author}{\bibfnamefont{A.}~\bibnamefont{Sammak}},
  \bibinfo{author}{\bibfnamefont{G.}~\bibnamefont{Scappucci}},
  \bibnamefont{and} \bibinfo{author}{\bibfnamefont{L.~M.}
  \bibnamefont{Vandersypen}}, \emph{\bibinfo{title}{Quantum logic with spin
  qubits crossing the surface code threshold}}, \bibinfo{journal}{Nature}
  \textbf{\bibinfo{volume}{601}}, \bibinfo{pages}{343} (\bibinfo{year}{2022}).

\bibitem[{\citenamefont{Place et~al.}(2021)\citenamefont{Place, Rodgers,
  Mundada, Smitham, Fitzpatrick, Leng, Premkumar, Bryon, Vrajitoarea, Sussman
  et~al.}}]{Place2021}
\bibinfo{author}{\bibfnamefont{A.~P.} \bibnamefont{Place}},
  \bibinfo{author}{\bibfnamefont{L.~V.} \bibnamefont{Rodgers}},
  \bibinfo{author}{\bibfnamefont{P.}~\bibnamefont{Mundada}},
  \bibinfo{author}{\bibfnamefont{B.~M.} \bibnamefont{Smitham}},
  \bibinfo{author}{\bibfnamefont{M.}~\bibnamefont{Fitzpatrick}},
  \bibinfo{author}{\bibfnamefont{Z.}~\bibnamefont{Leng}},
  \bibinfo{author}{\bibfnamefont{A.}~\bibnamefont{Premkumar}},
  \bibinfo{author}{\bibfnamefont{J.}~\bibnamefont{Bryon}},
  \bibinfo{author}{\bibfnamefont{A.}~\bibnamefont{Vrajitoarea}},
  \bibinfo{author}{\bibfnamefont{S.}~\bibnamefont{Sussman}},
  \bibnamefont{et~al.}, \emph{\bibinfo{title}{New material platform for
  superconducting transmon qubits with coherence times exceeding 0.3
  milliseconds}}, \bibinfo{journal}{Nature communications}
  \textbf{\bibinfo{volume}{12}}, \bibinfo{pages}{1} (\bibinfo{year}{2021}).

\bibitem[{\citenamefont{Reagor et~al.}(2016)\citenamefont{Reagor, Pfaff,
  Axline, Heeres, Ofek, Sliwa, Holland, Wang, Blumoff, Chou
  et~al.}}]{Reagor2016}
\bibinfo{author}{\bibfnamefont{M.}~\bibnamefont{Reagor}},
  \bibinfo{author}{\bibfnamefont{W.}~\bibnamefont{Pfaff}},
  \bibinfo{author}{\bibfnamefont{C.}~\bibnamefont{Axline}},
  \bibinfo{author}{\bibfnamefont{R.~W.} \bibnamefont{Heeres}},
  \bibinfo{author}{\bibfnamefont{N.}~\bibnamefont{Ofek}},
  \bibinfo{author}{\bibfnamefont{K.}~\bibnamefont{Sliwa}},
  \bibinfo{author}{\bibfnamefont{E.}~\bibnamefont{Holland}},
  \bibinfo{author}{\bibfnamefont{C.}~\bibnamefont{Wang}},
  \bibinfo{author}{\bibfnamefont{J.}~\bibnamefont{Blumoff}},
  \bibinfo{author}{\bibfnamefont{K.}~\bibnamefont{Chou}}, \bibnamefont{et~al.},
  \emph{\bibinfo{title}{{Quantum memory with millisecond coherence in circuit
  QED}}}, \bibinfo{journal}{Physical Review B} \textbf{\bibinfo{volume}{94}},
  \bibinfo{pages}{014506} (\bibinfo{year}{2016}).

\bibitem[{\citenamefont{Royer et~al.}(2020)\citenamefont{Royer, Singh, and
  Girvin}}]{Royer2020}
\bibinfo{author}{\bibfnamefont{B.}~\bibnamefont{Royer}},
  \bibinfo{author}{\bibfnamefont{S.}~\bibnamefont{Singh}}, \bibnamefont{and}
  \bibinfo{author}{\bibfnamefont{S.~M.} \bibnamefont{Girvin}},
  \emph{\bibinfo{title}{{Stabilization of finite-energy
  Gottesman-Kitaev-Preskill states}}}, \bibinfo{journal}{Physical Review
  Letters} \textbf{\bibinfo{volume}{125}}, \bibinfo{pages}{260509}
  (\bibinfo{year}{2020}).

\bibitem[{\citenamefont{Sutton and Barto}(2018)}]{Sutton2017}
\bibinfo{author}{\bibfnamefont{R.~S.} \bibnamefont{Sutton}} \bibnamefont{and}
  \bibinfo{author}{\bibfnamefont{A.~G.} \bibnamefont{Barto}},
  \emph{\bibinfo{title}{{Reinforcement learning: an introduction}}}
  (\bibinfo{publisher}{A Bradford Book}, \bibinfo{year}{2018}).

\bibitem[{\citenamefont{Schulman et~al.}(2017)\citenamefont{Schulman, Wolski,
  Dhariwal, Radford, and Klimov}}]{Schulman2017}
\bibinfo{author}{\bibfnamefont{J.}~\bibnamefont{Schulman}},
  \bibinfo{author}{\bibfnamefont{F.}~\bibnamefont{Wolski}},
  \bibinfo{author}{\bibfnamefont{P.}~\bibnamefont{Dhariwal}},
  \bibinfo{author}{\bibfnamefont{A.}~\bibnamefont{Radford}}, \bibnamefont{and}
  \bibinfo{author}{\bibfnamefont{O.}~\bibnamefont{Klimov}},
  \emph{\bibinfo{title}{{Proximal policy optimization algorithms}}},
  \bibinfo{journal}{arXiv:1707.06347}  (\bibinfo{year}{2017}).

\bibitem[{\citenamefont{Guadarrama et~al.}(2018)\citenamefont{Guadarrama,
  Korattikara, Ramirez, Castro, Holly, Fishman, Wang, Gonina, Wu, Kokiopoulou
  et~al.}}]{TFAgents}
\bibinfo{author}{\bibfnamefont{S.}~\bibnamefont{Guadarrama}},
  \bibinfo{author}{\bibfnamefont{A.}~\bibnamefont{Korattikara}},
  \bibinfo{author}{\bibfnamefont{O.}~\bibnamefont{Ramirez}},
  \bibinfo{author}{\bibfnamefont{P.}~\bibnamefont{Castro}},
  \bibinfo{author}{\bibfnamefont{E.}~\bibnamefont{Holly}},
  \bibinfo{author}{\bibfnamefont{S.}~\bibnamefont{Fishman}},
  \bibinfo{author}{\bibfnamefont{K.}~\bibnamefont{Wang}},
  \bibinfo{author}{\bibfnamefont{E.}~\bibnamefont{Gonina}},
  \bibinfo{author}{\bibfnamefont{N.}~\bibnamefont{Wu}},
  \bibinfo{author}{\bibfnamefont{E.}~\bibnamefont{Kokiopoulou}},
  \bibnamefont{et~al.}, \emph{\bibinfo{title}{{TF-Agents}: A library for
  reinforcement learning in tensorflow}},
  \bibinfo{howpublished}{\url{https://github.com/tensorflow/agents}}
  (\bibinfo{year}{2018}).

\bibitem[{\citenamefont{Lescanne et~al.}(2020)\citenamefont{Lescanne, Villiers,
  Peronnin, Sarlette, Delbecq, Huard, Kontos, Mirrahimi, and
  Leghtas}}]{Lescanne2020}
\bibinfo{author}{\bibfnamefont{R.}~\bibnamefont{Lescanne}},
  \bibinfo{author}{\bibfnamefont{M.}~\bibnamefont{Villiers}},
  \bibinfo{author}{\bibfnamefont{T.}~\bibnamefont{Peronnin}},
  \bibinfo{author}{\bibfnamefont{A.}~\bibnamefont{Sarlette}},
  \bibinfo{author}{\bibfnamefont{M.}~\bibnamefont{Delbecq}},
  \bibinfo{author}{\bibfnamefont{B.}~\bibnamefont{Huard}},
  \bibinfo{author}{\bibfnamefont{T.}~\bibnamefont{Kontos}},
  \bibinfo{author}{\bibfnamefont{M.}~\bibnamefont{Mirrahimi}},
  \bibnamefont{and} \bibinfo{author}{\bibfnamefont{Z.}~\bibnamefont{Leghtas}},
  \emph{\bibinfo{title}{{Exponential suppression of bit-flips in a qubit
  encoded in an oscillator}}}, \bibinfo{journal}{Nature Physics}
  \textbf{\bibinfo{volume}{16}}, \bibinfo{pages}{509} (\bibinfo{year}{2020}).

\bibitem[{\citenamefont{Lloyd and Viola}(2001)}]{Lloyd2001}
\bibinfo{author}{\bibfnamefont{S.}~\bibnamefont{Lloyd}} \bibnamefont{and}
  \bibinfo{author}{\bibfnamefont{L.}~\bibnamefont{Viola}},
  \emph{\bibinfo{title}{Engineering quantum dynamics}}, \bibinfo{journal}{Phys.
  Rev. A} \textbf{\bibinfo{volume}{65}}, \bibinfo{pages}{010101}
  (\bibinfo{year}{2001}).

\bibitem[{\citenamefont{Shen et~al.}(2017)\citenamefont{Shen, Noh, Albert,
  Krastanov, Devoret, Schoelkopf, Girvin, and Jiang}}]{Shen2017}
\bibinfo{author}{\bibfnamefont{C.}~\bibnamefont{Shen}},
  \bibinfo{author}{\bibfnamefont{K.}~\bibnamefont{Noh}},
  \bibinfo{author}{\bibfnamefont{V.~V.} \bibnamefont{Albert}},
  \bibinfo{author}{\bibfnamefont{S.}~\bibnamefont{Krastanov}},
  \bibinfo{author}{\bibfnamefont{M.~H.} \bibnamefont{Devoret}},
  \bibinfo{author}{\bibfnamefont{R.~J.} \bibnamefont{Schoelkopf}},
  \bibinfo{author}{\bibfnamefont{S.~M.} \bibnamefont{Girvin}},
  \bibnamefont{and} \bibinfo{author}{\bibfnamefont{L.}~\bibnamefont{Jiang}},
  \emph{\bibinfo{title}{{Quantum channel construction with circuit quantum
  electrodynamics}}}, \bibinfo{journal}{Physical Review B}
  \textbf{\bibinfo{volume}{95}}, \bibinfo{pages}{134501}
  (\bibinfo{year}{2017}).

\bibitem[{\citenamefont{Chen et~al.}(2016)\citenamefont{Chen, Kelly, Quintana,
  Barends, Campbell, Chen, Chiaro, Dunsworth, Fowler, Lucero
  et~al.}}]{Chen2016}
\bibinfo{author}{\bibfnamefont{Z.}~\bibnamefont{Chen}},
  \bibinfo{author}{\bibfnamefont{J.}~\bibnamefont{Kelly}},
  \bibinfo{author}{\bibfnamefont{C.}~\bibnamefont{Quintana}},
  \bibinfo{author}{\bibfnamefont{R.}~\bibnamefont{Barends}},
  \bibinfo{author}{\bibfnamefont{B.}~\bibnamefont{Campbell}},
  \bibinfo{author}{\bibfnamefont{Y.}~\bibnamefont{Chen}},
  \bibinfo{author}{\bibfnamefont{B.}~\bibnamefont{Chiaro}},
  \bibinfo{author}{\bibfnamefont{A.}~\bibnamefont{Dunsworth}},
  \bibinfo{author}{\bibfnamefont{A.~G.} \bibnamefont{Fowler}},
  \bibinfo{author}{\bibfnamefont{E.}~\bibnamefont{Lucero}},
  \bibnamefont{et~al.}, \emph{\bibinfo{title}{Measuring and suppressing quantum
  state leakage in a superconducting qubit}}, \bibinfo{journal}{Physical Review
  Letters} \textbf{\bibinfo{volume}{116}}, \bibinfo{pages}{020501}
  (\bibinfo{year}{2016}).

\bibitem[{\citenamefont{Eickbusch et~al.}(2021)\citenamefont{Eickbusch, Sivak,
  Ding, Elder, Jha, Venkatraman, Royer, Girvin, Schoelkopf, and
  Devoret}}]{Eickbusch2021}
\bibinfo{author}{\bibfnamefont{A.}~\bibnamefont{Eickbusch}},
  \bibinfo{author}{\bibfnamefont{V.}~\bibnamefont{Sivak}},
  \bibinfo{author}{\bibfnamefont{A.~Z.} \bibnamefont{Ding}},
  \bibinfo{author}{\bibfnamefont{S.~S.} \bibnamefont{Elder}},
  \bibinfo{author}{\bibfnamefont{S.~R.} \bibnamefont{Jha}},
  \bibinfo{author}{\bibfnamefont{J.}~\bibnamefont{Venkatraman}},
  \bibinfo{author}{\bibfnamefont{B.}~\bibnamefont{Royer}},
  \bibinfo{author}{\bibfnamefont{S.~M.} \bibnamefont{Girvin}},
  \bibinfo{author}{\bibfnamefont{R.~J.} \bibnamefont{Schoelkopf}},
  \bibnamefont{and} \bibinfo{author}{\bibfnamefont{M.~H.}
  \bibnamefont{Devoret}}, \emph{\bibinfo{title}{Fast universal control of an
  oscillator with weak dispersive coupling to a qubit}},
  \bibinfo{journal}{arXiv:2111.06414}  (\bibinfo{year}{2021}).

\bibitem[{\citenamefont{Kelly et~al.}(2014)\citenamefont{Kelly, Barends,
  Campbell, Chen, Chen, Chiaro, Dunsworth, Fowler, Hoi, Jeffrey
  et~al.}}]{Kelly2014}
\bibinfo{author}{\bibfnamefont{J.}~\bibnamefont{Kelly}},
  \bibinfo{author}{\bibfnamefont{R.}~\bibnamefont{Barends}},
  \bibinfo{author}{\bibfnamefont{B.}~\bibnamefont{Campbell}},
  \bibinfo{author}{\bibfnamefont{Y.}~\bibnamefont{Chen}},
  \bibinfo{author}{\bibfnamefont{Z.}~\bibnamefont{Chen}},
  \bibinfo{author}{\bibfnamefont{B.}~\bibnamefont{Chiaro}},
  \bibinfo{author}{\bibfnamefont{A.}~\bibnamefont{Dunsworth}},
  \bibinfo{author}{\bibfnamefont{A.~G.} \bibnamefont{Fowler}},
  \bibinfo{author}{\bibfnamefont{I.-C.} \bibnamefont{Hoi}},
  \bibinfo{author}{\bibfnamefont{E.}~\bibnamefont{Jeffrey}},
  \bibnamefont{et~al.}, \emph{\bibinfo{title}{Optimal quantum control using
  randomized benchmarking}}, \bibinfo{journal}{Physical Review Letters}
  \textbf{\bibinfo{volume}{112}}, \bibinfo{pages}{240504}
  (\bibinfo{year}{2014}).

\bibitem[{\citenamefont{Rol et~al.}(2017)\citenamefont{Rol, Bultink, O'Brien,
  de~Jong, Theis, Fu, Luthi, Vermeulen, de~Sterke, Bruno et~al.}}]{Rol2017}
\bibinfo{author}{\bibfnamefont{M.~A.} \bibnamefont{Rol}},
  \bibinfo{author}{\bibfnamefont{C.~C.} \bibnamefont{Bultink}},
  \bibinfo{author}{\bibfnamefont{T.~E.} \bibnamefont{O'Brien}},
  \bibinfo{author}{\bibfnamefont{S.~R.} \bibnamefont{de~Jong}},
  \bibinfo{author}{\bibfnamefont{L.~S.} \bibnamefont{Theis}},
  \bibinfo{author}{\bibfnamefont{X.}~\bibnamefont{Fu}},
  \bibinfo{author}{\bibfnamefont{F.}~\bibnamefont{Luthi}},
  \bibinfo{author}{\bibfnamefont{R.~F.~L.} \bibnamefont{Vermeulen}},
  \bibinfo{author}{\bibfnamefont{J.~C.} \bibnamefont{de~Sterke}},
  \bibinfo{author}{\bibfnamefont{A.}~\bibnamefont{Bruno}},
  \bibnamefont{et~al.}, \emph{\bibinfo{title}{Restless tuneup of high-fidelity
  qubit gates}}, \bibinfo{journal}{Physical Review Applied}
  \textbf{\bibinfo{volume}{7}}, \bibinfo{pages}{041001} (\bibinfo{year}{2017}).

\bibitem[{\citenamefont{Werninghaus et~al.}(2021)\citenamefont{Werninghaus,
  Egger, Roy, Machnes, Wilhelm, and Filipp}}]{Werninghaus2021}
\bibinfo{author}{\bibfnamefont{M.}~\bibnamefont{Werninghaus}},
  \bibinfo{author}{\bibfnamefont{D.~J.} \bibnamefont{Egger}},
  \bibinfo{author}{\bibfnamefont{F.}~\bibnamefont{Roy}},
  \bibinfo{author}{\bibfnamefont{S.}~\bibnamefont{Machnes}},
  \bibinfo{author}{\bibfnamefont{F.~K.} \bibnamefont{Wilhelm}},
  \bibnamefont{and} \bibinfo{author}{\bibfnamefont{S.}~\bibnamefont{Filipp}},
  \emph{\bibinfo{title}{{Leakage reduction in fast superconducting qubit gates
  via optimal control}}}, \bibinfo{journal}{npj Quantum Information}
  \textbf{\bibinfo{volume}{7}}, \bibinfo{pages}{14} (\bibinfo{year}{2021}).

\bibitem[{\citenamefont{Baum et~al.}(2021)\citenamefont{Baum, Amico, Howell,
  Hush, Liuzzi, Mundada, Merkh, Carvalho, and Biercuk}}]{Baum2021}
\bibinfo{author}{\bibfnamefont{Y.}~\bibnamefont{Baum}},
  \bibinfo{author}{\bibfnamefont{M.}~\bibnamefont{Amico}},
  \bibinfo{author}{\bibfnamefont{S.}~\bibnamefont{Howell}},
  \bibinfo{author}{\bibfnamefont{M.}~\bibnamefont{Hush}},
  \bibinfo{author}{\bibfnamefont{M.}~\bibnamefont{Liuzzi}},
  \bibinfo{author}{\bibfnamefont{P.}~\bibnamefont{Mundada}},
  \bibinfo{author}{\bibfnamefont{T.}~\bibnamefont{Merkh}},
  \bibinfo{author}{\bibfnamefont{A.~R.} \bibnamefont{Carvalho}},
  \bibnamefont{and} \bibinfo{author}{\bibfnamefont{M.~J.}
  \bibnamefont{Biercuk}}, \emph{\bibinfo{title}{Experimental deep reinforcement
  learning for error-robust gate-set design on a superconducting quantum
  computer}}, \bibinfo{journal}{PRX Quantum} \textbf{\bibinfo{volume}{2}},
  \bibinfo{pages}{040324} (\bibinfo{year}{2021}).

\bibitem[{\citenamefont{Sivak et~al.}(2022)\citenamefont{Sivak, Eickbusch, Liu,
  Royer, Tsioutsios, and Devoret}}]{Sivak2021}
\bibinfo{author}{\bibfnamefont{V.~V.} \bibnamefont{Sivak}},
  \bibinfo{author}{\bibfnamefont{A.}~\bibnamefont{Eickbusch}},
  \bibinfo{author}{\bibfnamefont{H.}~\bibnamefont{Liu}},
  \bibinfo{author}{\bibfnamefont{B.}~\bibnamefont{Royer}},
  \bibinfo{author}{\bibfnamefont{I.}~\bibnamefont{Tsioutsios}},
  \bibnamefont{and} \bibinfo{author}{\bibfnamefont{M.~H.}
  \bibnamefont{Devoret}}, \emph{\bibinfo{title}{Model-free quantum control with
  reinforcement learning}}, \bibinfo{journal}{Physical Review X}
  \textbf{\bibinfo{volume}{12}}, \bibinfo{pages}{011059}
  (\bibinfo{year}{2022}).

\bibitem[{\citenamefont{Terhal and Weigand}(2016)}]{Terhal2016}
\bibinfo{author}{\bibfnamefont{B.~M.} \bibnamefont{Terhal}} \bibnamefont{and}
  \bibinfo{author}{\bibfnamefont{D.}~\bibnamefont{Weigand}},
  \emph{\bibinfo{title}{{Encoding a qubit into a cavity mode in circuit QED
  using phase estimation}}}, \bibinfo{journal}{Physical Review A}
  \textbf{\bibinfo{volume}{93}}, \bibinfo{pages}{012315}
  (\bibinfo{year}{2016}).

\bibitem[{\citenamefont{Hastrup and Andersen}(2021)}]{Hastrup2021}
\bibinfo{author}{\bibfnamefont{J.}~\bibnamefont{Hastrup}} \bibnamefont{and}
  \bibinfo{author}{\bibfnamefont{U.~L.} \bibnamefont{Andersen}},
  \emph{\bibinfo{title}{{Improved readout of qubit-coupled
  Gottesman-Kitaev-Preskill states}}}, \bibinfo{journal}{Quantum Science and
  Technology} \textbf{\bibinfo{volume}{6}}, \bibinfo{pages}{035016}
  (\bibinfo{year}{2021}).

\bibitem[{\citenamefont{Klimov et~al.}(2018)\citenamefont{Klimov, Kelly, Chen,
  Neeley, Megrant, Burkett, Barends, Arya, Chiaro, Chen et~al.}}]{Klimov2018}
\bibinfo{author}{\bibfnamefont{P.~V.} \bibnamefont{Klimov}},
  \bibinfo{author}{\bibfnamefont{J.}~\bibnamefont{Kelly}},
  \bibinfo{author}{\bibfnamefont{Z.}~\bibnamefont{Chen}},
  \bibinfo{author}{\bibfnamefont{M.}~\bibnamefont{Neeley}},
  \bibinfo{author}{\bibfnamefont{A.}~\bibnamefont{Megrant}},
  \bibinfo{author}{\bibfnamefont{B.}~\bibnamefont{Burkett}},
  \bibinfo{author}{\bibfnamefont{R.}~\bibnamefont{Barends}},
  \bibinfo{author}{\bibfnamefont{K.}~\bibnamefont{Arya}},
  \bibinfo{author}{\bibfnamefont{B.}~\bibnamefont{Chiaro}},
  \bibinfo{author}{\bibfnamefont{Y.}~\bibnamefont{Chen}}, \bibnamefont{et~al.},
  \emph{\bibinfo{title}{Fluctuations of energy-relaxation times in
  superconducting qubits}}, \bibinfo{journal}{Physical Review Letters}
  \textbf{\bibinfo{volume}{121}}, \bibinfo{pages}{90502}
  (\bibinfo{year}{2018}).

\bibitem[{\citenamefont{Lisenfeld et~al.}(2019)\citenamefont{Lisenfeld, Bilmes,
  Megrant, Barends, Kelly, Klimov, Weiss, Martinis, and
  Ustinov}}]{Lisenfeld2019}
\bibinfo{author}{\bibfnamefont{J.}~\bibnamefont{Lisenfeld}},
  \bibinfo{author}{\bibfnamefont{A.}~\bibnamefont{Bilmes}},
  \bibinfo{author}{\bibfnamefont{A.}~\bibnamefont{Megrant}},
  \bibinfo{author}{\bibfnamefont{R.}~\bibnamefont{Barends}},
  \bibinfo{author}{\bibfnamefont{J.}~\bibnamefont{Kelly}},
  \bibinfo{author}{\bibfnamefont{P.}~\bibnamefont{Klimov}},
  \bibinfo{author}{\bibfnamefont{G.}~\bibnamefont{Weiss}},
  \bibinfo{author}{\bibfnamefont{J.~M.} \bibnamefont{Martinis}},
  \bibnamefont{and} \bibinfo{author}{\bibfnamefont{A.~V.}
  \bibnamefont{Ustinov}}, \emph{\bibinfo{title}{Electric field spectroscopy of
  material defects in transmon qubits}}, \bibinfo{journal}{npj Quantum
  Information} \textbf{\bibinfo{volume}{5}}, \bibinfo{pages}{1}
  (\bibinfo{year}{2019}).

\bibitem[{\citenamefont{Nielsen}(2002)}]{Nielsen2002}
\bibinfo{author}{\bibfnamefont{M.~A.} \bibnamefont{Nielsen}},
  \emph{\bibinfo{title}{{A simple formula for the average gate fidelity of a
  quantum dynamical operation}}}, \bibinfo{journal}{Physics Letters A}
  \textbf{\bibinfo{volume}{303}}, \bibinfo{pages}{249} (\bibinfo{year}{2002}).

\bibitem[{\citenamefont{Bravyi and Kitaev}(2005)}]{Bravyi2005}
\bibinfo{author}{\bibfnamefont{S.}~\bibnamefont{Bravyi}} \bibnamefont{and}
  \bibinfo{author}{\bibfnamefont{A.}~\bibnamefont{Kitaev}},
  \emph{\bibinfo{title}{{Universal quantum computation with ideal Clifford
  gates and noisy ancillas}}}, \bibinfo{journal}{Physical Review A}
  \textbf{\bibinfo{volume}{71}}, \bibinfo{pages}{022316}
  (\bibinfo{year}{2005}).

\bibitem[{\citenamefont{Grimm et~al.}(2020)\citenamefont{Grimm, Frattini, Puri,
  Mundhada, Touzard, Mirrahimi, Girvin, Shankar, and Devoret}}]{Grimm2019}
\bibinfo{author}{\bibfnamefont{A.}~\bibnamefont{Grimm}},
  \bibinfo{author}{\bibfnamefont{N.~E.} \bibnamefont{Frattini}},
  \bibinfo{author}{\bibfnamefont{S.}~\bibnamefont{Puri}},
  \bibinfo{author}{\bibfnamefont{S.~O.} \bibnamefont{Mundhada}},
  \bibinfo{author}{\bibfnamefont{S.}~\bibnamefont{Touzard}},
  \bibinfo{author}{\bibfnamefont{M.}~\bibnamefont{Mirrahimi}},
  \bibinfo{author}{\bibfnamefont{S.~M.} \bibnamefont{Girvin}},
  \bibinfo{author}{\bibfnamefont{S.}~\bibnamefont{Shankar}}, \bibnamefont{and}
  \bibinfo{author}{\bibfnamefont{M.~H.} \bibnamefont{Devoret}},
  \emph{\bibinfo{title}{{Stabilization and operation of a Kerr-cat qubit}}},
  \bibinfo{journal}{Nature} \textbf{\bibinfo{volume}{584}},
  \bibinfo{pages}{205} (\bibinfo{year}{2020}).

\bibitem[{\citenamefont{Chen et~al.}(2021)\citenamefont{Chen, Satzinger,
  Atalaya, Korotkov, Dunsworth, Sank, Quintana, McEwen, Barends, Klimov
  et~al.}}]{Chen2021}
\bibinfo{author}{\bibfnamefont{Z.}~\bibnamefont{Chen}},
  \bibinfo{author}{\bibfnamefont{K.~J.} \bibnamefont{Satzinger}},
  \bibinfo{author}{\bibfnamefont{J.}~\bibnamefont{Atalaya}},
  \bibinfo{author}{\bibfnamefont{A.~N.} \bibnamefont{Korotkov}},
  \bibinfo{author}{\bibfnamefont{A.}~\bibnamefont{Dunsworth}},
  \bibinfo{author}{\bibfnamefont{D.}~\bibnamefont{Sank}},
  \bibinfo{author}{\bibfnamefont{C.}~\bibnamefont{Quintana}},
  \bibinfo{author}{\bibfnamefont{M.}~\bibnamefont{McEwen}},
  \bibinfo{author}{\bibfnamefont{R.}~\bibnamefont{Barends}},
  \bibinfo{author}{\bibfnamefont{P.~V.} \bibnamefont{Klimov}},
  \bibnamefont{et~al.}, \emph{\bibinfo{title}{Exponential suppression of bit or
  phase errors with cyclic error correction}}, \bibinfo{journal}{Nature}
  \textbf{\bibinfo{volume}{595}}, \bibinfo{pages}{383} (\bibinfo{year}{2021}).

\bibitem[{\citenamefont{Andersen et~al.}(2020)\citenamefont{Andersen, Remm,
  Lazar, Krinner, Lacroix, Norris, Gabureac, Eichler, and
  Wallraff}}]{Andersen2019}
\bibinfo{author}{\bibfnamefont{C.~K.} \bibnamefont{Andersen}},
  \bibinfo{author}{\bibfnamefont{A.}~\bibnamefont{Remm}},
  \bibinfo{author}{\bibfnamefont{S.}~\bibnamefont{Lazar}},
  \bibinfo{author}{\bibfnamefont{S.}~\bibnamefont{Krinner}},
  \bibinfo{author}{\bibfnamefont{N.}~\bibnamefont{Lacroix}},
  \bibinfo{author}{\bibfnamefont{G.~J.} \bibnamefont{Norris}},
  \bibinfo{author}{\bibfnamefont{M.}~\bibnamefont{Gabureac}},
  \bibinfo{author}{\bibfnamefont{C.}~\bibnamefont{Eichler}}, \bibnamefont{and}
  \bibinfo{author}{\bibfnamefont{A.}~\bibnamefont{Wallraff}},
  \emph{\bibinfo{title}{Repeated quantum error detection in a surface code}},
  \bibinfo{journal}{Nature Physics} \textbf{\bibinfo{volume}{16}},
  \bibinfo{pages}{875} (\bibinfo{year}{2020}).

\bibitem[{\citenamefont{Ma et~al.}(2020)\citenamefont{Ma, Zhang, Wong, Noh,
  Rosenblum, Reinhold, Schoelkopf, and Jiang}}]{Ma2020}
\bibinfo{author}{\bibfnamefont{W.-L.} \bibnamefont{Ma}},
  \bibinfo{author}{\bibfnamefont{M.}~\bibnamefont{Zhang}},
  \bibinfo{author}{\bibfnamefont{Y.}~\bibnamefont{Wong}},
  \bibinfo{author}{\bibfnamefont{K.}~\bibnamefont{Noh}},
  \bibinfo{author}{\bibfnamefont{S.}~\bibnamefont{Rosenblum}},
  \bibinfo{author}{\bibfnamefont{P.}~\bibnamefont{Reinhold}},
  \bibinfo{author}{\bibfnamefont{R.~J.} \bibnamefont{Schoelkopf}},
  \bibnamefont{and} \bibinfo{author}{\bibfnamefont{L.}~\bibnamefont{Jiang}},
  \emph{\bibinfo{title}{Path-independent quantum gates with noisy ancilla}},
  \bibinfo{journal}{Physical Review Letters} \textbf{\bibinfo{volume}{125}},
  \bibinfo{pages}{110503} (\bibinfo{year}{2020}).

\bibitem[{\citenamefont{Rosenblum et~al.}(2018)\citenamefont{Rosenblum,
  Reinhold, Mirrahimi, Jiang, Frunzio, and Schoelkopf}}]{Rosenblum2018}
\bibinfo{author}{\bibfnamefont{S.}~\bibnamefont{Rosenblum}},
  \bibinfo{author}{\bibfnamefont{P.}~\bibnamefont{Reinhold}},
  \bibinfo{author}{\bibfnamefont{M.}~\bibnamefont{Mirrahimi}},
  \bibinfo{author}{\bibfnamefont{L.}~\bibnamefont{Jiang}},
  \bibinfo{author}{\bibfnamefont{L.}~\bibnamefont{Frunzio}}, \bibnamefont{and}
  \bibinfo{author}{\bibfnamefont{R.~J.} \bibnamefont{Schoelkopf}},
  \emph{\bibinfo{title}{{Fault-tolerant detection of a quantum error}}},
  \bibinfo{journal}{Science} \textbf{\bibinfo{volume}{361}},
  \bibinfo{pages}{266} (\bibinfo{year}{2018}).

\bibitem[{\citenamefont{Puri et~al.}(2019)\citenamefont{Puri, Grimm,
  Campagne-Ibarcq, Eickbusch, Noh, Roberts, Jiang, Mirrahimi, Devoret, and
  Girvin}}]{Puri2019}
\bibinfo{author}{\bibfnamefont{S.}~\bibnamefont{Puri}},
  \bibinfo{author}{\bibfnamefont{A.}~\bibnamefont{Grimm}},
  \bibinfo{author}{\bibfnamefont{P.}~\bibnamefont{Campagne-Ibarcq}},
  \bibinfo{author}{\bibfnamefont{A.}~\bibnamefont{Eickbusch}},
  \bibinfo{author}{\bibfnamefont{K.}~\bibnamefont{Noh}},
  \bibinfo{author}{\bibfnamefont{G.}~\bibnamefont{Roberts}},
  \bibinfo{author}{\bibfnamefont{L.}~\bibnamefont{Jiang}},
  \bibinfo{author}{\bibfnamefont{M.}~\bibnamefont{Mirrahimi}},
  \bibinfo{author}{\bibfnamefont{M.~H.} \bibnamefont{Devoret}},
  \bibnamefont{and} \bibinfo{author}{\bibfnamefont{S.~M.}
  \bibnamefont{Girvin}}, \emph{\bibinfo{title}{Stabilized cat in a driven
  nonlinear cavity: A fault-tolerant error syndrome detector}},
  \bibinfo{journal}{Phys. Rev. X} \textbf{\bibinfo{volume}{9}},
  \bibinfo{pages}{041009} (\bibinfo{year}{2019}).

\bibitem[{\citenamefont{Gross et~al.}(2018)\citenamefont{Gross, Caves, Milburn,
  and Combes}}]{Gross2018}
\bibinfo{author}{\bibfnamefont{J.~A.} \bibnamefont{Gross}},
  \bibinfo{author}{\bibfnamefont{C.~M.} \bibnamefont{Caves}},
  \bibinfo{author}{\bibfnamefont{G.~J.} \bibnamefont{Milburn}},
  \bibnamefont{and} \bibinfo{author}{\bibfnamefont{J.}~\bibnamefont{Combes}},
  \emph{\bibinfo{title}{Qubit models of weak continuous measurements: Markovian
  conditional and open-system dynamics}}, \bibinfo{journal}{Quantum Science and
  Technology} \textbf{\bibinfo{volume}{3}}, \bibinfo{pages}{024005}
  (\bibinfo{year}{2018}).

\bibitem[{\citenamefont{Pfaff et~al.}(2017)\citenamefont{Pfaff, Axline,
  Burkhart, Vool, Reinhold, Frunzio, Jiang, Devoret, and
  Schoelkopf}}]{Pfaff2017}
\bibinfo{author}{\bibfnamefont{W.}~\bibnamefont{Pfaff}},
  \bibinfo{author}{\bibfnamefont{C.~J.} \bibnamefont{Axline}},
  \bibinfo{author}{\bibfnamefont{L.~D.} \bibnamefont{Burkhart}},
  \bibinfo{author}{\bibfnamefont{U.}~\bibnamefont{Vool}},
  \bibinfo{author}{\bibfnamefont{P.}~\bibnamefont{Reinhold}},
  \bibinfo{author}{\bibfnamefont{L.}~\bibnamefont{Frunzio}},
  \bibinfo{author}{\bibfnamefont{L.}~\bibnamefont{Jiang}},
  \bibinfo{author}{\bibfnamefont{M.~H.} \bibnamefont{Devoret}},
  \bibnamefont{and} \bibinfo{author}{\bibfnamefont{R.~J.}
  \bibnamefont{Schoelkopf}}, \emph{\bibinfo{title}{{Controlled release of
  multiphoton quantum states from a microwave cavity memory}}},
  \bibinfo{journal}{Nature Physics} \textbf{\bibinfo{volume}{13}},
  \bibinfo{pages}{882} (\bibinfo{year}{2017}).

\end{thebibliography}


\begin{thebibliography}{40}
\expandafter\ifx\csname natexlab\endcsname\relax\def\natexlab#1{#1}\fi
\expandafter\ifx\csname bibnamefont\endcsname\relax
  \def\bibnamefont#1{#1}\fi
\expandafter\ifx\csname bibfnamefont\endcsname\relax
  \def\bibfnamefont#1{#1}\fi
\expandafter\ifx\csname citenamefont\endcsname\relax
  \def\citenamefont#1{#1}\fi
\expandafter\ifx\csname url\endcsname\relax
  \def\url#1{\texttt{#1}}\fi
\expandafter\ifx\csname urlprefix\endcsname\relax\def\urlprefix{URL }\fi
\providecommand{\bibinfo}[2]{#2}
\providecommand{\eprint}[2][]{\url{#2}}

\bibitem[{\citenamefont{Axline et~al.}(2016)\citenamefont{Axline, Reagor,
  Heeres, Reinhold, Wang, Shain, Pfaff, Chu, Frunzio, and
  Schoelkopf}}]{Axline2016}
\bibinfo{author}{\bibfnamefont{C.}~\bibnamefont{Axline}},
  \bibinfo{author}{\bibfnamefont{M.}~\bibnamefont{Reagor}},
  \bibinfo{author}{\bibfnamefont{R.}~\bibnamefont{Heeres}},
  \bibinfo{author}{\bibfnamefont{P.}~\bibnamefont{Reinhold}},
  \bibinfo{author}{\bibfnamefont{C.}~\bibnamefont{Wang}},
  \bibinfo{author}{\bibfnamefont{K.}~\bibnamefont{Shain}},
  \bibinfo{author}{\bibfnamefont{W.}~\bibnamefont{Pfaff}},
  \bibinfo{author}{\bibfnamefont{Y.}~\bibnamefont{Chu}},
  \bibinfo{author}{\bibfnamefont{L.}~\bibnamefont{Frunzio}}, \bibnamefont{and}
  \bibinfo{author}{\bibfnamefont{R.~J.} \bibnamefont{Schoelkopf}},
  \emph{\bibinfo{title}{{An architecture for integrating planar and 3D cQED
  devices}}}, \bibinfo{journal}{Applied Physics Letters}
  \textbf{\bibinfo{volume}{109}}, \bibinfo{pages}{042601}
  (\bibinfo{year}{2016}).

\bibitem[{\citenamefont{Rosenblum et~al.}(2018)\citenamefont{Rosenblum,
  Reinhold, Mirrahimi, Jiang, Frunzio, and Schoelkopf}}]{Rosenblum2018}
\bibinfo{author}{\bibfnamefont{S.}~\bibnamefont{Rosenblum}},
  \bibinfo{author}{\bibfnamefont{P.}~\bibnamefont{Reinhold}},
  \bibinfo{author}{\bibfnamefont{M.}~\bibnamefont{Mirrahimi}},
  \bibinfo{author}{\bibfnamefont{L.}~\bibnamefont{Jiang}},
  \bibinfo{author}{\bibfnamefont{L.}~\bibnamefont{Frunzio}}, \bibnamefont{and}
  \bibinfo{author}{\bibfnamefont{R.~J.} \bibnamefont{Schoelkopf}},
  \emph{\bibinfo{title}{{Fault-tolerant detection of a quantum error}}},
  \bibinfo{journal}{Science} \textbf{\bibinfo{volume}{361}},
  \bibinfo{pages}{266} (\bibinfo{year}{2018}).

\bibitem[{\citenamefont{Place et~al.}(2021)\citenamefont{Place, Rodgers,
  Mundada, Smitham, Fitzpatrick, Leng, Premkumar, Bryon, Vrajitoarea, Sussman
  et~al.}}]{Place2021}
\bibinfo{author}{\bibfnamefont{A.~P.} \bibnamefont{Place}},
  \bibinfo{author}{\bibfnamefont{L.~V.} \bibnamefont{Rodgers}},
  \bibinfo{author}{\bibfnamefont{P.}~\bibnamefont{Mundada}},
  \bibinfo{author}{\bibfnamefont{B.~M.} \bibnamefont{Smitham}},
  \bibinfo{author}{\bibfnamefont{M.}~\bibnamefont{Fitzpatrick}},
  \bibinfo{author}{\bibfnamefont{Z.}~\bibnamefont{Leng}},
  \bibinfo{author}{\bibfnamefont{A.}~\bibnamefont{Premkumar}},
  \bibinfo{author}{\bibfnamefont{J.}~\bibnamefont{Bryon}},
  \bibinfo{author}{\bibfnamefont{A.}~\bibnamefont{Vrajitoarea}},
  \bibinfo{author}{\bibfnamefont{S.}~\bibnamefont{Sussman}},
  \bibnamefont{et~al.}, \emph{\bibinfo{title}{New material platform for
  superconducting transmon qubits with coherence times exceeding 0.3
  milliseconds}}, \bibinfo{journal}{Nature communications}
  \textbf{\bibinfo{volume}{12}}, \bibinfo{pages}{1} (\bibinfo{year}{2021}).

\bibitem[{\citenamefont{Read et~al.}(2022)\citenamefont{Read, Chapman, Lei,
  Curtis, Ganjam, Krayzman, Frunzio, and Schoelkopf}}]{Read2022}
\bibinfo{author}{\bibfnamefont{A.~P.} \bibnamefont{Read}},
  \bibinfo{author}{\bibfnamefont{B.~J.} \bibnamefont{Chapman}},
  \bibinfo{author}{\bibfnamefont{C.~U.} \bibnamefont{Lei}},
  \bibinfo{author}{\bibfnamefont{J.~C.} \bibnamefont{Curtis}},
  \bibinfo{author}{\bibfnamefont{S.}~\bibnamefont{Ganjam}},
  \bibinfo{author}{\bibfnamefont{L.}~\bibnamefont{Krayzman}},
  \bibinfo{author}{\bibfnamefont{L.}~\bibnamefont{Frunzio}}, \bibnamefont{and}
  \bibinfo{author}{\bibfnamefont{R.~J.} \bibnamefont{Schoelkopf}},
  \emph{\bibinfo{title}{Precision measurement of the microwave dielectric loss
  of sapphire in the quantum regime with parts-per-billion sensitivity}},
  \bibinfo{journal}{arXiv:2206.14334}  (\bibinfo{year}{2022}).

\bibitem[{\citenamefont{Ofek et~al.}(2016)\citenamefont{Ofek, Petrenko, Heeres,
  Reinhold, Leghtas, Vlastakis, Liu, Frunzio, Girvin, Jiang
  et~al.}}]{Ofek2016a}
\bibinfo{author}{\bibfnamefont{N.}~\bibnamefont{Ofek}},
  \bibinfo{author}{\bibfnamefont{A.}~\bibnamefont{Petrenko}},
  \bibinfo{author}{\bibfnamefont{R.}~\bibnamefont{Heeres}},
  \bibinfo{author}{\bibfnamefont{P.}~\bibnamefont{Reinhold}},
  \bibinfo{author}{\bibfnamefont{Z.}~\bibnamefont{Leghtas}},
  \bibinfo{author}{\bibfnamefont{B.}~\bibnamefont{Vlastakis}},
  \bibinfo{author}{\bibfnamefont{Y.}~\bibnamefont{Liu}},
  \bibinfo{author}{\bibfnamefont{L.}~\bibnamefont{Frunzio}},
  \bibinfo{author}{\bibfnamefont{S.~M.} \bibnamefont{Girvin}},
  \bibinfo{author}{\bibfnamefont{L.}~\bibnamefont{Jiang}},
  \bibnamefont{et~al.}, \emph{\bibinfo{title}{{Extending the lifetime of a
  quantum bit with error correction in superconducting circuits}}},
  \bibinfo{journal}{Nature} \textbf{\bibinfo{volume}{536}},
  \bibinfo{pages}{441} (\bibinfo{year}{2016}).

\bibitem[{\citenamefont{Hu et~al.}(2019)\citenamefont{Hu, Ma, Cai, Mu, Xu,
  Wang, Wu, Wang, Song, Zou et~al.}}]{Hu2019a}
\bibinfo{author}{\bibfnamefont{L.}~\bibnamefont{Hu}},
  \bibinfo{author}{\bibfnamefont{Y.}~\bibnamefont{Ma}},
  \bibinfo{author}{\bibfnamefont{W.}~\bibnamefont{Cai}},
  \bibinfo{author}{\bibfnamefont{X.}~\bibnamefont{Mu}},
  \bibinfo{author}{\bibfnamefont{Y.}~\bibnamefont{Xu}},
  \bibinfo{author}{\bibfnamefont{W.}~\bibnamefont{Wang}},
  \bibinfo{author}{\bibfnamefont{Y.}~\bibnamefont{Wu}},
  \bibinfo{author}{\bibfnamefont{H.}~\bibnamefont{Wang}},
  \bibinfo{author}{\bibfnamefont{Y.~P.} \bibnamefont{Song}},
  \bibinfo{author}{\bibfnamefont{C.-L.} \bibnamefont{Zou}},
  \bibnamefont{et~al.}, \emph{\bibinfo{title}{{Quantum error correction and
  universal gate set operation on a binomial bosonic logical qubit}}},
  \bibinfo{journal}{Nature Physics} \textbf{\bibinfo{volume}{15}},
  \bibinfo{pages}{503} (\bibinfo{year}{2019}).

\bibitem[{\citenamefont{Serniak et~al.}(2019)\citenamefont{Serniak, Diamond,
  Hays, Fatemi, Shankar, Frunzio, Schoelkopf, and Devoret}}]{Serniak2019}
\bibinfo{author}{\bibfnamefont{K.}~\bibnamefont{Serniak}},
  \bibinfo{author}{\bibfnamefont{S.}~\bibnamefont{Diamond}},
  \bibinfo{author}{\bibfnamefont{M.}~\bibnamefont{Hays}},
  \bibinfo{author}{\bibfnamefont{V.}~\bibnamefont{Fatemi}},
  \bibinfo{author}{\bibfnamefont{S.}~\bibnamefont{Shankar}},
  \bibinfo{author}{\bibfnamefont{L.}~\bibnamefont{Frunzio}},
  \bibinfo{author}{\bibfnamefont{R.}~\bibnamefont{Schoelkopf}},
  \bibnamefont{and} \bibinfo{author}{\bibfnamefont{M.}~\bibnamefont{Devoret}},
  \emph{\bibinfo{title}{Direct dispersive monitoring of charge parity in
  offset-charge-sensitive transmons}}, \bibinfo{journal}{Physical Review
  Applied} \textbf{\bibinfo{volume}{12}}, \bibinfo{pages}{014052}
  (\bibinfo{year}{2019}).

\bibitem[{\citenamefont{Sivak et~al.}(2022)\citenamefont{Sivak, Eickbusch, Liu,
  Royer, Tsioutsios, and Devoret}}]{Sivak2021}
\bibinfo{author}{\bibfnamefont{V.~V.} \bibnamefont{Sivak}},
  \bibinfo{author}{\bibfnamefont{A.}~\bibnamefont{Eickbusch}},
  \bibinfo{author}{\bibfnamefont{H.}~\bibnamefont{Liu}},
  \bibinfo{author}{\bibfnamefont{B.}~\bibnamefont{Royer}},
  \bibinfo{author}{\bibfnamefont{I.}~\bibnamefont{Tsioutsios}},
  \bibnamefont{and} \bibinfo{author}{\bibfnamefont{M.~H.}
  \bibnamefont{Devoret}}, \emph{\bibinfo{title}{Model-free quantum control with
  reinforcement learning}}, \bibinfo{journal}{Physical Review X}
  \textbf{\bibinfo{volume}{12}}, \bibinfo{pages}{011059}
  (\bibinfo{year}{2022}).

\bibitem[{\citenamefont{Schuster et~al.}(2007)\citenamefont{Schuster, Houck,
  Schreier, Wallraff, Gambetta, Blais, Frunzio, Majer, Johnson, Devoret
  et~al.}}]{Schuster2007a}
\bibinfo{author}{\bibfnamefont{D.~I.} \bibnamefont{Schuster}},
  \bibinfo{author}{\bibfnamefont{A.~A.} \bibnamefont{Houck}},
  \bibinfo{author}{\bibfnamefont{J.~A.} \bibnamefont{Schreier}},
  \bibinfo{author}{\bibfnamefont{A.}~\bibnamefont{Wallraff}},
  \bibinfo{author}{\bibfnamefont{J.~M.} \bibnamefont{Gambetta}},
  \bibinfo{author}{\bibfnamefont{A.}~\bibnamefont{Blais}},
  \bibinfo{author}{\bibfnamefont{L.}~\bibnamefont{Frunzio}},
  \bibinfo{author}{\bibfnamefont{J.}~\bibnamefont{Majer}},
  \bibinfo{author}{\bibfnamefont{B.}~\bibnamefont{Johnson}},
  \bibinfo{author}{\bibfnamefont{M.~H.} \bibnamefont{Devoret}},
  \bibnamefont{et~al.}, \emph{\bibinfo{title}{{Resolving photon number states
  in a superconducting circuit}}}, \bibinfo{journal}{Nature}
  \textbf{\bibinfo{volume}{445}}, \bibinfo{pages}{515} (\bibinfo{year}{2007}).

\bibitem[{\citenamefont{Eickbusch et~al.}(2021)\citenamefont{Eickbusch, Sivak,
  Ding, Elder, Jha, Venkatraman, Royer, Girvin, Schoelkopf, and
  Devoret}}]{Eickbusch2021}
\bibinfo{author}{\bibfnamefont{A.}~\bibnamefont{Eickbusch}},
  \bibinfo{author}{\bibfnamefont{V.}~\bibnamefont{Sivak}},
  \bibinfo{author}{\bibfnamefont{A.~Z.} \bibnamefont{Ding}},
  \bibinfo{author}{\bibfnamefont{S.~S.} \bibnamefont{Elder}},
  \bibinfo{author}{\bibfnamefont{S.~R.} \bibnamefont{Jha}},
  \bibinfo{author}{\bibfnamefont{J.}~\bibnamefont{Venkatraman}},
  \bibinfo{author}{\bibfnamefont{B.}~\bibnamefont{Royer}},
  \bibinfo{author}{\bibfnamefont{S.~M.} \bibnamefont{Girvin}},
  \bibinfo{author}{\bibfnamefont{R.~J.} \bibnamefont{Schoelkopf}},
  \bibnamefont{and} \bibinfo{author}{\bibfnamefont{M.~H.}
  \bibnamefont{Devoret}}, \emph{\bibinfo{title}{Fast universal control of an
  oscillator with weak dispersive coupling to a qubit}},
  \bibinfo{journal}{arXiv:2111.06414}  (\bibinfo{year}{2021}).

\bibitem[{\citenamefont{Gambetta et~al.}(2006)\citenamefont{Gambetta, Blais,
  Schuster, Wallraff, Frunzio, Majer, Devoret, Girvin, and
  Schoelkopf}}]{Gambetta2006}
\bibinfo{author}{\bibfnamefont{J.}~\bibnamefont{Gambetta}},
  \bibinfo{author}{\bibfnamefont{A.}~\bibnamefont{Blais}},
  \bibinfo{author}{\bibfnamefont{D.~I.} \bibnamefont{Schuster}},
  \bibinfo{author}{\bibfnamefont{A.}~\bibnamefont{Wallraff}},
  \bibinfo{author}{\bibfnamefont{L.}~\bibnamefont{Frunzio}},
  \bibinfo{author}{\bibfnamefont{J.}~\bibnamefont{Majer}},
  \bibinfo{author}{\bibfnamefont{M.~H.} \bibnamefont{Devoret}},
  \bibinfo{author}{\bibfnamefont{S.~M.} \bibnamefont{Girvin}},
  \bibnamefont{and} \bibinfo{author}{\bibfnamefont{R.~J.}
  \bibnamefont{Schoelkopf}}, \emph{\bibinfo{title}{Qubit-photon interactions in
  a cavity: Measurement-induced dephasing and number splitting}},
  \bibinfo{journal}{Physical Review A} \textbf{\bibinfo{volume}{74}},
  \bibinfo{pages}{042318} (\bibinfo{year}{2006}).

\bibitem[{\citenamefont{Walter et~al.}(2017)\citenamefont{Walter, Kurpiers,
  Gasparinetti, Magnard, Poto{\v{c}}nik, Salath{\'{e}}, Pechal, Mondal,
  Oppliger, Eichler et~al.}}]{Walter2017}
\bibinfo{author}{\bibfnamefont{T.}~\bibnamefont{Walter}},
  \bibinfo{author}{\bibfnamefont{P.}~\bibnamefont{Kurpiers}},
  \bibinfo{author}{\bibfnamefont{S.}~\bibnamefont{Gasparinetti}},
  \bibinfo{author}{\bibfnamefont{P.}~\bibnamefont{Magnard}},
  \bibinfo{author}{\bibfnamefont{A.}~\bibnamefont{Poto{\v{c}}nik}},
  \bibinfo{author}{\bibfnamefont{Y.}~\bibnamefont{Salath{\'{e}}}},
  \bibinfo{author}{\bibfnamefont{M.}~\bibnamefont{Pechal}},
  \bibinfo{author}{\bibfnamefont{M.}~\bibnamefont{Mondal}},
  \bibinfo{author}{\bibfnamefont{M.}~\bibnamefont{Oppliger}},
  \bibinfo{author}{\bibfnamefont{C.}~\bibnamefont{Eichler}},
  \bibnamefont{et~al.}, \emph{\bibinfo{title}{{Rapid high-fidelity single-shot
  dispersive readout of superconducting qubits}}}, \bibinfo{journal}{Physical
  Review Applied} \textbf{\bibinfo{volume}{7}}, \bibinfo{pages}{054020}
  (\bibinfo{year}{2017}).

\bibitem[{\citenamefont{McClure et~al.}(2016)\citenamefont{McClure, Paik,
  Bishop, Steffen, Chow, and Gambetta}}]{McClure2016}
\bibinfo{author}{\bibfnamefont{D.~T.} \bibnamefont{McClure}},
  \bibinfo{author}{\bibfnamefont{H.}~\bibnamefont{Paik}},
  \bibinfo{author}{\bibfnamefont{L.~S.} \bibnamefont{Bishop}},
  \bibinfo{author}{\bibfnamefont{M.}~\bibnamefont{Steffen}},
  \bibinfo{author}{\bibfnamefont{J.~M.} \bibnamefont{Chow}}, \bibnamefont{and}
  \bibinfo{author}{\bibfnamefont{J.~M.} \bibnamefont{Gambetta}},
  \emph{\bibinfo{title}{Rapid driven reset of a qubit readout resonator}},
  \bibinfo{journal}{Physical Review Applied} \textbf{\bibinfo{volume}{5}},
  \bibinfo{pages}{011001} (\bibinfo{year}{2016}).

\bibitem[{\citenamefont{Campagne-Ibarcq
  et~al.}(2020)\citenamefont{Campagne-Ibarcq, Eickbusch, Touzard, Zalys-Geller,
  Frattini, Sivak, Reinhold, Puri, Shankar, Schoelkopf
  et~al.}}]{Campagne-Ibarcq2020}
\bibinfo{author}{\bibfnamefont{P.}~\bibnamefont{Campagne-Ibarcq}},
  \bibinfo{author}{\bibfnamefont{A.}~\bibnamefont{Eickbusch}},
  \bibinfo{author}{\bibfnamefont{S.}~\bibnamefont{Touzard}},
  \bibinfo{author}{\bibfnamefont{E.}~\bibnamefont{Zalys-Geller}},
  \bibinfo{author}{\bibfnamefont{N.~E.} \bibnamefont{Frattini}},
  \bibinfo{author}{\bibfnamefont{V.~V.} \bibnamefont{Sivak}},
  \bibinfo{author}{\bibfnamefont{P.}~\bibnamefont{Reinhold}},
  \bibinfo{author}{\bibfnamefont{S.}~\bibnamefont{Puri}},
  \bibinfo{author}{\bibfnamefont{S.}~\bibnamefont{Shankar}},
  \bibinfo{author}{\bibfnamefont{R.~J.} \bibnamefont{Schoelkopf}},
  \bibnamefont{et~al.}, \emph{\bibinfo{title}{{Quantum error correction of a
  qubit encoded in grid states of an oscillator}}}, \bibinfo{journal}{Nature}
  \textbf{\bibinfo{volume}{584}}, \bibinfo{pages}{368} (\bibinfo{year}{2020}).

\bibitem[{\citenamefont{Touzard et~al.}(2019)\citenamefont{Touzard, Kou,
  Frattini, Sivak, Puri, Grimm, Frunzio, Shankar, and Devoret}}]{Touzard2018}
\bibinfo{author}{\bibfnamefont{S.}~\bibnamefont{Touzard}},
  \bibinfo{author}{\bibfnamefont{A.}~\bibnamefont{Kou}},
  \bibinfo{author}{\bibfnamefont{N.~E.} \bibnamefont{Frattini}},
  \bibinfo{author}{\bibfnamefont{V.~V.} \bibnamefont{Sivak}},
  \bibinfo{author}{\bibfnamefont{S.}~\bibnamefont{Puri}},
  \bibinfo{author}{\bibfnamefont{A.}~\bibnamefont{Grimm}},
  \bibinfo{author}{\bibfnamefont{L.}~\bibnamefont{Frunzio}},
  \bibinfo{author}{\bibfnamefont{S.}~\bibnamefont{Shankar}}, \bibnamefont{and}
  \bibinfo{author}{\bibfnamefont{M.~H.} \bibnamefont{Devoret}},
  \emph{\bibinfo{title}{{Gated conditional displacement readout of
  superconducting qubits}}}, \bibinfo{journal}{Physical Review Letters}
  \textbf{\bibinfo{volume}{122}}, \bibinfo{pages}{080502}
  (\bibinfo{year}{2019}).

\bibitem[{\citenamefont{Sank et~al.}(2016)\citenamefont{Sank, Chen, Khezri,
  Kelly, Barends, Campbell, Chen, Chiaro, Dunsworth, Fowler et~al.}}]{Sank2016}
\bibinfo{author}{\bibfnamefont{D.}~\bibnamefont{Sank}},
  \bibinfo{author}{\bibfnamefont{Z.}~\bibnamefont{Chen}},
  \bibinfo{author}{\bibfnamefont{M.}~\bibnamefont{Khezri}},
  \bibinfo{author}{\bibfnamefont{J.}~\bibnamefont{Kelly}},
  \bibinfo{author}{\bibfnamefont{R.}~\bibnamefont{Barends}},
  \bibinfo{author}{\bibfnamefont{B.}~\bibnamefont{Campbell}},
  \bibinfo{author}{\bibfnamefont{Y.}~\bibnamefont{Chen}},
  \bibinfo{author}{\bibfnamefont{B.}~\bibnamefont{Chiaro}},
  \bibinfo{author}{\bibfnamefont{A.}~\bibnamefont{Dunsworth}},
  \bibinfo{author}{\bibfnamefont{A.}~\bibnamefont{Fowler}},
  \bibnamefont{et~al.}, \emph{\bibinfo{title}{Measurement-induced state
  transitions in a superconducting qubit: beyond the rotating wave
  approximation}}, \bibinfo{journal}{Physical Review Letters}
  \textbf{\bibinfo{volume}{117}}, \bibinfo{pages}{190503}
  (\bibinfo{year}{2016}).

\bibitem[{\citenamefont{Shillito et~al.}(2022)\citenamefont{Shillito, Petrescu,
  Cohen, Beall, Hauru, Ganahl, Lewis, Vidal, and Blais}}]{Shillito2022}
\bibinfo{author}{\bibfnamefont{R.}~\bibnamefont{Shillito}},
  \bibinfo{author}{\bibfnamefont{A.}~\bibnamefont{Petrescu}},
  \bibinfo{author}{\bibfnamefont{J.}~\bibnamefont{Cohen}},
  \bibinfo{author}{\bibfnamefont{J.}~\bibnamefont{Beall}},
  \bibinfo{author}{\bibfnamefont{M.}~\bibnamefont{Hauru}},
  \bibinfo{author}{\bibfnamefont{M.}~\bibnamefont{Ganahl}},
  \bibinfo{author}{\bibfnamefont{A.~G.} \bibnamefont{Lewis}},
  \bibinfo{author}{\bibfnamefont{G.}~\bibnamefont{Vidal}}, \bibnamefont{and}
  \bibinfo{author}{\bibfnamefont{A.}~\bibnamefont{Blais}},
  \emph{\bibinfo{title}{Dynamics of transmon ionization}},
  \bibinfo{journal}{Physical Review Applied} \textbf{\bibinfo{volume}{18}},
  \bibinfo{pages}{034031} (\bibinfo{year}{2022}).

\bibitem[{\citenamefont{Klimov et~al.}(2018)\citenamefont{Klimov, Kelly, Chen,
  Neeley, Megrant, Burkett, Barends, Arya, Chiaro, Chen et~al.}}]{Klimov2018}
\bibinfo{author}{\bibfnamefont{P.~V.} \bibnamefont{Klimov}},
  \bibinfo{author}{\bibfnamefont{J.}~\bibnamefont{Kelly}},
  \bibinfo{author}{\bibfnamefont{Z.}~\bibnamefont{Chen}},
  \bibinfo{author}{\bibfnamefont{M.}~\bibnamefont{Neeley}},
  \bibinfo{author}{\bibfnamefont{A.}~\bibnamefont{Megrant}},
  \bibinfo{author}{\bibfnamefont{B.}~\bibnamefont{Burkett}},
  \bibinfo{author}{\bibfnamefont{R.}~\bibnamefont{Barends}},
  \bibinfo{author}{\bibfnamefont{K.}~\bibnamefont{Arya}},
  \bibinfo{author}{\bibfnamefont{B.}~\bibnamefont{Chiaro}},
  \bibinfo{author}{\bibfnamefont{Y.}~\bibnamefont{Chen}}, \bibnamefont{et~al.},
  \emph{\bibinfo{title}{Fluctuations of energy-relaxation times in
  superconducting qubits}}, \bibinfo{journal}{Physical Review Letters}
  \textbf{\bibinfo{volume}{121}}, \bibinfo{pages}{90502}
  (\bibinfo{year}{2018}).

\bibitem[{\citenamefont{Reagor et~al.}(2016)\citenamefont{Reagor, Pfaff,
  Axline, Heeres, Ofek, Sliwa, Holland, Wang, Blumoff, Chou
  et~al.}}]{Reagor2016}
\bibinfo{author}{\bibfnamefont{M.}~\bibnamefont{Reagor}},
  \bibinfo{author}{\bibfnamefont{W.}~\bibnamefont{Pfaff}},
  \bibinfo{author}{\bibfnamefont{C.}~\bibnamefont{Axline}},
  \bibinfo{author}{\bibfnamefont{R.~W.} \bibnamefont{Heeres}},
  \bibinfo{author}{\bibfnamefont{N.}~\bibnamefont{Ofek}},
  \bibinfo{author}{\bibfnamefont{K.}~\bibnamefont{Sliwa}},
  \bibinfo{author}{\bibfnamefont{E.}~\bibnamefont{Holland}},
  \bibinfo{author}{\bibfnamefont{C.}~\bibnamefont{Wang}},
  \bibinfo{author}{\bibfnamefont{J.}~\bibnamefont{Blumoff}},
  \bibinfo{author}{\bibfnamefont{K.}~\bibnamefont{Chou}}, \bibnamefont{et~al.},
  \emph{\bibinfo{title}{{Quantum memory with millisecond coherence in circuit
  QED}}}, \bibinfo{journal}{Physical Review B} \textbf{\bibinfo{volume}{94}},
  \bibinfo{pages}{014506} (\bibinfo{year}{2016}).

\bibitem[{\citenamefont{de~Neeve et~al.}(2022)\citenamefont{de~Neeve, Nguyen,
  Behrle, and Home}}]{DeNeeve2020}
\bibinfo{author}{\bibfnamefont{B.}~\bibnamefont{de~Neeve}},
  \bibinfo{author}{\bibfnamefont{T.-L.} \bibnamefont{Nguyen}},
  \bibinfo{author}{\bibfnamefont{T.}~\bibnamefont{Behrle}}, \bibnamefont{and}
  \bibinfo{author}{\bibfnamefont{J.~P.} \bibnamefont{Home}},
  \emph{\bibinfo{title}{Error correction of a logical grid state qubit by
  dissipative pumping}}, \bibinfo{journal}{Nature Physics}
  \textbf{\bibinfo{volume}{18}}, \bibinfo{pages}{296} (\bibinfo{year}{2022}).

\bibitem[{\citenamefont{Gross et~al.}(2018)\citenamefont{Gross, Caves, Milburn,
  and Combes}}]{Gross2018}
\bibinfo{author}{\bibfnamefont{J.~A.} \bibnamefont{Gross}},
  \bibinfo{author}{\bibfnamefont{C.~M.} \bibnamefont{Caves}},
  \bibinfo{author}{\bibfnamefont{G.~J.} \bibnamefont{Milburn}},
  \bibnamefont{and} \bibinfo{author}{\bibfnamefont{J.}~\bibnamefont{Combes}},
  \emph{\bibinfo{title}{Qubit models of weak continuous measurements: Markovian
  conditional and open-system dynamics}}, \bibinfo{journal}{Quantum Science and
  Technology} \textbf{\bibinfo{volume}{3}}, \bibinfo{pages}{024005}
  (\bibinfo{year}{2018}).

\bibitem[{\citenamefont{Royer et~al.}(2020)\citenamefont{Royer, Singh, and
  Girvin}}]{Royer2020}
\bibinfo{author}{\bibfnamefont{B.}~\bibnamefont{Royer}},
  \bibinfo{author}{\bibfnamefont{S.}~\bibnamefont{Singh}}, \bibnamefont{and}
  \bibinfo{author}{\bibfnamefont{S.~M.} \bibnamefont{Girvin}},
  \emph{\bibinfo{title}{{Stabilization of finite-energy
  Gottesman-Kitaev-Preskill states}}}, \bibinfo{journal}{Physical Review
  Letters} \textbf{\bibinfo{volume}{125}}, \bibinfo{pages}{260509}
  (\bibinfo{year}{2020}).

\bibitem[{Cho()}]{Chollet2015}
\bibinfo{howpublished}{\url{https://keras.io}}.

\bibitem[{Siv()}]{Sivak2022github}
\bibinfo{howpublished}{\url{https://github.com/v-sivak/tf_quantum_simulator}}.

\bibitem[{\citenamefont{Schulman et~al.}(2017)\citenamefont{Schulman, Wolski,
  Dhariwal, Radford, and Klimov}}]{Schulman2017}
\bibinfo{author}{\bibfnamefont{J.}~\bibnamefont{Schulman}},
  \bibinfo{author}{\bibfnamefont{F.}~\bibnamefont{Wolski}},
  \bibinfo{author}{\bibfnamefont{P.}~\bibnamefont{Dhariwal}},
  \bibinfo{author}{\bibfnamefont{A.}~\bibnamefont{Radford}}, \bibnamefont{and}
  \bibinfo{author}{\bibfnamefont{O.}~\bibnamefont{Klimov}},
  \emph{\bibinfo{title}{{Proximal policy optimization algorithms}}},
  \bibinfo{journal}{arXiv:1707.06347}  (\bibinfo{year}{2017}).

\bibitem[{\citenamefont{Guadarrama et~al.}(2018)\citenamefont{Guadarrama,
  Korattikara, Ramirez, Castro, Holly, Fishman, Wang, Gonina, Wu, Kokiopoulou
  et~al.}}]{TFAgents}
\bibinfo{author}{\bibfnamefont{S.}~\bibnamefont{Guadarrama}},
  \bibinfo{author}{\bibfnamefont{A.}~\bibnamefont{Korattikara}},
  \bibinfo{author}{\bibfnamefont{O.}~\bibnamefont{Ramirez}},
  \bibinfo{author}{\bibfnamefont{P.}~\bibnamefont{Castro}},
  \bibinfo{author}{\bibfnamefont{E.}~\bibnamefont{Holly}},
  \bibinfo{author}{\bibfnamefont{S.}~\bibnamefont{Fishman}},
  \bibinfo{author}{\bibfnamefont{K.}~\bibnamefont{Wang}},
  \bibinfo{author}{\bibfnamefont{E.}~\bibnamefont{Gonina}},
  \bibinfo{author}{\bibfnamefont{N.}~\bibnamefont{Wu}},
  \bibinfo{author}{\bibfnamefont{E.}~\bibnamefont{Kokiopoulou}},
  \bibnamefont{et~al.}, \emph{\bibinfo{title}{{TF-Agents}: A library for
  reinforcement learning in tensorflow}},
  \bibinfo{howpublished}{\url{https://github.com/tensorflow/agents}}
  (\bibinfo{year}{2018}).

\bibitem[{\citenamefont{Tzitrin et~al.}(2020)\citenamefont{Tzitrin, Bourassa,
  Menicucci, and Sabapathy}}]{Tzitrin2019}
\bibinfo{author}{\bibfnamefont{I.}~\bibnamefont{Tzitrin}},
  \bibinfo{author}{\bibfnamefont{J.~E.} \bibnamefont{Bourassa}},
  \bibinfo{author}{\bibfnamefont{N.~C.} \bibnamefont{Menicucci}},
  \bibnamefont{and} \bibinfo{author}{\bibfnamefont{K.~K.}
  \bibnamefont{Sabapathy}}, \emph{\bibinfo{title}{{Progress towards practical
  qubit computation using approximate Gottesman-Kitaev-Preskill codes}}},
  \bibinfo{journal}{Physical Review A} \textbf{\bibinfo{volume}{101}},
  \bibinfo{pages}{032315} (\bibinfo{year}{2020}).

\bibitem[{\citenamefont{Knill and Laflamme}(1997)}]{Knill1997}
\bibinfo{author}{\bibfnamefont{E.}~\bibnamefont{Knill}} \bibnamefont{and}
  \bibinfo{author}{\bibfnamefont{R.}~\bibnamefont{Laflamme}},
  \emph{\bibinfo{title}{Theory of quantum error-correcting codes}},
  \bibinfo{journal}{Phys. Rev. A} \textbf{\bibinfo{volume}{55}},
  \bibinfo{pages}{900} (\bibinfo{year}{1997}).

\bibitem[{\citenamefont{Gottesman}(1996)}]{Gottesman1996}
\bibinfo{author}{\bibfnamefont{D.}~\bibnamefont{Gottesman}},
  \emph{\bibinfo{title}{{Class of quantum error-correcting codes saturating the
  quantum Hamming bound}}}, \bibinfo{journal}{Phys. Rev. A}
  \textbf{\bibinfo{volume}{54}}, \bibinfo{pages}{1862} (\bibinfo{year}{1996}).

\bibitem[{\citenamefont{Murch et~al.}(2012)\citenamefont{Murch, Vool, Zhou,
  Weber, Girvin, and Siddiqi}}]{Murch2012}
\bibinfo{author}{\bibfnamefont{K.~W.} \bibnamefont{Murch}},
  \bibinfo{author}{\bibfnamefont{U.}~\bibnamefont{Vool}},
  \bibinfo{author}{\bibfnamefont{D.}~\bibnamefont{Zhou}},
  \bibinfo{author}{\bibfnamefont{S.~J.} \bibnamefont{Weber}},
  \bibinfo{author}{\bibfnamefont{S.~M.} \bibnamefont{Girvin}},
  \bibnamefont{and} \bibinfo{author}{\bibfnamefont{I.}~\bibnamefont{Siddiqi}},
  \emph{\bibinfo{title}{{Cavity-assisted quantum bath engineering}}},
  \bibinfo{journal}{Physical Review Letters} \textbf{\bibinfo{volume}{109}},
  \bibinfo{pages}{1} (\bibinfo{year}{2012}).

\bibitem[{\citenamefont{Geerlings et~al.}(2013)\citenamefont{Geerlings,
  Leghtas, Pop, Shankar, Frunzio, Schoelkopf, Mirrahimi, and
  Devoret}}]{Geerlings2013}
\bibinfo{author}{\bibfnamefont{K.}~\bibnamefont{Geerlings}},
  \bibinfo{author}{\bibfnamefont{Z.}~\bibnamefont{Leghtas}},
  \bibinfo{author}{\bibfnamefont{I.~M.} \bibnamefont{Pop}},
  \bibinfo{author}{\bibfnamefont{S.}~\bibnamefont{Shankar}},
  \bibinfo{author}{\bibfnamefont{L.}~\bibnamefont{Frunzio}},
  \bibinfo{author}{\bibfnamefont{R.~J.} \bibnamefont{Schoelkopf}},
  \bibinfo{author}{\bibfnamefont{M.}~\bibnamefont{Mirrahimi}},
  \bibnamefont{and} \bibinfo{author}{\bibfnamefont{M.~H.}
  \bibnamefont{Devoret}}, \emph{\bibinfo{title}{{Demonstrating a driven reset
  protocol for a superconducting qubit}}}, \bibinfo{journal}{Physical Review
  Letters} \textbf{\bibinfo{volume}{110}}, \bibinfo{pages}{1}
  (\bibinfo{year}{2013}).

\bibitem[{\citenamefont{Magnard et~al.}(2018)\citenamefont{Magnard, Kurpiers,
  Royer, Walter, Besse, Gasparinetti, Pechal, Heinsoo, Storz, Blais
  et~al.}}]{Magnard2018}
\bibinfo{author}{\bibfnamefont{P.}~\bibnamefont{Magnard}},
  \bibinfo{author}{\bibfnamefont{P.}~\bibnamefont{Kurpiers}},
  \bibinfo{author}{\bibfnamefont{B.}~\bibnamefont{Royer}},
  \bibinfo{author}{\bibfnamefont{T.}~\bibnamefont{Walter}},
  \bibinfo{author}{\bibfnamefont{J.~C.} \bibnamefont{Besse}},
  \bibinfo{author}{\bibfnamefont{S.}~\bibnamefont{Gasparinetti}},
  \bibinfo{author}{\bibfnamefont{M.}~\bibnamefont{Pechal}},
  \bibinfo{author}{\bibfnamefont{J.}~\bibnamefont{Heinsoo}},
  \bibinfo{author}{\bibfnamefont{S.}~\bibnamefont{Storz}},
  \bibinfo{author}{\bibfnamefont{A.}~\bibnamefont{Blais}},
  \bibnamefont{et~al.}, \emph{\bibinfo{title}{{Fast and unconditional
  all-microwave reset of a superconducting qubit}}}, \bibinfo{journal}{Physical
  Review Letters} \textbf{\bibinfo{volume}{121}}, \bibinfo{pages}{60502}
  (\bibinfo{year}{2018}), \eprint{1801.07689}.

\bibitem[{\citenamefont{Chen et~al.}(2016)\citenamefont{Chen, Kelly, Quintana,
  Barends, Campbell, Chen, Chiaro, Dunsworth, Fowler, Lucero
  et~al.}}]{Chen2016}
\bibinfo{author}{\bibfnamefont{Z.}~\bibnamefont{Chen}},
  \bibinfo{author}{\bibfnamefont{J.}~\bibnamefont{Kelly}},
  \bibinfo{author}{\bibfnamefont{C.}~\bibnamefont{Quintana}},
  \bibinfo{author}{\bibfnamefont{R.}~\bibnamefont{Barends}},
  \bibinfo{author}{\bibfnamefont{B.}~\bibnamefont{Campbell}},
  \bibinfo{author}{\bibfnamefont{Y.}~\bibnamefont{Chen}},
  \bibinfo{author}{\bibfnamefont{B.}~\bibnamefont{Chiaro}},
  \bibinfo{author}{\bibfnamefont{A.}~\bibnamefont{Dunsworth}},
  \bibinfo{author}{\bibfnamefont{A.~G.} \bibnamefont{Fowler}},
  \bibinfo{author}{\bibfnamefont{E.}~\bibnamefont{Lucero}},
  \bibnamefont{et~al.}, \emph{\bibinfo{title}{Measuring and suppressing quantum
  state leakage in a superconducting qubit}}, \bibinfo{journal}{Physical Review
  Letters} \textbf{\bibinfo{volume}{116}}, \bibinfo{pages}{020501}
  (\bibinfo{year}{2016}).

\bibitem[{\citenamefont{McEwen et~al.}(2021)\citenamefont{McEwen, Kafri, Chen,
  Atalaya, Satzinger, Quintana, Klimov, Sank, Gidney, Fowler
  et~al.}}]{McEwen2021}
\bibinfo{author}{\bibfnamefont{M.}~\bibnamefont{McEwen}},
  \bibinfo{author}{\bibfnamefont{D.}~\bibnamefont{Kafri}},
  \bibinfo{author}{\bibfnamefont{Z.}~\bibnamefont{Chen}},
  \bibinfo{author}{\bibfnamefont{J.}~\bibnamefont{Atalaya}},
  \bibinfo{author}{\bibfnamefont{K.}~\bibnamefont{Satzinger}},
  \bibinfo{author}{\bibfnamefont{C.}~\bibnamefont{Quintana}},
  \bibinfo{author}{\bibfnamefont{P.~V.} \bibnamefont{Klimov}},
  \bibinfo{author}{\bibfnamefont{D.}~\bibnamefont{Sank}},
  \bibinfo{author}{\bibfnamefont{C.}~\bibnamefont{Gidney}},
  \bibinfo{author}{\bibfnamefont{A.}~\bibnamefont{Fowler}},
  \bibnamefont{et~al.}, \emph{\bibinfo{title}{Removing leakage-induced
  correlated errors in superconducting quantum error correction}},
  \bibinfo{journal}{Nature communications} \textbf{\bibinfo{volume}{12}},
  \bibinfo{pages}{1} (\bibinfo{year}{2021}).

\bibitem[{\citenamefont{Vlastakis et~al.}(2013)\citenamefont{Vlastakis,
  Kirchmair, Leghtas, Nigg, Frunzio, Girvin, Mirrahimi, Devoret, and
  Schoelkopf}}]{Vlastakis2013}
\bibinfo{author}{\bibfnamefont{B.}~\bibnamefont{Vlastakis}},
  \bibinfo{author}{\bibfnamefont{G.}~\bibnamefont{Kirchmair}},
  \bibinfo{author}{\bibfnamefont{Z.}~\bibnamefont{Leghtas}},
  \bibinfo{author}{\bibfnamefont{S.~E.} \bibnamefont{Nigg}},
  \bibinfo{author}{\bibfnamefont{L.}~\bibnamefont{Frunzio}},
  \bibinfo{author}{\bibfnamefont{S.~M.} \bibnamefont{Girvin}},
  \bibinfo{author}{\bibfnamefont{M.}~\bibnamefont{Mirrahimi}},
  \bibinfo{author}{\bibfnamefont{M.~H.} \bibnamefont{Devoret}},
  \bibnamefont{and} \bibinfo{author}{\bibfnamefont{R.~J.}
  \bibnamefont{Schoelkopf}}, \emph{\bibinfo{title}{{Deterministically encoding
  quantum information using 100-photon Schrodinger cat states}}},
  \bibinfo{journal}{Science} \textbf{\bibinfo{volume}{342}},
  \bibinfo{pages}{607} (\bibinfo{year}{2013}).

\bibitem[{\citenamefont{Clerk et~al.}(2010)\citenamefont{Clerk, Devoret,
  Girvin, Marquardt, and Schoelkopf}}]{Clerk2010}
\bibinfo{author}{\bibfnamefont{A.~A.} \bibnamefont{Clerk}},
  \bibinfo{author}{\bibfnamefont{M.~H.} \bibnamefont{Devoret}},
  \bibinfo{author}{\bibfnamefont{S.~M.} \bibnamefont{Girvin}},
  \bibinfo{author}{\bibfnamefont{F.}~\bibnamefont{Marquardt}},
  \bibnamefont{and} \bibinfo{author}{\bibfnamefont{R.~J.}
  \bibnamefont{Schoelkopf}}, \emph{\bibinfo{title}{{Introduction to quantum
  noise, measurement, and amplification}}}, \bibinfo{journal}{Reviews of Modern
  Physics} \textbf{\bibinfo{volume}{82}}, \bibinfo{pages}{1155}
  (\bibinfo{year}{2010}).

\bibitem[{\citenamefont{Carroll et~al.}(2021)\citenamefont{Carroll, Rosenblatt,
  Jurcevic, Lauer, and Kandala}}]{Carroll2021}
\bibinfo{author}{\bibfnamefont{M.}~\bibnamefont{Carroll}},
  \bibinfo{author}{\bibfnamefont{S.}~\bibnamefont{Rosenblatt}},
  \bibinfo{author}{\bibfnamefont{P.}~\bibnamefont{Jurcevic}},
  \bibinfo{author}{\bibfnamefont{I.}~\bibnamefont{Lauer}}, \bibnamefont{and}
  \bibinfo{author}{\bibfnamefont{A.}~\bibnamefont{Kandala}},
  \emph{\bibinfo{title}{{Dynamics of superconducting qubit relaxation times}}},
  \bibinfo{journal}{arXiv:2105.15201}  (\bibinfo{year}{2021}).

\bibitem[{\citenamefont{Greenbaum}(2015)}]{Greenbaum2015}
\bibinfo{author}{\bibfnamefont{D.}~\bibnamefont{Greenbaum}},
  \emph{\bibinfo{title}{{Introduction to quantum gate set tomography}}},
  \bibinfo{journal}{arXiv:1509.02921}  (\bibinfo{year}{2015}).

\bibitem[{\citenamefont{Nielsen}(2002)}]{Nielsen2002}
\bibinfo{author}{\bibfnamefont{M.~A.} \bibnamefont{Nielsen}},
  \emph{\bibinfo{title}{{A simple formula for the average gate fidelity of a
  quantum dynamical operation}}}, \bibinfo{journal}{Physics Letters A}
  \textbf{\bibinfo{volume}{303}}, \bibinfo{pages}{249} (\bibinfo{year}{2002}).

\bibitem[{\citenamefont{Gertler et~al.}(2021)\citenamefont{Gertler, Baker, Li,
  Shirol, Koch, and Wang}}]{Gertler2020}
\bibinfo{author}{\bibfnamefont{J.~M.} \bibnamefont{Gertler}},
  \bibinfo{author}{\bibfnamefont{B.}~\bibnamefont{Baker}},
  \bibinfo{author}{\bibfnamefont{J.}~\bibnamefont{Li}},
  \bibinfo{author}{\bibfnamefont{S.}~\bibnamefont{Shirol}},
  \bibinfo{author}{\bibfnamefont{J.}~\bibnamefont{Koch}}, \bibnamefont{and}
  \bibinfo{author}{\bibfnamefont{C.}~\bibnamefont{Wang}},
  \emph{\bibinfo{title}{{Protecting a bosonic qubit with autonomous quantum
  error correction}}}, \bibinfo{journal}{Nature}
  \textbf{\bibinfo{volume}{590}}, \bibinfo{pages}{243} (\bibinfo{year}{2021}).

\end{thebibliography}
\end{document}